\documentclass[letterpaper,twocolumn,10pt]{article}
\usepackage{usenix2020}

\usepackage{algpseudocode,algorithm,algorithmicx}
\usepackage[T1]{fontenc}
\usepackage{amsmath,amssymb,amsfonts}
\usepackage{amsthm}
\usepackage{bm}
\usepackage{bbm}
\usepackage{booktabs}
\usepackage{enumerate}
\usepackage{graphicx}
\usepackage{multirow}
\usepackage{subcaption}
\usepackage{url}
\usepackage{xcolor}
\usepackage{xspace}
\usepackage{enumitem}
\usepackage{txfonts}
\usepackage{csquotes}

\newcommand{\eg}{\textit{e}.\textit{g}.\@\xspace}
\newcommand{\ie}{\textit{i}.\textit{e}.\@\xspace}
\newcommand{\etal}{\textit{et al}.\@\xspace}

\begin{document}
\title{\Large \bf Membership Inference Attacks and Defenses in Neural Network Pruning}
\author{\rm Xiaoyong Yuan, Lan Zhang\\
Michigan Technological University}
\maketitle
\begin{abstract}
Neural network pruning has been an essential technique to reduce the computation and memory requirements for using deep neural networks for resource-constrained devices. Most existing research focuses primarily on balancing the sparsity and accuracy of a pruned neural network by strategically removing insignificant parameters and retraining the pruned model. Such efforts on reusing training samples pose serious privacy risks due to increased memorization, which, however, has not been investigated yet. 

In this paper, we conduct the first analysis of privacy risks in neural network pruning. Specifically, we investigate the impacts of neural network pruning on training data privacy, \ie, membership inference attacks. We first explore the impact of neural network pruning on prediction divergence, where the pruning process disproportionately affects the pruned model’s behavior for members and non-members. Meanwhile, the influence of divergence even varies among different classes in a fine-grained manner. Enlightened by such divergence, we proposed a self-attention membership inference attack against the pruned neural networks. Extensive experiments are conducted to rigorously evaluate the privacy impacts of different pruning approaches, sparsity levels, and adversary knowledge. The proposed attack shows the higher attack performance on the pruned models when compared with eight existing membership inference attacks. 
In addition, we propose a new defense mechanism to protect the pruning process by mitigating the prediction divergence based on KL-divergence distance, whose effectiveness has been experimentally demonstrated to effectively mitigate the privacy risks while maintaining the sparsity and accuracy of the pruned models.

\end{abstract}
\section{Introduction}
Much of the progress in artificial intelligence over the past decade has been the result of deep neural networks (DNNs). The powerful DNNs with a large number of parameters consume considerable storage and memory bandwidth, which makes it challenging to deploy the state-of-the-art neural networks on resource-constrained devices. 
To address this issue, neural network pruning as one of the most popular compression technologies has attracted great attention~\cite{mozer1989skeletonization,han2015deep}. 
By removing insignificant parameters from a DNN, recent research has shown that neural network pruning can substantially reduce the size of a DNN and speedup the inference process without largely compromising prediction accuracy~\cite{han2015deep,li2016pruning,liu2017learning,blalock2020state}. 
In general, neural network pruning includes three main stages: 1) train an original DNN; 2) remove the insignificant parameters; 3) fine-tune the remaining parameters with the training dataset. Most existing research on neural network pruning has focused on improving the trade-off between accuracy and sparsity by strategically designing the last two stages~\cite{han2015deep,li2016pruning,liu2017learning,blalock2020state}. However, such efforts on reusing training samples pose serious privacy risks of the pruned neural networks due to the potentially increased memorization of training samples. %

\begin{figure}[!t]
    \centering
    \begin{subfigure}[b]{0.49\linewidth}
    \centering
    \includegraphics[width=\linewidth]{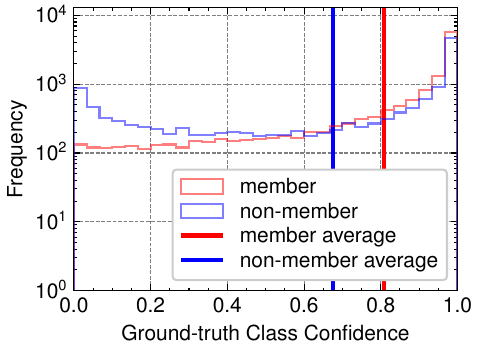}
    \caption{Original model confidence}
    \label{fig:big_conf}
    \end{subfigure}
    \begin{subfigure}[b]{0.49\linewidth}
    \centering
    \includegraphics[width=\linewidth]{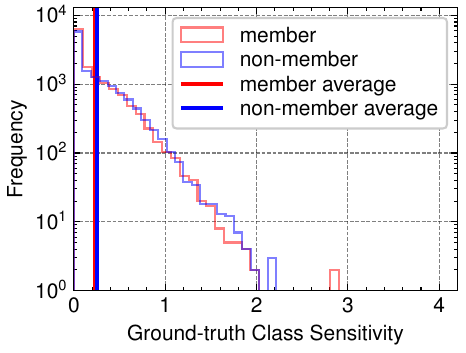}
    \caption{Original model sensitivity}
    \label{fig:big_sens}
    \end{subfigure}
    \begin{subfigure}[b]{0.49\linewidth}
    \centering
    \includegraphics[width=\linewidth]{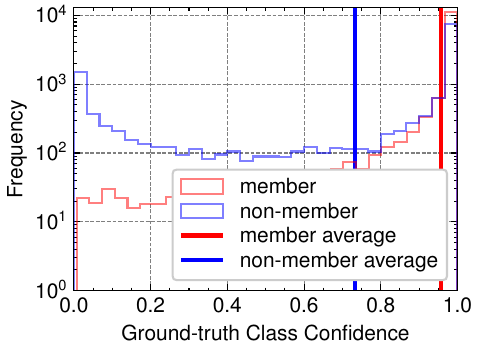}
    \caption{Pruned model confidence}
    \label{fig:prune_conf}
    \end{subfigure}
    \begin{subfigure}[b]{0.49\linewidth}
    \centering
    \includegraphics[width=\linewidth]{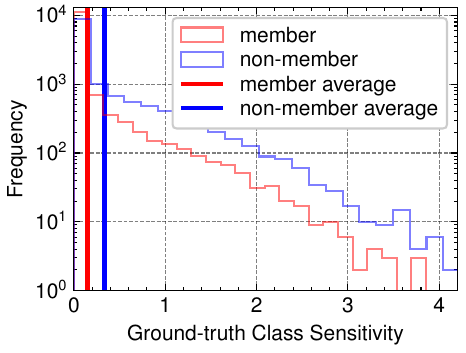}
    \caption{Pruned model sensitivity}
    \label{fig:prune_sens}
    \end{subfigure}
    
    \caption{Histograms of the prediction confidences and the prediction sensitivity of the ground-truth label. We remove 70\% of the parameters in the original DenseNet121 model using l1 unstructured pruning on the CIFAR10 dataset. The figures show the frequency of prediction confidence (a) and (c) and prediction sensitivity (b) and (d) belonging to the ground-truth class on the training and test data. The vertical lines indicate the average values of training data, \ie, members (black), and test data, \ie, non-members (red), respectively. In both prediction confidence and sensitivity measurements, neural network pruning makes the distances between the two vertical lines in the pruned model larger than that in the original model, which indicates a larger confidence gap and sensitivity gap between members and non-members due to pruning.}
    \label{fig:histogram}
    \vspace{-1em}

\end{figure}

The privacy risks of DNNs have already been pointed out, where a DNN is prone to memorizing sensitive information of the training dataset~\cite{nasr2019comprehensive,hayes2019logan,chen2020gan,li2020membership}. Taking the membership inference attack (MIA) as an example, an adversary can infer whether a given data sample was used to train a DNN, seriously threatening individual privacy. For instance, an adversary can infer an individual was a confirmed case, if it is known that the individual's record was used to train an infectious disease model. The MIA was first proposed against black-box models in~\cite{shokri2017membership}, where the adversary only has access to the data sample and predictions of the target model. Later on, more attention has been attracted against various DNN models, such as generative models~\cite{hayes2019logan,chen2020gan}, graph models~\cite{olatunji2101membership}, machine translation~\cite{hisamoto2020membership}, text generation~\cite{song2019auditing}, genomic analysis~\cite{chen2020differential}, and transfer learning~\cite{zou2020privacy}. Although extensive analysis has been conducted, none of the existing efforts have been put into analyzing MIAs against pruned neural networks. 

In view of this, the paper focuses on one fundamental question: {\textit{comparing with original deep neural networks, are the pruned networks more vulnerable to membership inference attacks?}} Specifically, most MIAs infer a sample's membership based on the different behaviors of a target model between members (\ie, training samples) and non-members (\ie, test samples), such as the different prediction confidences~\cite{shokri2017membership,li2020membership}. Since most neural network pruning approaches rely on reusing the training dataset to fine-tune the parameters after pruning the insignificant parameters, the additional training at the pruned neural network inevitably increases its memorization of the training samples. Moreover, the pruned neural network enforces a small number of parameters to achieve similar prediction capabilities, which also increases the memorization of training data and makes the pruned model more sensitive to the training data. Hence, such increased memorization can intuitively lead to a larger divergence of the prediction confidences and sensitivities between members and non-members. 
Figure~\ref{fig:histogram} illustrates the prediction confidence and the prediction sensitivity\footnote{The definitions of the prediction confidence and the prediction sensitivity are detailed respectively in Section~\ref{4.1}.} of members and non-members in the original DNN and the pruned network, respectively. 
The larger divergence of the confidences and the sensitivities in the pruned model at (c) and (d) confirms our intuition: \textit{neural network pruning can aggravate the privacy issues of the original deep neural network}. Therefore, in the following paper, we conduct a comprehensive analysis to reveal the impacts of neural network pruning on training data privacy, \ie, MIAs. Specifically, we first explore the impact of neural network pruning on prediction divergence: \textit{the pruning process disproportionately affects the pruned model's behavior for members and non-members}.
Enlightened by this insight, a new MIA is proposed against the pruned neural networks. In addition, with the proposed new attack, we propose a new defense mechanism to protect the fine-tuning process by mitigating the prediction divergence based on KL-divergence distance. Extensive experiments are conducted to rigorously evaluate our proposals. To the best of our knowledge, this is the first study to investigate the privacy risks of neural network pruning. Our main contributions are summarized below:

\begin{itemize}\setlength{\itemsep}{0.1em} \setlength{\parskip}{0.1em}
\item We investigate the privacy risk of neural network pruning and propose a new MIA: self-attention membership inference attack (SAMIA). By exploring the impacts of neural network pruning on prediction divergence, the proposed attack results in high attack accuracy of revealing the membership status from the pruned models. In particular, SAMIA has advantages in identifying the pruned models' prediction divergence by using finer-grained prediction metrics. We recommend SAMIA as a competitive baseline attack model for future privacy risk study of neural network pruning.
\item To rigorously evaluate the privacy impacts of different pruning approaches, sparsity levels, and adversary knowledge, we conduct extensive experiments on seven commonly used datasets, four neural network architectures, four pruning approaches, five sparsity levels, and 255 pruned models in total. Experimental results demonstrate the effectiveness of the proposed attacks against pruned neural networks, which further indicates that neural network pruning can aggravate the privacy issues of the original DNN. The adversary can successfully reveal the membership status, even without the knowledge of the pruning approach used in the target model. Furthermore, we evaluate the privacy impacts of different pruning approaches and various sparsity levels. 
\item %
To defend the pruned models against MIAs, we propose a new defense mechanism: pair-based posterior balancing (PPB). PPB protects the fine-tuning process of neural network pruning by narrowing down the divergences of posterior predictions and reducing the prediction sensitivities based on their KL-divergence distances. 
Experimental results demonstrate the effectiveness of the PPB mechanism, which significantly mitigates the privacy risks while maintaining the sparsity and accuracy of the pruned model. 
Besides, compared with the state-of-the-art defenses, PPB achieves a better trade-off between prediction performance and privacy in most cases.
\end{itemize}

\section{Background and Related Work}
\label{sec:back}

\subsection{Neural Network Pruning}
The state-of-the-art neural networks are usually deep and resource hungry, requiring large amounts of computation and memory, which becomes a particular challenge on resource-constrained end devices. As one of the most popular network compression approaches, neural network pruning has attracted great attention in recent years~\cite{han2015deep,li2016pruning,liu2017learning,blalock2020state}. In general, most network pruning studies follow the pruning workflow: "train-prune-finetuning." For example, Han~\etal~\cite{han2015deep} proposed to remove the individual parameters with the lowest magnitude. Randomly removing individual parameters reduces the model size, but may not be efficient to facilitate hardware optimization and accelerate the neural network computation. Therefore, many methods were proposed to remove parameters in an organized way by removing a group of parameters (\ie, structured pruning). For example, Li~\etal~\cite{li2016pruning} removed the entire filters with the lowest magnitude in the neural network, which leads to significant speedup compared with the unstructured pruning.
Liu~\etal~\cite{liu2017learning} removed the entire channels according to the corresponding scaling factors in the followed batch normalization layers. In this paper, we investigate the privacy risks of both unstructured and structured pruning approaches.

More recently, new pruning approaches have been proposed, which prune parameters by searching the optimal neural architecture~\cite{cai2019once,li2020eagleeye} or fine-tune the pruned model by rewinding the parameters to the previous states~\cite{frankle2018lottery,renda2020comparing}. The privacy risks discussed in this paper might exist in these new pruning approaches. 
We will investigate their privacy risks in our future work.

On the other hand, recent efforts have been put into neural network pruning from other important perspectives. Paganini~\cite{paganini2020prune} investigated the unfairness and systematic biases in the pruned models. Hooker \etal~\cite{hooker2019compressed} demonstrated the biased performance on different groups and classes after pruning. Given the potential of pervasively implementing neural network pruning, this work targets another critical and urgent aspect regarding neural network pruning, \ie,  training data privacy. %

\vspace{-0.5em}
\subsection{Membership Inference Attacks (MIAs)}
Membership inference attacks have raised serious privacy threats by determining if a record was in the training dataset of a neural network model via querying that model. Given a target neural network model $f: \mathbb{R}^n \rightarrow \mathbb{R}$, the process of MIA can be formally defined as:
\vspace{-0.5em}
\begin{equation}
    \mathcal{A}: \bm{x}, f \rightarrow \{0, 1\},\vspace{-0.5em}
\end{equation}
where $\mathcal{A}$ denotes the attack model, which is a binary classifier. If the data sample $\bm{x}$ is used to train the target model $f$ , the attack model $\mathcal{A}$ outputs $1$ (\ie, member), and $0$ otherwise (\ie, non-member).

Due to the practical consideration, most MIAs focused on the black-box setting, where an adversary only has access to the target model's outputs. By leveraging the target model's prediction confidences, Shokri~\etal,~\cite{shokri2015privacy} proposed a black-box MIA. They constructed several shadow models to mimic the behavior of a target model. The well-established shadow models will then be used to generate data to train a neural network-based binary classifier to determine the membership of a record against the target model, \ie, whether a record belongs to the target model's training dataset or not. Salem~\etal,~\cite{salem2019ml} further boosted this attack successfully by only using a single shadow model. To further improve the attack accuracy, Nasr~\etal,~\cite{nasr2018machine} included more features, such as the class labels of data samples, to train the binary classifier. In addition to the aforementioned neural network-based binary classifier, Leino~\etal,~\cite{leino2020stolen}, Yeom~\etal,~\cite{yeom2018privacy}, and Song~\etal,~\cite{song2019privacy,song2020systematic} proposed the metric-based binary classifier, where the membership of a record is directly determined by a predefined threshold based on the metrics, such as the prediction confidences, entropy, or modified entropy of the record. 
Song and Mittal showed that by setting a class-dependent threshold, the metric-based classifier could achieve comparable or even better inference performance compared with the neural network-based classifier~\cite{song2020systematic}.
Despite the extensive research on MIAs, none of them is designed towards pruned models. Therefore, we propose SAMIA to investigate the privacy risks of pruned models.

\vspace{-0.5em}
\subsection{Defenses against MIAs}
Recent efforts have been made to defend against MIAs. As one of the most popular privacy-preserving techniques, differential privacy (DP) provides provable defense against MIAs by adding noise to the gradient or parameter during model training~\cite{dwork2008differential,dwork2006calibrating,abadi2016deep}. However, DP usually requires a large magnitude of noises to achieve a meaningful privacy guarantee, which seriously degrades the performance of the protected models~\cite{jayaraman2019evaluating}. 
On the other hand, regularization~\cite{shokri2017membership}, dropout, and model stacking~\cite{salem2019ml} have been used in model training to reduce the privacy risks caused by overfitting. Although these approaches reduced the vulnerability by bridging the generalization gap between member and non-member data samples, in many cases, the privacy risks after applying these approaches are still high.
Recent adversarial learning techniques~\cite{goodfellow2014explaining,yuan2019adversarial} 
have been introduced in defending against MIAs by adding noises to the prediction confidences for misleading the adversary~\cite{nasr2018machine,jia2019memguard}.
In a recent analysis of the defense mechanisms, Song and Mittal showed that the early stopping mechanism achieved comparable performance with most defenses~\cite{song2020systematic}. 
In this paper, we provide a comprehensive analysis of defenses in neural network pruning, including our proposed PPB defense along with the existing defense mechanisms.

\section{System Overview} %

\begin{figure}[!t]
    \centering
    \includegraphics[width=\linewidth]{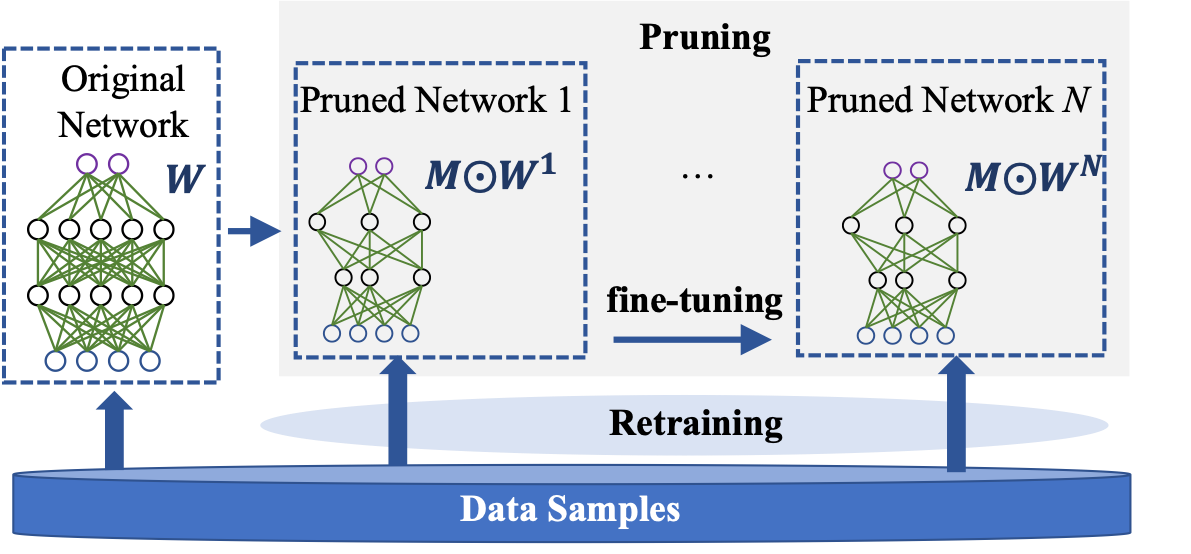}
    
    \caption{{A typical workflow of neural network pruning}.}
    \label{fig:pipeline}
    
\end{figure}

\subsection{Neural Network Pruning Workflow}\label{S3.1}
This paper is focused on a general neural network pruning process, whose workflow includes three key stages: original network training, coarse pruning, and fine-tuning, as illustrated in Figure~\ref{fig:pipeline}. Specifically,
\vspace{-0.3em}
\begin{enumerate}[leftmargin=*]\setlength{\itemsep}{0.1em} \setlength{\parskip}{0.1em}
\item \textit{Original network training}: A large size original neural network model $f(\bm{x}; \bm{W})$ (sometime over-parameterized) is first trained at this stage, where $\bm{x}$ is the training data and $\bm{W}$ is the model parameters;
\item \textit{Pruning}: Upon the original network, the pruning is conducted by removing insignificant parameters or groups of parameters according to a specific criterion. The pruned network can be given by $f(\bm{x}; \bm{M}\odot \bm{W})$, where $\bm{M}\in {0, 1}^{|\bm{W}|}$ denotes the binary mask that can set a parameter to be 0, and $\odot$ denotes the element-wise multiplication; 
\item \textit{Fine-tuning}: To recover the performance loss due to pruning, a pruned network can be fine-tuned by reusing the training data. After $N$-epoch fine-tuning, a pruned network can be given by $f(\bm{x}; \bm{M}\odot \bm{W}^N)$. 
\end{enumerate}
\vspace{-0.3em}

For the sake of simplicity, we use $f$ to denote the original model $f(\bm{x}; \bm{W})$ and $f_p$ to denote the pruned model $f(\bm{x}; M\odot \bm{W}^N)$ in the following paper.

\subsection{Adversarial Knowledge}
The goal of MIAs is to find the membership of a data sample, \ie, whether the sample is used to train a target model or not.  In this paper, we assume the adversary of the MIAs against a pruned neural network has the following knowledge.

\begin{itemize}[leftmargin=*] \setlength{\itemsep}{0.1em} \setlength{\parskip}{0.1em}
\item \textit{Access to query the pruned network}. The pruned model is made available to the public, \ie, queryable. Due to practical considerations, the original model is assumed not published and inaccessible.
\item \textit{Access to the prediction confidences}. We consider the practical black-box MIAs~\cite{nasr2019comprehensive}. The adversary can only acquire the output, \ie, the prediction confidences, of the pruned network. Any internal information about the pruned model and the original model, such as the network architecture and activation functions, are inaccessible to the adversary. %
\item \textit{Access to the pruning approach and the sparsity level}. We consider two different types of adversaries with or without knowledge of the pruning approach and the sparsity level. 
\item \textit{Access to the defense approach}.
The arms race between attacks and defenses is one main challenge in machine learning privacy. If the defense mechanisms are designed without considering the adversary's knowledge, their performance might be substantially degraded when adaptive attacks are used against those defensive mechanisms~\cite{jia2019memguard,song2020systematic}. Hence, we consider both non-adaptive and adaptive attacks to evaluate defense mechanisms: 1) non-adaptive attacks, \ie, the adversary has no access to the defense mechanisms; 2) adaptive attacks, where the adversary has full knowledge of the defense mechanisms and performs the MIAs by taking the defensive mechanisms into account.

\end{itemize}

\section{MIA against Neural Network Pruning}
\label{sec:attack}
\begin{figure}[!t]
    \centering
    \includegraphics[width=0.95\linewidth]{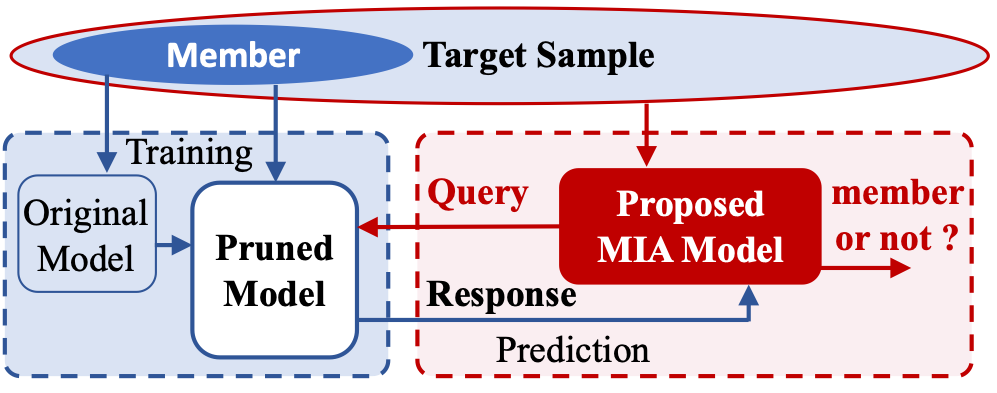}
    
    \caption{Framework of membership inference attacks (MIA) against neural network pruning.}
    \label{fig:attack_pipeline}
    
\end{figure}

Given the workflow of neural network pruning presented in Section~\ref{S3.1}, this section focuses on investigating the privacy risks introduced by the pruning process. A general framework of MIAs against the pruned model is illustrated in Figure~\ref{fig:attack_pipeline}. Specifically, to extract the membership information from the pruned model, the adversary first derives the predictions of the given input sample by querying the target pruned model. The adversary then feeds the predictions into the trained attack model and provides the binary classification of the membership status.
The attack model is derived following the shadow-training technique, which was originally proposed by Shokri \etal~\cite{shokri2015privacy} and is widely used in MIAs~\cite{salem2019ml,nasr2018machine}: a shadow model is trained and pruned to imitate the behavior of the target pruned model. The adversary trains an attack model based on the pruned shadow model's predictions over shadow training and test data.

\begin{figure}[!t]
\centering
\begin{subfigure}[b]{0.48\linewidth}
\includegraphics[width=1.0\linewidth]{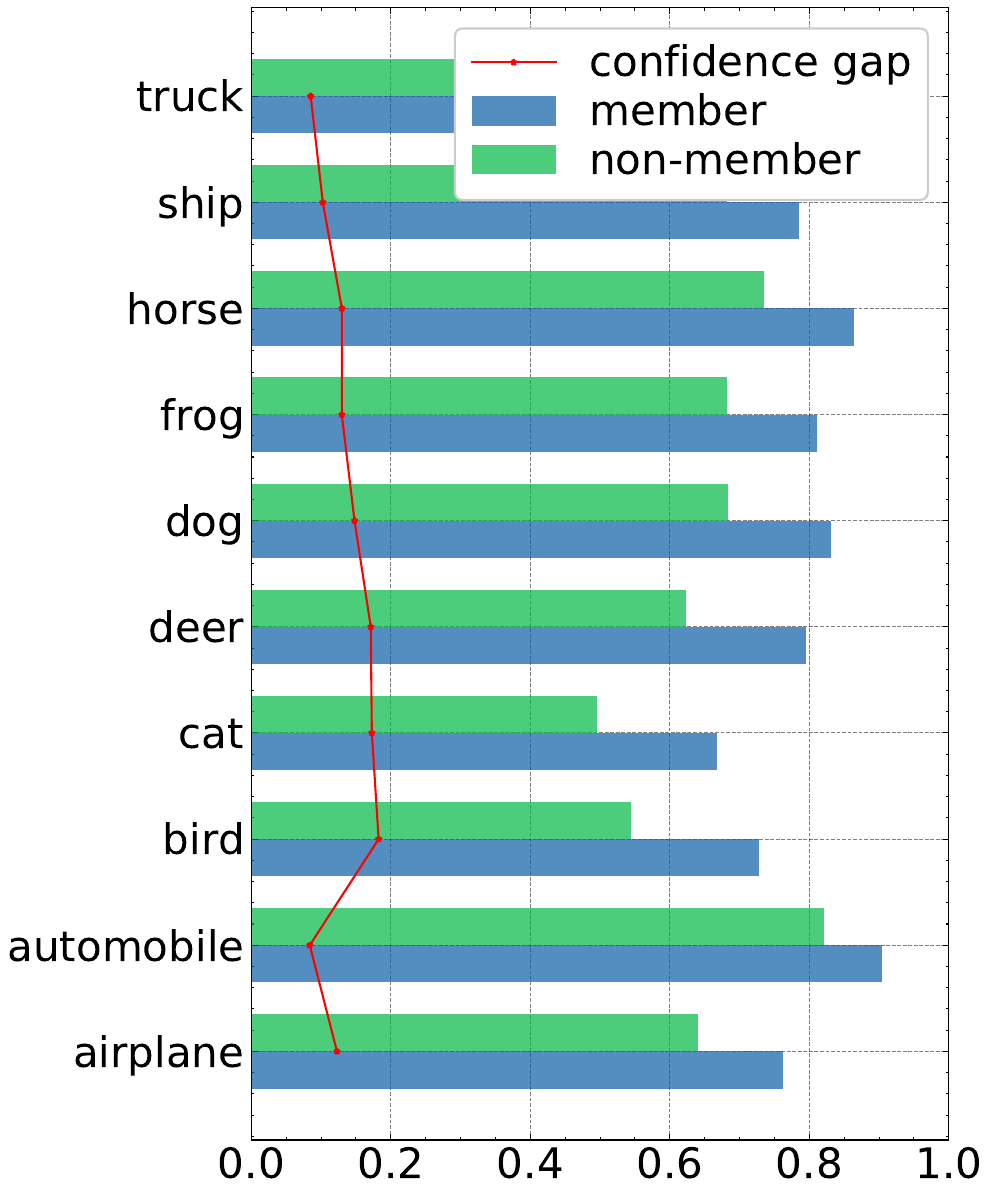}

\caption{Original Model Confidence}
\end{subfigure}
\begin{subfigure}[b]{0.48\linewidth}
\includegraphics[width=1.0\linewidth]{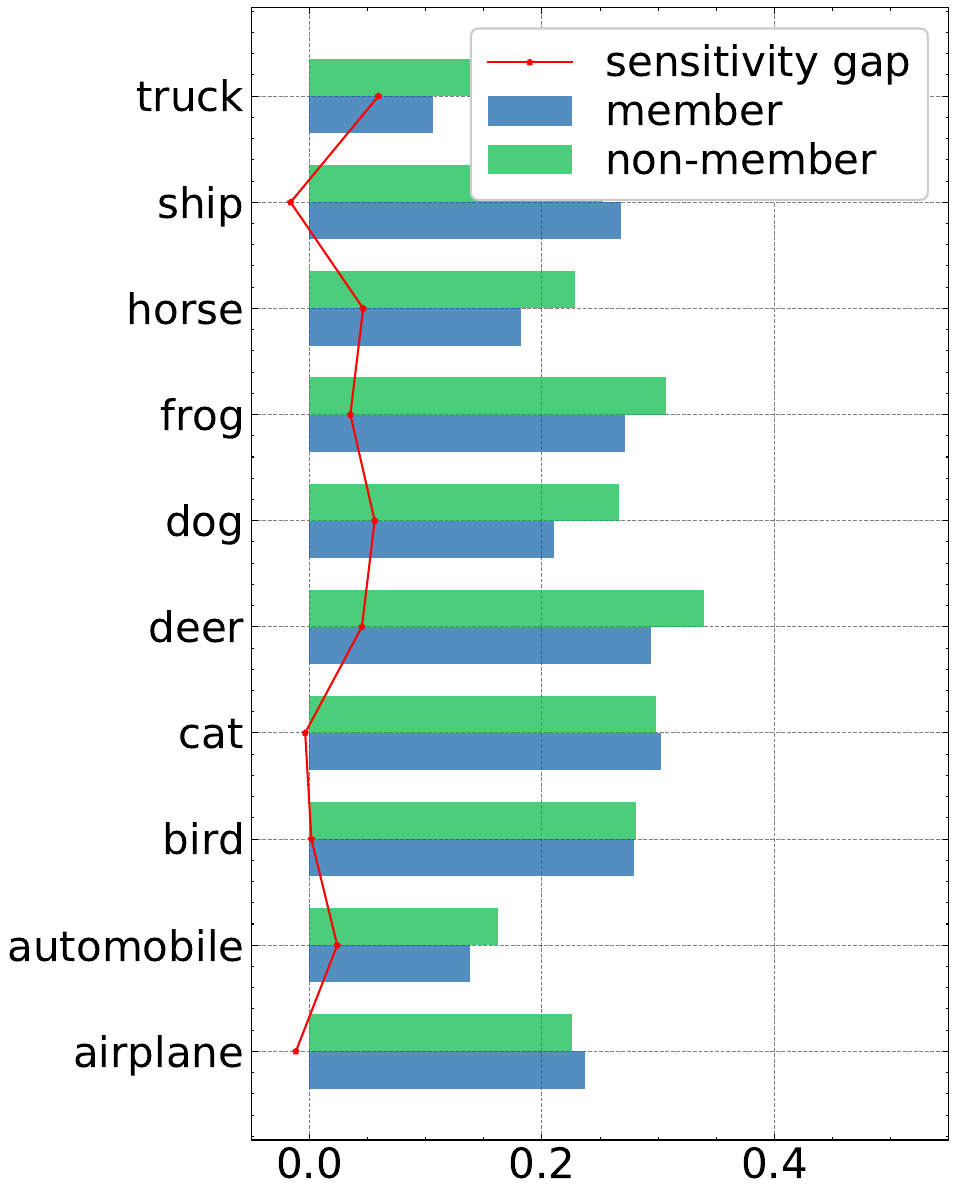}

\caption{Original Model Sensitivity}
\end{subfigure}
\begin{subfigure}[b]{0.48\linewidth}
\includegraphics[width=1.0\linewidth]{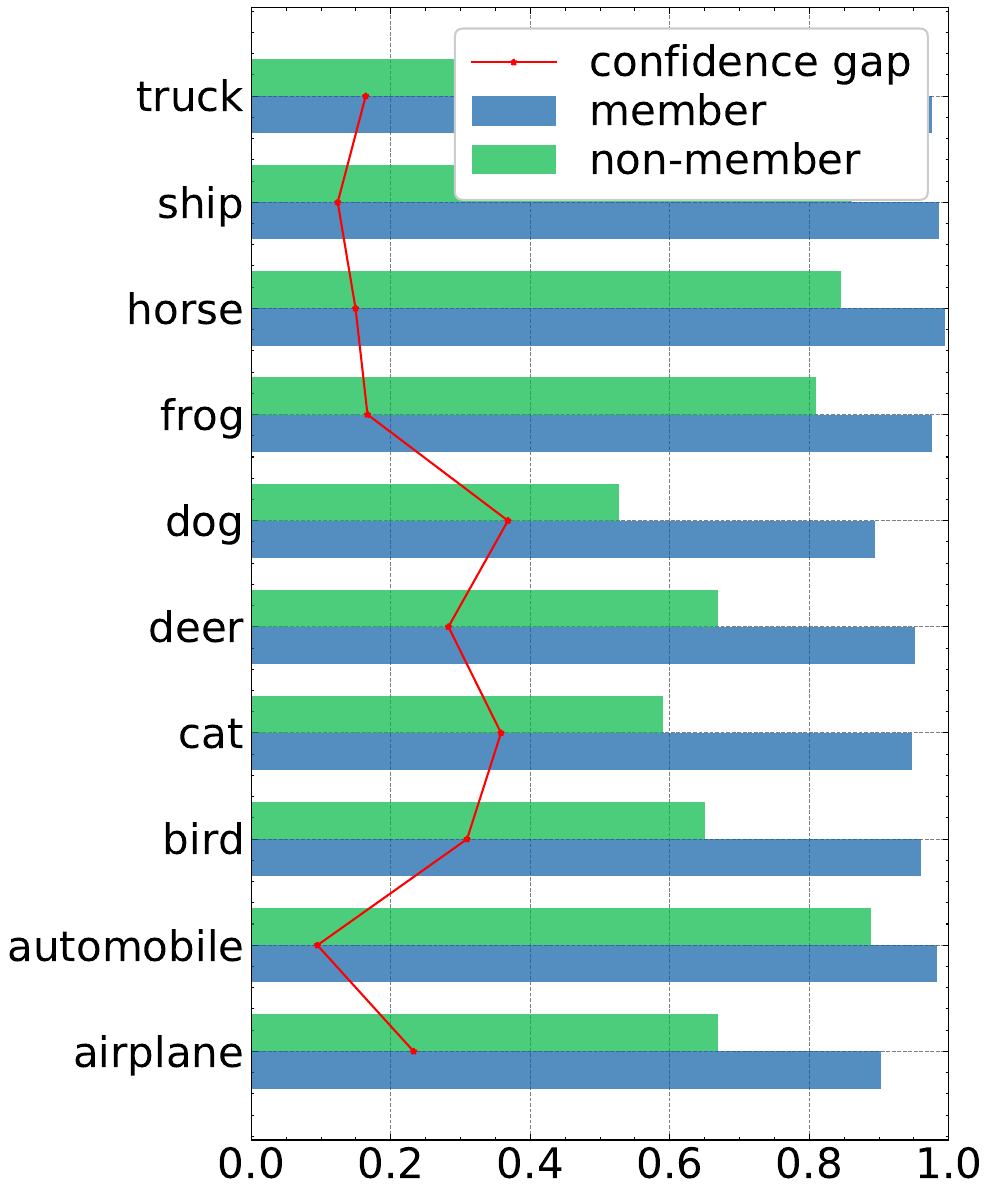}

\caption{{Pruned Model Confidence}}
\end{subfigure}
\begin{subfigure}[b]{0.48\linewidth}
\includegraphics[width=1.0\linewidth]{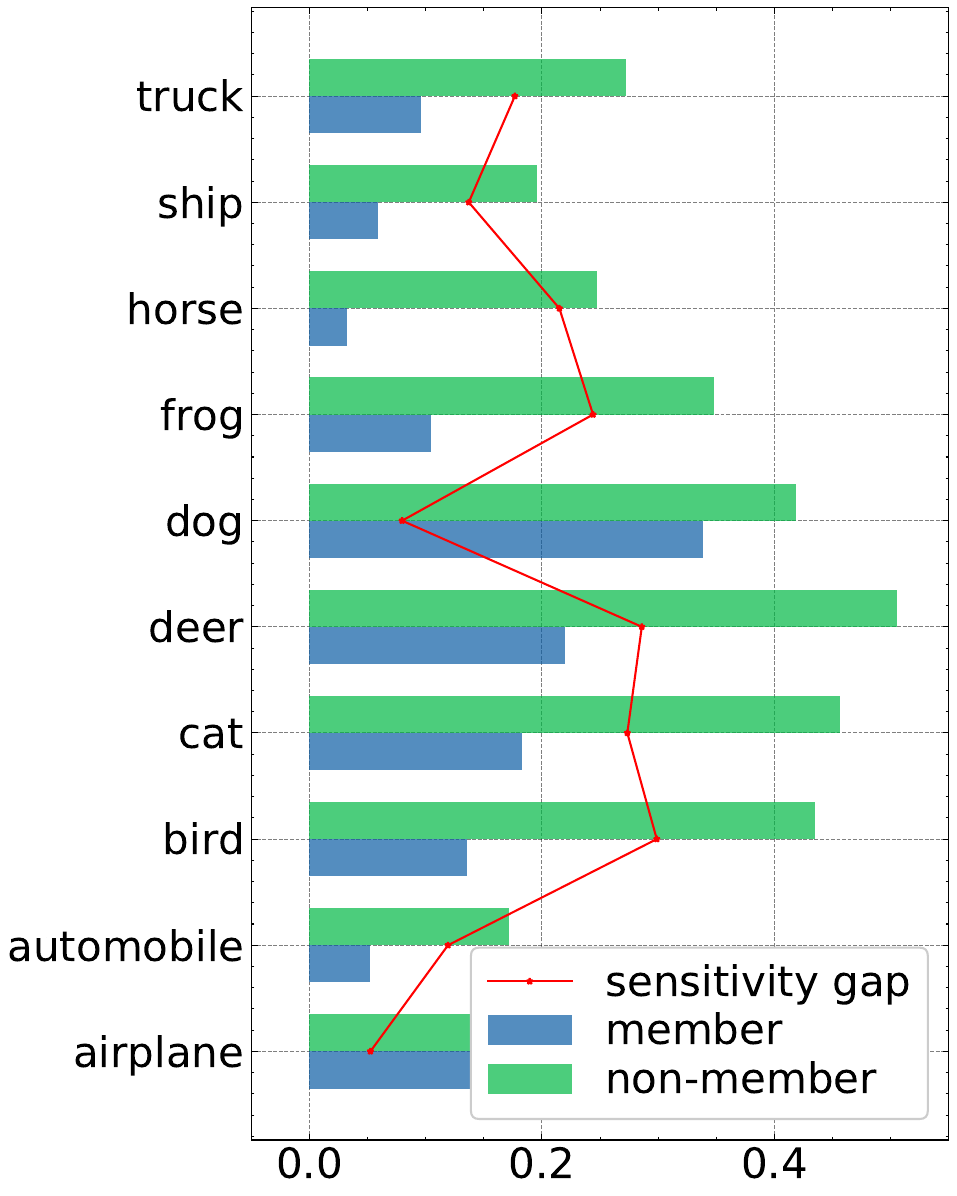}

\caption{Pruned Model Sensitivity}
\end{subfigure}

\caption{Divergence of the pruned model's prediction confidences and prediction sensitivities over different classes, respectively. We prune DenseNet121 models with 70\% sparsity on the CIFAR10 datasets. The blue bar indicates the average prediction confidence/sensitivity of members in different classes. The green bar indicates the average prediction confidence/sensitivity of non-members. The divergence of confidence (a) and sensitivity (b) between members and non-members is increased after pruning as shown in (c) and (d), respectively. Such divergence also differs among classes.}
\label{fig:cls_div}
\end{figure}

\subsection{Divergence of Prediction Behaviors}\label{4.1}
To investigate the prediction behaviors of pruned neural networks, we first introduce two metrics: prediction confidence and prediction sensitivity. Specifically, given an input sample $\bm{x}$ and a pruned model $f_p$, the prediction confidence is defined as $\mathrm{PC} = f_p(\bm{x})$. 
In addition, to further measure the prediction behavior changes in terms of slight input change, we introduce prediction sensitivity, which is defined as\vspace{-0.2em}
\begin{equation}\label{eq:sensitivity}
    \mathrm{PS} = \frac{1}{n} \sum_{i=1}^n  \frac{|f_p(\bm{x} + \epsilon\bm{\delta_i}) - f_p(\bm{x})|}{\epsilon},
\end{equation}
where $\bm{\delta_i} \sim \mathcal{N}(0, 1)$ is a random Gaussian noise vector added to the input data $\bm{x}$, and  $\epsilon$ controls the magnitude of input changes. %
A similar idea has been used in the gradient estimation for black-box adversarial attacks~\cite{chen2017zoo}. It has been shown that a small number of noise vectors can achieve a good estimation of prediction changes, so that we set a small query budget in the evaluation ($n=10$)~\cite{chen2017zoo,bhagoji2018practical}. 
Accordingly, we use the confidence and sensitivity to measure the divergence between members and non-members. We define the confidence gap as\vspace{-0.2em}
\begin{equation}
    \frac{1}{|\mathcal{D}_{train}|}\sum_{(\bm{x_i}, y_i)\in \mathcal{D}_{train}}f_p^{y_i}({\bm{x_i}}) - \frac{1}{|\mathcal{D}_{test}|}\sum_{(\bm{x_i}, y_i)\in \mathcal{D}_{test}}f_p^{y_i}(\bm{x_i}),
\end{equation}
where $f_p^{y_i}$ denotes the prediction confidence of ground-truth class $y_i$. 
Confidence gap calculates the difference of average confidence between members and non-members in the ground-truth class.
Similarly, we define the sensitivity gap as\vspace{-0.2em}
\begin{equation}
    \frac{1}{|\mathcal{D}_{train}|}\sum_{(\bm{x_i}, y_i)\in \mathcal{D}_{train}}\mathrm{PS}^{y_i}(\bm{x_i}) - \frac{1}{|\mathcal{D}_{test}|}\sum_{(\bm{x_i}, y_i)\in \mathcal{D}_{test}}\mathrm{PS}^{y_i}(\bm{x_i}),
\end{equation}
where $\mathrm{PS}^{y_i}$ denotes the prediction sensitivity (Eq.~\ref{eq:sensitivity}) of ground-truth class $y_i$. The sensitivity gap calculates the difference of average sensitivity between members and non-members in the ground-truth class.

As illustrated in Figure~\ref{fig:histogram}, the divergence of prediction confidences and prediction sensitivities is increased due to neural network pruning, which introduces the new attack vectors for MIAs and thus makes the pruned models more vulnerable. Moreover, the divergences of prediction confidences and sensitivities from the pruned model vary widely among the different classes of training and test data. 
Figure~\ref{fig:cls_div}  shows that the divergences of the pruned models' prediction behavior (confidence and sensitivity) over members and non-members are significantly different among classes. Similar observations of prediction confidences on different classes after pruning have been made in other fields such as model fairness and transparency~\cite{paganini2020prune,hooker2019compressed,hooker2020characterising}.

\subsection{SAMIA: Self-Attention MIA} 
Upon the above observations, we propose one hypothesis: \textit{the divergences among classes, \ie, confidence gap and sensitivity gap, can provide fine-grained ``evidence'' for MIAs, leading to serious privacy leakage}. In addition, most existing MIA research only considers the confidence gap and a single threshold of the ground-truth class, which may underestimate the privacy risks of MIAs in neural network pruning. Hence, we propose SAMIA, a self-attention MIA, to fully utilize the increased divergence information along with the class information to conduct a finer-grained analysis. Specifically, self-attention is a neural network module to capture global dependencies among inputs and allows the inputs to interact with each other. Despite the recent success of self-attention mechanism in many areas, such as natural language processing~\cite{vaswani2017attention,devlin2019bert} and computer vision~\cite{zhang2019self,wang2018non,fu2019dual}, it has not been well exploited in the research of privacy attacks yet. 

In SAMIA, we leverage the self-attention mechanism to automatically extract the finer-grained ``thresholds'' from different classes by capturing the dependency between predicted information (confidence and sensitivity) and class information and allowing them to interact with each other. Specifically, SAMIA takes the pruned model's prediction confidence and sensitivity and ground-truth labels as inputs.  Given a specific class, the self-attention mechanism finds out the specific confidence information and sensitivity information that the attack ``threshold'' should pay more ``attention'' to. 

\begin{figure}[!t]
\centering
\includegraphics[width=0.7\linewidth]{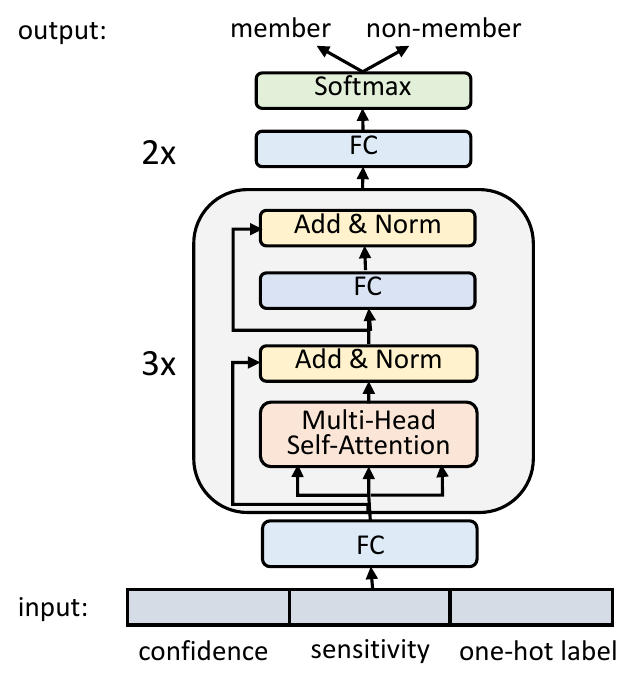}
\vspace{-0.5em}
\caption{Attack model architecture in SAMIA.}
\label{fig:transformer}
\vspace{-0.5em}
\end{figure}

Figure~\ref{fig:transformer} illustrates the network architecture of the attack model used in SAMIA, enlightened by the idea of Transformer~\cite{vaswani2017attention}, \ie, one of the most widely used self-attention architectures. We first convert the ground-truth label into a one-hot vector and then feed both pruned model's prediction confidence, sensitivity, and the one-hot vector into the attack model as the input features. 
The input features are encoded into a vector using a Fully Connected (FC) layer, which is then fed into the multi-head self-attention modules. In each module, we encode the features as query $Q$, key $K$, and value $V$ vectors using a linear function following the self-attention strategy. The attention module calculates the attention scores of the subgroups in a scaled dot-product way:
$\mathrm{Attention}(Q, K, V) = \mathrm{softmax}(QK^T)V,$
where $\mathrm{softmax}()$ denotes the softmax function to make the attention scores sum up to $1$.
The output of the attention module is the weighted sum of the value vector, where the weight assigned to each value is derived by the attention scores $\mathrm{softmax}(QK^T)$. In addition, we calculate four attention scores (\ie, multi-head attention) to capture the different attention strategies. Followed by the attention module, we add the result to the input features and apply the layer normalization~\cite{ba2016layer} to stabilize the attack model training. The result will be fed into another FC layer with layer normalization. We consider these operations as a block and repeat the block three times, followed by two fully connected layers. 
A non-linear activation function, ReLU is applied to the output of the first few FC layers. A softmax function is applied to the last FC layer to provide the binary prediction on the membership status. 

Compared with existing MIAs that learn a single threshold of prediction confidence to determine the membership, the proposed SAMIA captures the information of confidences and sensitivities and intuitively better learns the diverse thresholds to multiple classes. Our evaluation results demonstrate that SAMIA leads to higher attack accuracy compared with the state-of-the-art attacks.

\section{Attack Evaluation}
\label{sec:eval}
This section conducts comprehensive experiments\footnote{Due to the space limit, we only present the major results in this paper. More details can be found in the extended version \url{https://arxiv.org/abs/2202.03335}.} to thoroughly investigate the privacy risks of the proposed MIAs against neural network pruning. In the following, we first introduce the experimental setup, and then evaluate the privacy risks of the pruned models by comparing them with those of original models. Next, we investigate the impact of the confidence gap, sensitivity gap, and generalization gap, respectively. Finally, we evaluate the privacy risks without the knowledge of pruning approaches and sparsity levels.

\subsection{Evaluation Setup}

In the evaluation, we consider the most widely used datasets, neural network architectures, and optimization approaches following recent research of MIAs~\cite{shokri2017membership,hui2019practical,salem2019ml,song2020systematic}. 

\begin{figure*}[!t]
\centering
\begin{subfigure}[b]{0.24\linewidth}
\centering
\includegraphics[width=\linewidth]{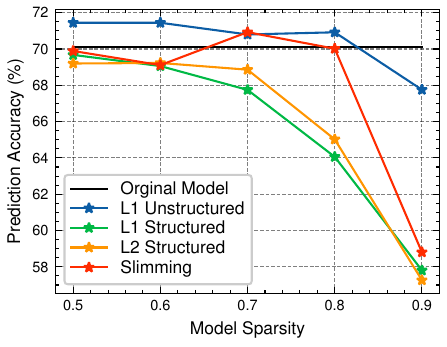}
\caption{CIFAR10, ResNet18}
\end{subfigure}
\begin{subfigure}[b]{0.24\linewidth}
\centering
\includegraphics[width=\linewidth]{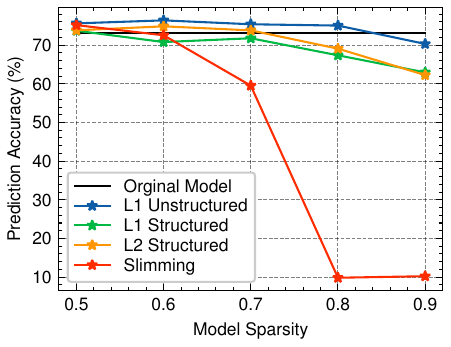}
\caption{CIFAR10, DenseNet121}
\end{subfigure}
\begin{subfigure}[b]{0.24\linewidth}
\centering
\includegraphics[width=\linewidth]{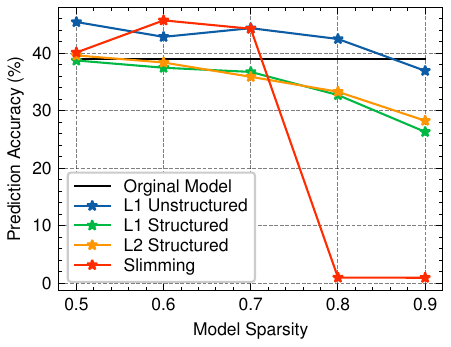}
\caption{CIFAR100, DenseNet121}
\label{fig:prune_acc_cifar100_densenet}
\end{subfigure}
\begin{subfigure}[b]{0.24\linewidth}
\includegraphics[width=\linewidth]{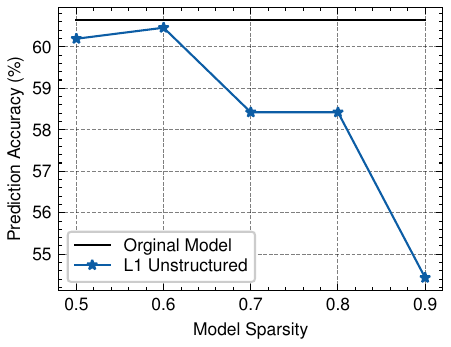}
\caption{Location, FC}
\end{subfigure}
\caption{Prediction accuracy (test accuracy) of the pruned models using different pruning approaches and sparsity levels. Each point indicates the prediction accuracy achieved by the pruned model with a certain pruning approach and sparsity level. The black line indicates the prediction accuracy of the original models.}
\label{fig:prune_acc}
\end{figure*}

\subsubsection{Datasets}
We consider seven popular datasets in the experiments: CIFAR10, CIFAR100, CHMNIST, SVHN, Texas, Location, and Purchase. 

\begin{itemize}[leftmargin=*] \setlength{\itemsep}{0.1em} \setlength{\parskip}{0.1em}
    \item {\textit{CIFAR10}} and \textit{CIFAR100}~\cite{krizhevsky2009learning}. These are two benchmark datasets for image classification. CIFAR10 dataset contains 60,000 $32\times 32$ color images in 10 classes, with 6,000 images per class. CIFAR100 dataset contains 60,000 color images in 100 classes, with 600 images per class.
    \item \textit{CHMNIST}~\cite{kather2016multi}. This dataset consists of 5,000 histological images of human colorectal cancer containing 10 classes of tissues. We resize all images to $32\times 32$, the same dimension as CIFAR10 and CIFAR100.
    \item \textit{SVHN}~\cite{netzer2011reading}. This dataset consists of 99,289 $32\times 32$ color images from house numbers in the Google Street View dataset, containing 10 classes from 0 to 9.
    \item \textit{Location}~\cite{yang2016participatory,yang2015nationtelescope}. This dataset contains location ``check-in'' records of mobile users in the Foursquare social network, restricted to the Bangkok area. The dataset is used to predict users' geosocial type based on the geographical history record features: whether the user visited a certain region or location type. We use the preprocessed purchase dataset provided by Shokri \etal~\cite{shokri2017membership}, which contains 5,010 data samples, 446 binary features, and 30 classes. 
    \item \textit{Texas}~\cite{hospital2020}. This dataset is presented in the Hospital Discharge Data Public Use Data File provided by the Texas Department of State Health Services. The dataset is used to predict the types of patient's main procedure based on a wide range of features, such as external causes of injury, diagnosis of the patient, procedures the patient underwent, and other generic information. We use the preprocessed purchase dataset provided by~\cite{shokri2017membership}, which contains 67,330 data samples, 6,169 binary features, and 100 classes. 
    \item \textit{Purchase}~\cite{acquire2014}. This dataset is presented in Acquire Valued Shoppers Challenge to predict which shoppers will become repeat buyers based on the purchase history. We use the preprocessed purchase dataset provided by Shokri \etal~\cite{shokri2017membership}, which contains 197,324 data samples, 600 binary features, and 100 classes. 

\end{itemize}

Each above dataset is first randomly and equally split into two parts: one for target model, one for shadow model. In each part, we split the data into three datasets: training (45\%), validation (10\%), and test (45\%). We use the validation dataset to determine if the model needs to stop training or fine-tuning for early stopping.
Therefore, the membership inference via random guessing results in 50\% attack accuracy.
Due to the space limit, we only show the results of the CIFAR10 and Purchase datasets. The rest results are presented in the Appendix.

\subsubsection{Neural Network Architectures}
For the four image datasets, \ie, CIFAR10, CIFAR100, CHMNIST, and SVHN, we consider three representative neural network architectures: ResNet18, VGG16, and DenseNet121\footnote{All neural networks are trained using \url{https://github.com/huyvnphan/PyTorch_CIFAR10}}. For the other three datasets, \ie, Texas, Purchase, and Location, we implement fully connected (FC) neural networks with two layers, and the numbers of neurons for each layer are 256 and 128, respectively. All the FC layers except the last one are followed by ReLU activation functions.  In addition, Adam optimizer~\cite{kingma2014adam} is implemented with a learning rate of 0.001 and the batch size of 128 to train all the original models and fine-tune all the models after pruning.

\subsubsection{Neural Network Pruning Approaches}
Four representative neural network pruning approaches are considered, including L1 unstructured pruning,  L1 structured pruning, L2 structured pruning, and Network slimming.

\begin{itemize}[leftmargin=*]\setlength{\itemsep}{0.1em} \setlength{\parskip}{0.1em}
    \item \textit{L1 unstructured pruning}~\cite{han2015deep} (L1 unstructured), which removes the weights with the lowest absolute values individually. This pruning approach can produce a sparse neural network with a small size, but may not improve efficiency given the existing hardware and software optimization. 
    \item \textit{L1 structured pruning}~\cite{li2016pruning} (L1 structured), which removes the entire filters with the lowest absolute values from the convolution layers. By removing the entire filters, this method leads to significant speedup compared with the unstructured pruning since optimization for dense matrix can be applied for efficient computation.
    \item \textit{L2 structured pruning} (L2 structured), which removes the entire filters with the lowest L2 norm values from the convolution layers, similar to L1 structured pruning. 
    \item \textit{Network slimming}~\cite{liu2017learning} (Slimming), which associates scaling factors used in the batch normalization layer with each channel and removes the entire channels with the lowest scaling factors. This method automatically identifies the insignificant channels and finds the target architectures.
    
\end{itemize}

We apply the L1 unstructured pruning to all models. 
Since structured pruning approaches, \ie, L1 structured and L2 structured pruning and Slimming, can only be applied to pruning convolution layers, we evaluate the structured pruning approaches on the ResNet18, VGG16, and DenseNet121 models trained on CIFAR10, CIFAR100, SVHN datasets. In addition, five sparsity levels $\gamma=\{0.5, 0.6, 0.7, 0.8, 0.9\}$ are investigated for all pruning approaches, which denote the portions of the removed parameters\footnote{Since structured pruning only removes the parameters in the convolution layers, the sparsity levels for structured pruning only count the removed parameters in the convolution layers instead of the entire neural network.}. We follow typical pruning procedures: train the original model, prune the model using the above approaches, and finally fine-tune the pruned model.

Figure~\ref{fig:prune_acc} shows the prediction accuracy of the original model and the pruned models with different pruning approaches and sparsity levels. We observe that the pruned models achieve close performance compared to the original model if the sparsity level is not high. The accuracy of pruned models is reduced with the increase of the pruning sparsity. Sometimes, pruned models can achieve higher accuracy than the original models, which has been shown in recent studies of neural network pruning~\cite{blalock2020state}. Unstructured pruning usually performs better than structured pruning in the evaluation, since structured pruning forces the removed parameter in a restricted way, which limits the performance of the pruned model but increases the speed of model inference.

\subsubsection{State-of-the-art MIAs}
\label{sec:state_mia}
To thoroughly evaluate the proposed SAMIA, we investigate eight state-of-the-art MIAs along with SAMIA.
\footnote{We implement Conf, Xent, Mentr, and Top1-Conf attacks based on \url{https://github.com/inspire-group/membership-inference-evaluation} and BlindMI attack based on \url{https://github.com/hyhmia/BlindMI/blob/master/BlindMI_Diff_W.py}. }.
\begin{itemize}[leftmargin=*]\setlength{\itemsep}{0.1em} \setlength{\parskip}{0.1em}
    \item \textit{Ground-truth class confidence-based threshold attack (Conf)}. Yeom \etal used the prediction confidence of ground-truth class to identify membership status~\cite{yeom2018privacy}. The adversary learns a threshold to determine the membership of a data sample based on the confidence of ground-truth class. Given an input sample $\bm{x}$, its class $y$, and the pruned model $f_p$, the attack function is defined as $I_{\mathrm{conf}}(f_p, (\bm{x},y)) = \mathbbm{1}\{f_p^{(y)}(\bm{x})\geq\zeta_y\}$, where $f_p^{(y)}$ is the prediction confidence of class $y$ and $\zeta_y$ is the threshold of class $y$ derived from the shadow pruned model.
    \item \textit{Cross-Entropy-based threshold attack (Xent)}. The entropy loss can be used to derive the threshold from the shadow pruned model~\cite{yeom2018privacy}. The attack function is defined as $I_{\mathrm{xent}}(f_p, (\bm{x},y)) = \mathbbm{1}\{\mathrm{xent}(f_p^{(y)}(\bm{x}))\geq\zeta_y\}$, where $\mathrm{xent}$ denotes the cross entropy loss.
    \item \textit{Modified-entropy-based threshold attack (Mentr)}. Song and Mittal proposed modified entropy by including the information about the ground-truth class, which achieved better performance than using prediction confidence~\cite{song2020systematic}. The attack function is defined as $I_{\mathrm{mentr}}(f_p, (\bm{x},y)) = \mathbbm{1}\{\mathrm{mentr}(f_p^{(y)}(\bm{x}))\geq\zeta_y\}$, where $\mathrm{mentr}(f_p(\bm{x}), y) = - (1 - f_p^{(y)}(\bm{x}))\log (f_p^{(y)}(\bm{x})) - \sum_{t\neq y}f_p^{(t)}(\bm{x})\log(1-f_p^{(t)}(\bm{x}))$.
    \item \textit{Top1 Confidence-based threshold attack (Top1-Conf)}. Salem~\etal proposed to derive the threshold from the highest prediction confidence~\cite{salem2019ml}. The attack function is defined as $I_{\mathrm{top1}}(f_p, (\bm{x})) = \mathbbm{1}\{\mathrm{top1}(f_p(\bm{x}))\geq\zeta_y\}$, where $\mathrm{top1}$ calculates the highest value from the prediction confidence.
    \item \textit{Confidence-based Neural Network attack (NN)}. Shokri \etal proposed to use prediction confidence as features to train a neural network from the shadow model~\cite{shokri2017membership}, which is used to distinguish member and non-member data.
    \item \textit{Top-3 Confidence-based Neural Network attack (Top3-NN)}. Salem~\etal proposed to use the top-3 prediction confidences as features~\cite{salem2019ml} to train a neural network classifier.
    \item \textit{Confidence-based Neural Network attack with ground-truth class (NNCls)}. Nasr~\etal combined one-hot encoded class labels with the prediction confidence as features to train a neural network classifier~\cite{nasr2018machine}.
    \item \textit{Blind Membership Inference Attack (BlindMI)}. Hui~\etal proposed to determine the membership of a data sample by moving it to a non-member set and check if the moving operation increases the distance between member and non-member sets~\cite{hui2019practical}. BlindMI considers the data sample as a non-member if the distance is increased. We use the default BlindMI attack provided in~\cite{hui2019practical}.
\end{itemize}

In the main paper, we present the results of five attacks, that achieve the highest attack accuracies in most experiments, \ie, Conf, Mentr, NNCls, BlindMI, and SAMIA. The results of the rest of the attacks are reported in the Appendix.

Besides, it should be mentioned that to provide a practical analysis of privacy risks, we adopt early stopping and l2 regularization as a baseline defense mechanism and apply it to all the following experiments of membership inference attacks. Other defense mechanisms will be discussed in Section~\ref{sec:defense}.

\subsubsection{SAMIA Settings}
Following the experimental setting in~\cite{shokri2017membership}, we first train five shadow models and their pruned models. The predictions of the shadow models on shadow training and shadow test datasets are used to train an attack model. In the attack model, we use four attention heads, 64 neural units, and GeLU activation function~\cite{hendrycks2016gaussian} in each self-attention module, with a 20\% dropout rate.
We use SGD optimizer~\cite{saad1998online} to train the attack models for 100 epochs with batch size 128. The learning rate of the SGD optimizer is set as 0.01 and reduced to 0.001 and 0.0001 at the 1/2 and 3/4 of the training process (\ie, the 50th and 75th epoch). 
Due to the large number of settings evaluated in the attacks and defenses and the high computational cost in each setting, we conduct all the experiments only once. Thus experimental variation may be observed due to the randomness in neural network pruning and membership inference attacks (\eg, parameter initialization, dataset shuffling).

\begin{figure*}[!t]
    \centering
    \begin{subfigure}[b]{0.24\textwidth}
    \includegraphics[width=\linewidth]{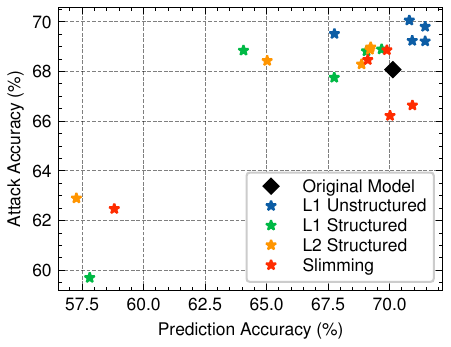}
    \caption{CIFAR10, ResNet18}
    \end{subfigure}
    \begin{subfigure}[b]{0.24\textwidth}
    \includegraphics[width=\linewidth]{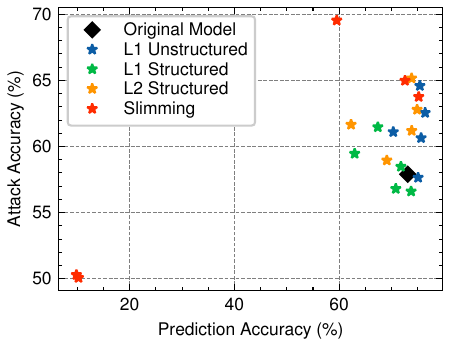}
    \caption{CIFAR10, DenseNet121}
    \end{subfigure}
    \begin{subfigure}[b]{0.24\textwidth}
    \includegraphics[width=\linewidth]{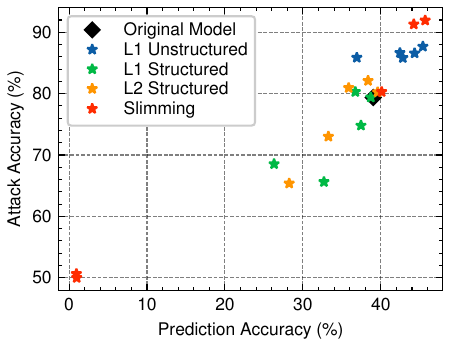}
    \caption{CIFAR100, DenseNet121}
    \end{subfigure}
    \begin{subfigure}[b]{0.24\textwidth}
    \includegraphics[width=\linewidth]{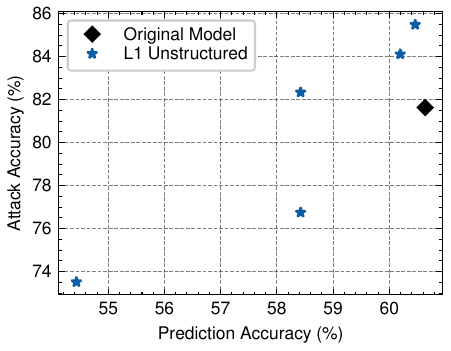}
    \caption{Location, FC}
    \end{subfigure}
    \caption{Privacy Risks of Neural Network Pruning (w.r.t. prediction accuracy). Most pruning approaches result in a higher attack accuracy when considering a similar prediction accuracy, compared with the original models. We present the attack accuracy of SAMIA for pruned models and the attack accuracy of Conf attack for the original models.}
    \label{fig:acc_attack}
\end{figure*}

\begin{figure*}[!t]
    \centering
    \begin{subfigure}[b]{0.24\textwidth}
    \includegraphics[width=\linewidth]{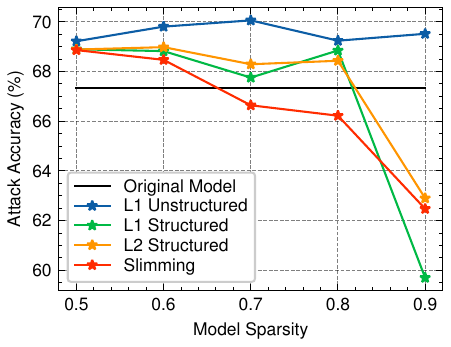}
    \caption{CIFAR10, ResNet18}
    \end{subfigure}
    \begin{subfigure}[b]{0.24\textwidth}
    \includegraphics[width=\linewidth]{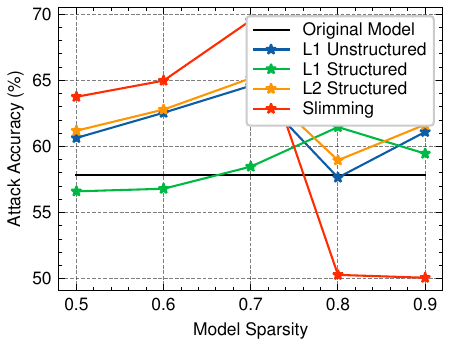}
    \caption{CIFAR10, DenseNet121}
    \end{subfigure}
    \begin{subfigure}[b]{0.24\textwidth}
    \includegraphics[width=\linewidth]{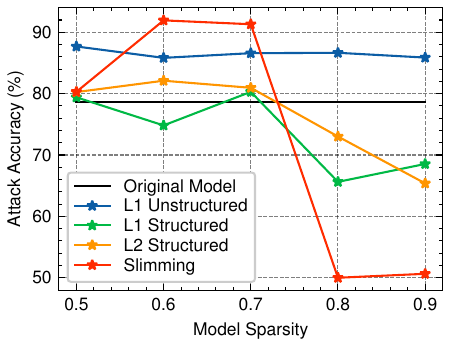}
    \caption{CIFAR100, DenseNet121}
    \end{subfigure}
    \begin{subfigure}[b]{0.24\textwidth}
    \includegraphics[width=\linewidth]{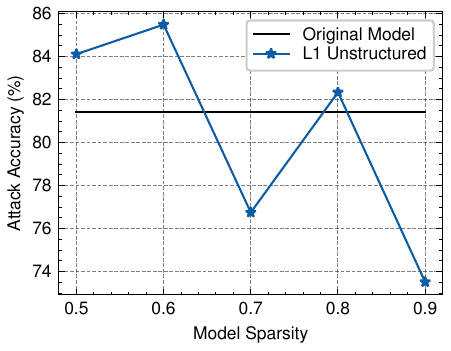}
    \caption{Location, FC}
    \end{subfigure}
    \caption{Privacy Risks of Neural Network Pruning (w.r.t. model sparsity). We present the attack accuracy of SAMIA for pruned models and the attack accuracy of Conf attack for the original models.}
    \label{fig:mia_sparsity}
\end{figure*}

\begin{figure*}[!t]
    \centering
    \begin{subfigure}[b]{0.24\textwidth}
    \includegraphics[width=\linewidth]{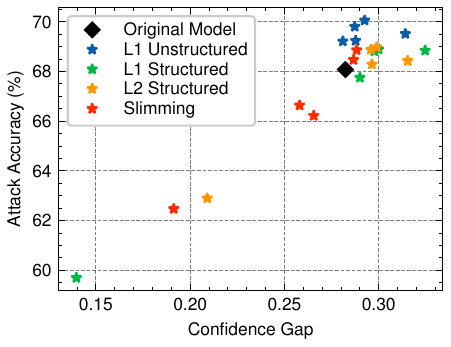}
    \caption{Confidence gap}
    \label{fig:conf_gap_cifar10_resnet18}
    \end{subfigure}
    \begin{subfigure}[b]{0.24\textwidth}
    \includegraphics[width=\linewidth]{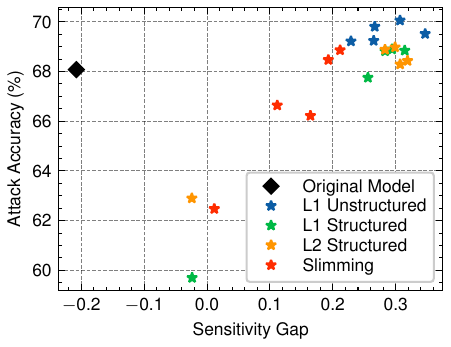}
    \caption{Sensitivity gap}
    \label{fig:sens_gap_cifar10_resnet}
    \end{subfigure}
    \begin{subfigure}[b]{0.24\textwidth}
    \includegraphics[width=\linewidth]{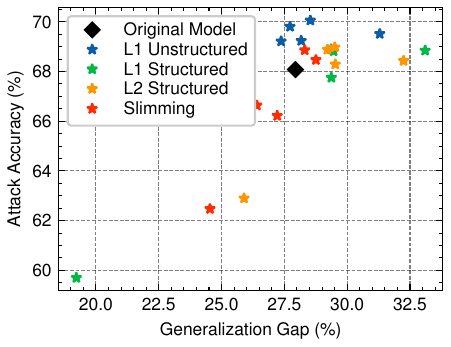}
    \caption{Generalization gap}
    \end{subfigure}
    \caption{Impact of confidence gap, sensitivity gap, and generalization gap (CIFAR10, ResNet18). We present the relationship between the gap and the attack accuracy of SAMIA.}
    \label{fig:gap_cifar10_resnet18}
\end{figure*}

\begin{figure*}[!t]
\centering
\begin{subfigure}[b]{0.24\linewidth}
\includegraphics[width=0.95\linewidth]{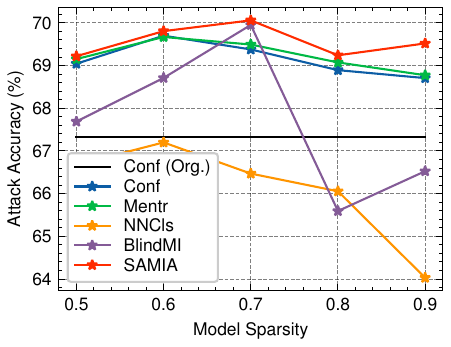}
\caption{L1 unstructured}
\end{subfigure}
\begin{subfigure}[b]{0.24\linewidth}
\includegraphics[width=0.95\linewidth]{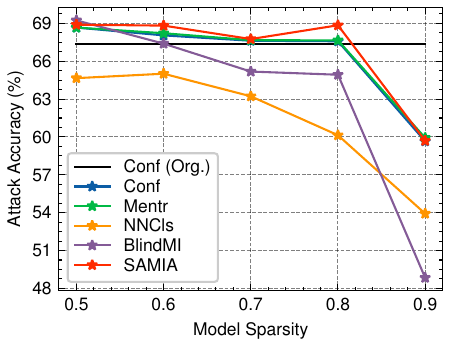}
\caption{L1 structured}
\end{subfigure}
\begin{subfigure}[b]{0.24\linewidth}
\includegraphics[width=0.95\linewidth]{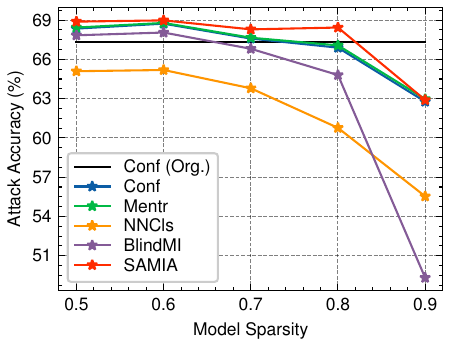}
\caption{L2 structured}
\end{subfigure}
\begin{subfigure}[b]{0.24\linewidth}
\includegraphics[width=0.95\linewidth]{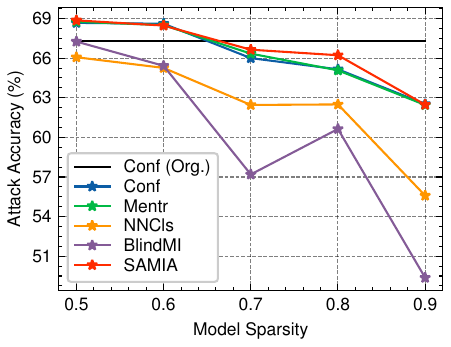}
\caption{Slimming}
\end{subfigure}
\begin{subfigure}[b]{0.24\linewidth}
\includegraphics[width=0.95\linewidth]{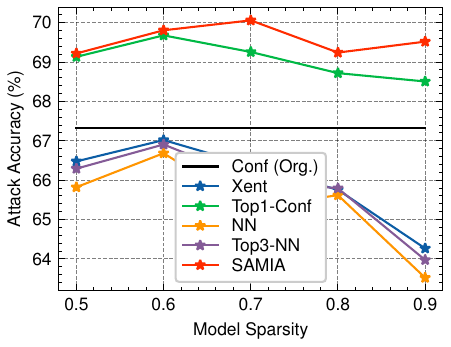}
\caption{L1 unstructured}
\end{subfigure}
\begin{subfigure}[b]{0.24\linewidth}
\includegraphics[width=0.95\linewidth]{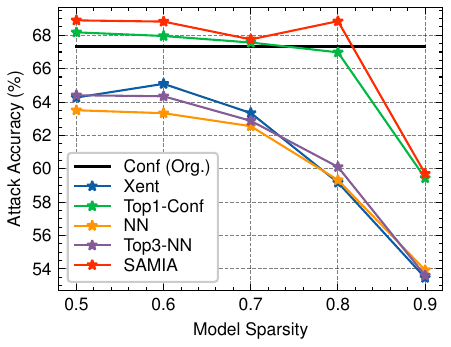}
\caption{L1 structured}
\end{subfigure}
\begin{subfigure}[b]{0.24\linewidth}
\includegraphics[width=0.95\linewidth]{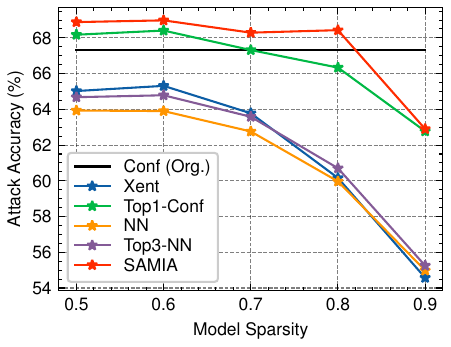}
\caption{L2 structured}
\end{subfigure}
\begin{subfigure}[b]{0.24\linewidth}
\includegraphics[width=0.95\linewidth]{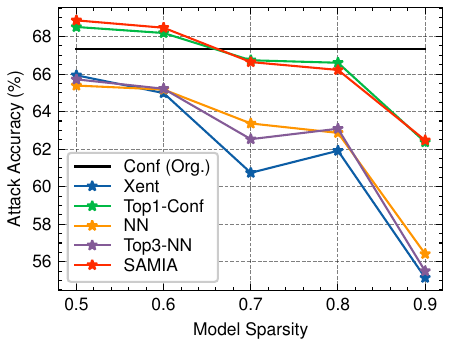}
\caption{Slimming}
\end{subfigure}
\caption{Attack performance comparison of MIAs (CIFAR10, ResNet18). We present the attack accuracy of state-of-the-art membership inference attacks and compare them with the proposed SAMIA. Four pruning approaches are used on CIFAR10 ResNet18 models. The black line presents the attack accuracy of original models using Conf attack, \ie, Conf (Org.).}
\label{fig:mias_cifar10_resnet}
\end{figure*}

\subsection{Privacy Risk Discussions}

In this section, we evaluate the privacy risks of the pruned models and compare them with the original models and then investigate several key factors on privacy risks of neural network pruning. Additionally, we investigate the privacy risks of different pruning approaches and  discuss the effectiveness of the proposed SAMIA and the impact of unknown sparsity levels and pruning approaches.

\subsubsection{Privacy Risks of Neural Network Pruning}\label{sec:5.2.1}
Since different pruning approaches and sparsity levels may achieve distinct prediction accuracy, to make a fair comparison, we evaluate the privacy risks of pruning by taking the prediction accuracy into consideration. 
Figure~\ref{fig:acc_attack} shows the relationship between prediction accuracy and (SAMIA) attack accuracy when we apply different pruning approaches and sparsity levels on the CIFAR10, CIFAR100, and Location datasets.
We observe that \textit{when the pruned model achieves a comparable prediction accuracy with the original model, most pruning approaches result in an increased attack accuracy (\ie, privacy risk).
}
The attack accuracy may be decreased with the loss of prediction accuracy, as the pruned model becomes less effective for both prediction and attack.
However, we still observe that in most cases, when the pruned model performs worse than the original model, the pruned model's attack accuracy remains higher than the original one's. 
Therefore, the pruned models become more vulnerable to membership inference attacks than the original models.

When a low sparsity level is used, we always observe the increased privacy risk of the pruned model (Figure~\ref{fig:mia_sparsity}). Since with a low sparsity level, the pruned model is more likely to achieve a comparable or even higher prediction accuracy compared with the original model, which increases the accuracy of prediction confidence used in MIAs and further increases the privacy risk.

\begin{figure}[!t]
\centering
\begin{subfigure}[b]{0.49\linewidth}
\includegraphics[width=0.95\linewidth]{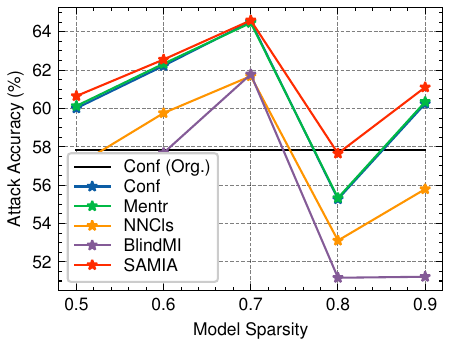}
\caption{CIFAR10, DenseNet121}
\end{subfigure}
\begin{subfigure}[b]{0.49\linewidth}
\includegraphics[width=0.95\linewidth]{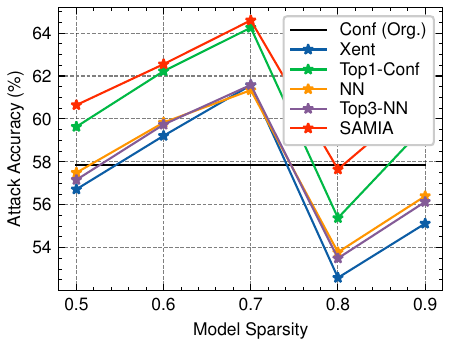}
\caption{CIFAR10, DenseNet121}
\end{subfigure}
\begin{subfigure}[b]{0.49\linewidth}
\includegraphics[width=0.95\linewidth]{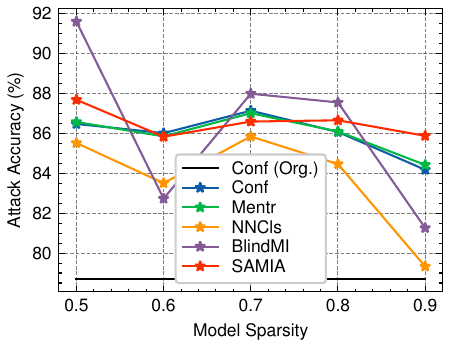}
\caption{CIFAR100, DenseNet121}
\label{fig:mias_cifar100_densenet1}
\end{subfigure}
\begin{subfigure}[b]{0.49\linewidth}
\includegraphics[width=0.95\linewidth]{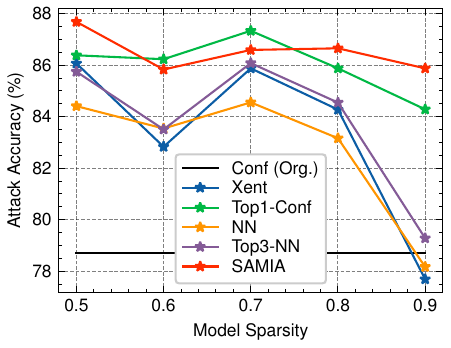}
\caption{CIFAR100, DenseNet121}
\label{fig:mias_cifar100_densenet2}
\end{subfigure}
\begin{subfigure}[b]{0.49\linewidth}
\includegraphics[width=0.95\linewidth]{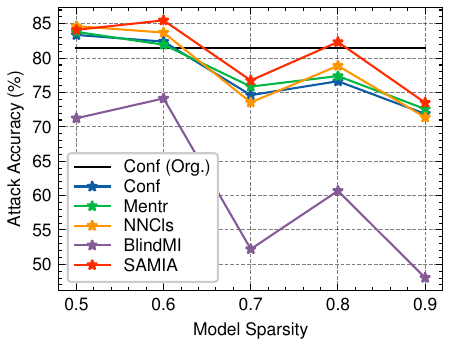}
\caption{Location, FC}
\end{subfigure}
\begin{subfigure}[b]{0.49\linewidth}
\includegraphics[width=0.95\linewidth]{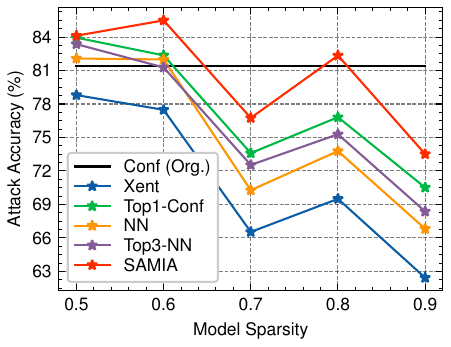}
\caption{Location, FC}
\end{subfigure}
\caption{Attack performance comparison of MIAs on different datasets (L1 Unstructured). We present the attack accuracy of state-of-the-art membership inference attacks and compared them with the proposed SAMIA.  We present the attack accuracy of three models (CIFAR10 DenseNet121, CIFAR100 DenseNet121, Location FC) pruned by L1 unstructured pruning. The black line presents the attack accuracy of original models using Conf attack, \ie, Conf (Org.).}
\label{fig:mias_dataset}
\end{figure}

\subsubsection{Impact of Confidence, Sensitivity, and Generalization Gap.}
\label{sec:impact_gap}
As aforementioned, we hypothesize that neural network pruning leads to the increased \textit{confidence gap} and \textit{sensitivity gap} (in ground-truth class) of pruned models, thus increasing their membership inference risks. 
Meanwhile, overtraining is considered as one of the key causes of membership leakage in previous research~\cite{yeom2018privacy,song2020systematic}, leading to our evaluation on \textit{generalization gap}, \ie, the difference between training accuracy and testing accuracy. 
From Figure~\ref{fig:gap_cifar10_resnet18}, we observe that {neural network pruning increases the gaps between members and non-members, \ie, confidence gap, sensitivity gap, and generalization gap, in most settings}. Further, with the increase of gaps, we observe the increase of attack accuracy, which indicates  \textit{the strong correlation between the gaps, \ie, confidence gap, sensitivity gap, and generalization gap, and the increased privacy risk.}

The strong correlation of confidence gap and sensitivity gap validates our intuition that these gaps can be leveraged by the adversary to infer the membership status, introducing a new attack surface in neural network pruning. 
By investigating the attack results, we find that the confidence gap plays the most important role in the privacy risk. L1 unstructured and slimming pruning usually lead to an increased confidence gap, which will be leveraged by the adversary and result in a higher attack accuracy. Additionally, sensitivity gap can also leak the membership information. For example, the confidence gap of L1 unstructured pruning on a CIFAR10 ResNet18 model (Figure~\ref{fig:conf_gap_cifar10_resnet18}) is close to the original model, but due to the increased sensitivity gap (Figure~\ref{fig:sens_gap_cifar10_resnet}), the pruning still results in an increased attack accuracy.

\subsubsection{Privacy Risks of Pruning Approaches.}
Following the same settings above, we investigate the privacy risks of different pruning approaches by comparing the attack accuracy under the similar prediction accuracy.
As shown in Figure~\ref{fig:acc_attack} and~\ref{fig:mia_sparsity}, given the similar prediction accuracy of the pruned models, L1 unstructured and slimming pruning result in the highest attack accuracy. 
Besides, L1 structured pruning achieves the lowest attack accuracy among all pruning approaches, but still in some cases, the attack accuracy is higher than the original model, even with the similar or lower prediction accuracy.
The structured constraint used in L1 structured pruning regularizes the model in the fine-tuning and thereby reduces the privacy risk.

\subsubsection{Effectiveness of SAMIA}
To investigate the effectiveness of the proposed SAMIA, we compare SAMIA with the state-of-the-art MIAs in terms of attack accuracy. As shown in Figure~\ref{fig:mias_cifar10_resnet}, we observe that \textit{our proposed SAMIA achieves the highest attack accuracy in most cases compared with baseline attacks}, which is mainly due to the fact that SAMIA best leverages both confidence gap and sensitivity gap (in ground-truth class) introduced in pruning. Besides, Top1-conf and Mentr attacks are also effective, as both attacks take advantage of the confidence gap, the most important factor for model privacy. 
We also observe that when the pruning introduces a high generalization gap , all attacks can achieve a high attack accuracy (\eg, CIFAR100 DenseNet121 in Figure~\ref{fig:mias_cifar100_densenet1},~\ref{fig:mias_cifar100_densenet2} and Appendix Figure~\ref{fig:gen_gap_cifar100_densenet})), which has been discussed in the previous MIA research.

\subsubsection{Unknown Sparsity Level and Pruning Approach}
\label{section:unknownSLPA}
In the evaluation above, we assume the adversary has the knowledge of sparsity levels and pruning approaches used in network pruning. 
In this section, we explore the privacy risks of a more realistic scenario, \ie, when the adversary has no prior knowledge of the sparsity levels and the pruning approaches. 

\begin{figure*}[!t]
\centering
\begin{subfigure}[b]{0.24\linewidth}
\includegraphics[width=\linewidth]{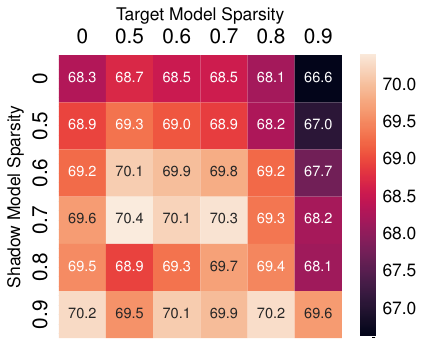}
\caption{L1 Unstructured}
\end{subfigure}
\begin{subfigure}[b]{0.24\linewidth}
\includegraphics[width=\linewidth]{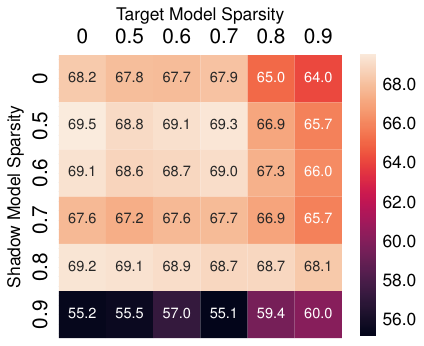}
\caption{L1 Structured}
\end{subfigure}
\begin{subfigure}[b]{0.24\linewidth}
\includegraphics[width=\linewidth]{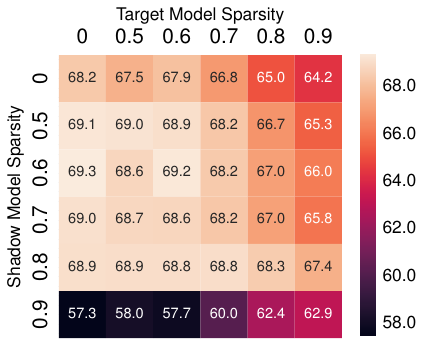}
\caption{L2 Structured}
\end{subfigure}
\begin{subfigure}[b]{0.24\linewidth}
\includegraphics[width=\linewidth]{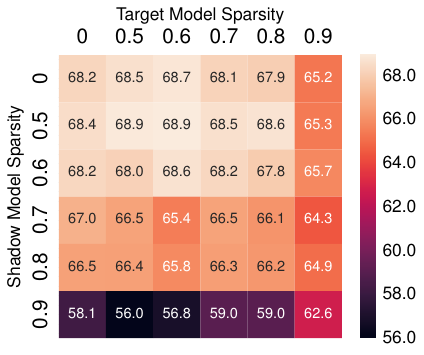}
\caption{Slimming}   
\end{subfigure}
\caption{Attack accuracy with unknown sparsity levels (CIFAR10, ResNet18).}
\label{fig:unknown_sparsity_cifar10_resnet}
\end{figure*}

\begin{figure*}[!t]
\centering
\begin{subfigure}[b]{0.24\linewidth}
\includegraphics[width=\linewidth]{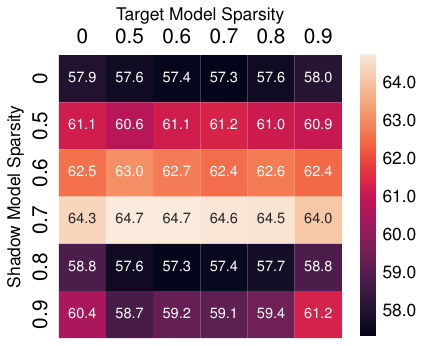}
\caption{CIFAR10, DenseNet121}
\end{subfigure}
\begin{subfigure}[b]{0.24\linewidth}
\includegraphics[width=\linewidth]{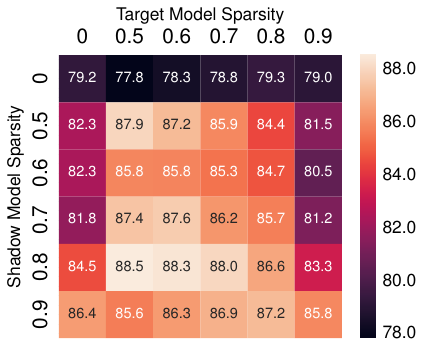}
\caption{CIFAR100, DenseNet121}
\end{subfigure}
\begin{subfigure}[b]{0.24\linewidth}
\includegraphics[width=\linewidth]{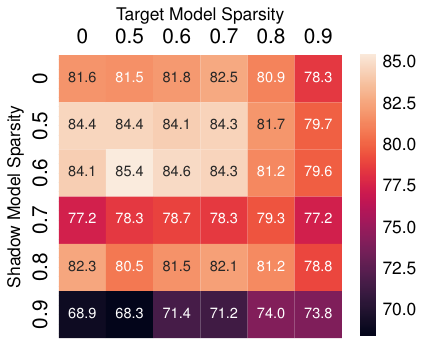}
\caption{Location, FC}
\end{subfigure}
\caption{Attack accuracy with unknown sparsity levels (L1 Unstructured).}
\label{fig:unknown_sparsity_dataset}
\end{figure*}

\vspace{0.2em}
\noindent
\textbf{Unknown sparsity level.} We assume the adversary only knows the pruning approach but not the sparsity level that is the major factor of model efficiency. We evaluate the attack accuracy of SAMIA when the adversary prunes target models and shadow models using different sparsity levels. We also consider the case when the target model is not pruned, \ie, sparsity level $= 0$. As shown in Figure~\ref{fig:unknown_sparsity_cifar10_resnet} and~\ref{fig:unknown_sparsity_dataset}, the attack accuracy is not affected too much due to the different sparsity levels between target models and shadow models. In some cases, using a different sparsity level in pruning shadow models can even increase the attack accuracy. 
\textit{The attack accuracy mainly depends on the performance of the shadow model}, and thus the adversary can attack victim models with higher attack accuracy by selecting a good pruned shadow model. 
For instance, the adversary can use each shadow model to attack other shadow models with different sparsity levels and select the one with the highest attack accuracy.

\vspace{0.2em}
\noindent

\textbf{Unknown sparsity level and pruning approach.} 
Since we assume the adversary has no prior knowledge of the sparsity level and pruning approach, the adversary may randomly pick a sparsity level and a pruning approach to prune a shadow model for attacks. To evaluate the attack accuracy, we conduct 20 experiments for the aforementioned four image datasets and the corresponding neural networks. In each experiment, we randomly select the sparsity levels and pruning approaches for target models and shadow models, respectively. For example, the target model uses L1 Structured pruning with 0.5 sparsity level while the shadow model uses Slimming pruning with 0.8 sparsity level. The sparsity levels are selected from the set of $\{0.5, 0.6, 0.7, 0.8, 0.9\}$ and the pruning approaches are selected from the four pruning approaches.
To measure the privacy risks, we define the attack accuracy loss as $(acc_{known} - acc_{unknown}) /acc_{known}$, where $acc_{known}$ denotes the attack accuracy when the adversary knows all the pruning information and $acc_{unknown}$ denotes the attack accuracy without knowing any pruning information. Table~\ref{tab:acc_loss_unknown} shows the average attack accuracy loss over 20 experiments for each dataset and model. 
We observe that \textit{without knowing the sparsity levels and pruning approaches, the attack is still effective in most cases except the CIFAR10 VGG16 and CIFAR100 ResNet18 models}.
The poor attack performance in these two models is due to the ineffectiveness of shadow models using specific sparsity levels and pruning approaches for attacks.
For example, we observe a significant drop of attack accuracy (from 90\% to 50\%) when applying L1 structured and slimming pruning with sparsity levels 0.7 to 0.9, on the CIFAR10 VGG16 model (Figure~\ref{fig:mia_cifar10_vgg16} in Appendix). The large gap makes the shadow models pruned using these settings ineffective in attacking the unknown victim model.

\begin{table}[!tb]
\centering
\caption{Attack accuracy loss with unknown sparsity levels and pruning approaches.}
\label{tab:acc_loss_unknown}
\small
\begin{tabular}{@{}llr@{}}
\toprule
Dataset                                      & Model       & Attack Acc Loss \\ \midrule
\multirow{3}{*}{CIFAR10} & ResNet18    & 4.77\%          \\
                         & DenseNet121 & 1.63\%          \\
                         & VGG16       & 26.83\%         \\\midrule
\multirow{3}{*}{CIFAR100}                    & ResNet18    & 12.41\%         \\
                                             & DenseNet121 & 6.90\%          \\
                                             & VGG16       & 2.43\%          \\\midrule
\multirow{3}{*}{SVHN}                        & ResNet18    & 0.60\%          \\
                                             & DenseNet121 & 0.12\%          \\
                                             & VGG16       & 0.05\%          \\\midrule
\multirow{3}{*}{CHMNIST}                     & ResNet18    & 0.78\%          \\
                                             & DenseNet121 & 0.52\%          \\
                                             & VGG16       & -0.58\%         \\ \bottomrule 
\end{tabular}%
\end{table}

\section{Defenses against MIAs} %
\label{sec:defense}
Given the privacy risks of pruned neural networks, this section focuses on defenses against the proposed SAMIA. 
We first present the design principle of defenses for pruned neural networks, then describe the proposed defensive design, and lastly compare the performance of the proposed defense with the state-of-the-art defenses.
In addition, to rigorously evaluate the defense performance, we consider the defenses against both the non-adaptive attacks and adaptive attacks, where the adversary of adaptive attacks is put into the last step of the arms race between privacy attacks and defenses (\ie, the adversary knows all the details of defense mechanism and performs adaptive attacks against the defended models)~\cite{athalye2018obfuscated,song2020systematic}. Extensive experiments are conducted to evaluate our defensive proposals.

\subsection{Design Principles of Defenses}

Two major design principles are considered for the defenses of pruned neural networks. On the one hand, effective defenses should be able to reduce the behavior discrepancy introduced by pruning. The above attack evaluation has demonstrated that the privacy risks introduced by MIAs in the pruned models are due to the increased divergence of prediction confidences and sensitivities. Hence, it is essential to reduce such divergence between members and non-members of the pruned neural networks for defense. On the other hand, the defenses need to take into consideration the resource constraints imposed by low-end devices. Neural network pruning aims to reduce the computational cost during inference. 
Such cost cannot be increased by the defenses. Therefore, the defenses should be designed to mitigate the privacy risks of pruned models before deploying them on devices, thus without introducing additional defense costs in the inference phase.

\begin{figure}[!tb]
\centering
\begin{subfigure}[b]{0.49\linewidth}
\centering
\includegraphics[width=\linewidth]{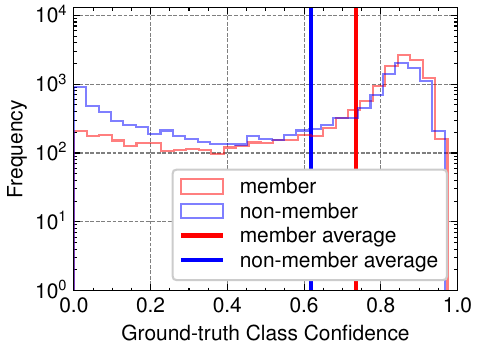}
\caption{Confidence gap}
\end{subfigure}    
\begin{subfigure}[b]{0.49\linewidth}
\centering
\includegraphics[width=\linewidth]{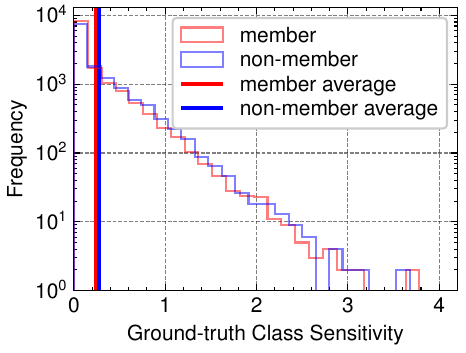}
\caption{Sensitivity gap}
\end{subfigure}
\caption{Divergence of the pruned model's prediction confidences and sensitivities using PPB defense (CIFAR10, DenseNet121).} %
\label{fig:hist_defense}
\end{figure}

\begin{figure}[!tb]
\centering
\begin{subfigure}[b]{0.49\linewidth}
\centering
\includegraphics[width=\linewidth]{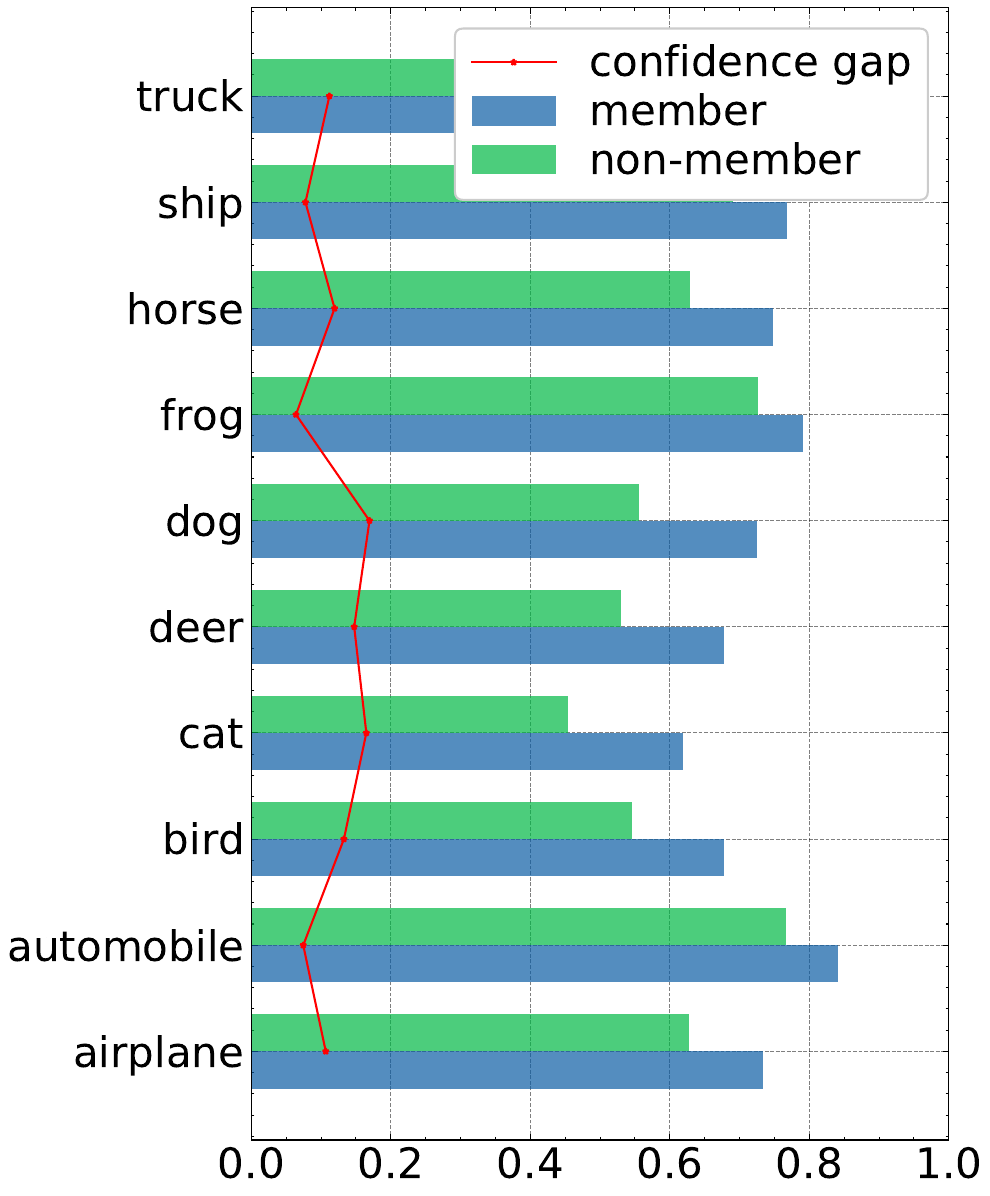}
\caption{Confidence gap}
\end{subfigure}    
\begin{subfigure}[b]{0.48\linewidth}
\centering
\includegraphics[width=\linewidth]{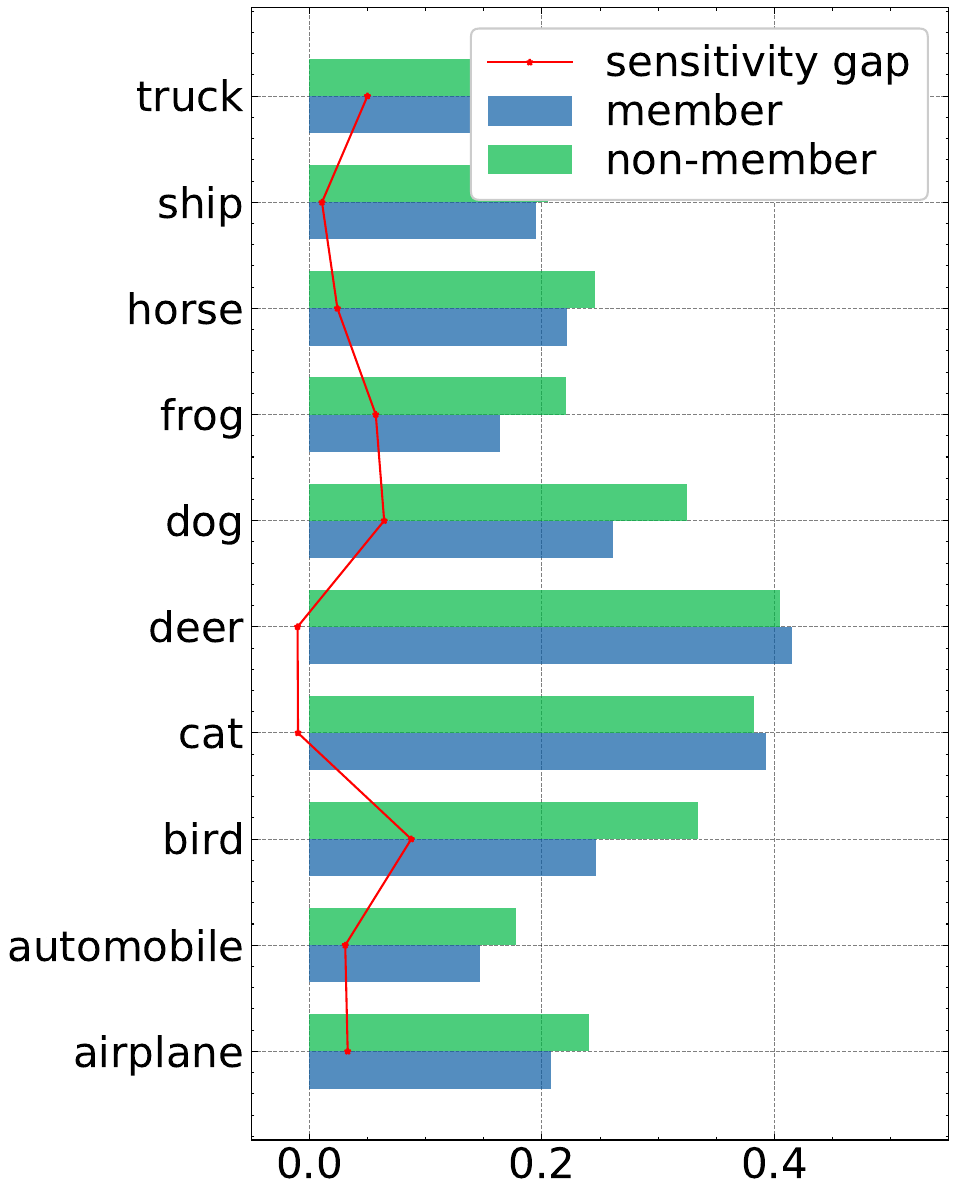}
\caption{Sensitivity gap}
\end{subfigure}
\caption{Divergence of the pruned model's prediction confidences and sensitivities over different classes with PPB defense (CIFAR10, DenseNet121). }
\label{fig:div_cls_defense}
\end{figure}

\subsection{Proposed Defense: PPB} 
Following the two design principles, we propose a countermeasure approach named by pair-based posterior balancing (PPB). 
The main idea of PPB defense is to mitigate the new prediction behaviors on prediction confidence and sensitivity by aligning the posterior predictions of different input samples. In this way, PPB can reduce the divergence of prediction confidence between members and non-members as well as the degree of sensitivities. Specifically, given any pair of two input samples, we try to make the distributions of their ranked posterior predictions as close as possible.
The difference between ranked posteriors' distributions is measured by the Kullback–Leibler divergence (KL divergence)~\cite{kullback1951information}. Give two posterior predictions $P$ and $Q$, the KL divergence is defined as:
\begin{equation}
    \mathcal{L}_{\mathrm{KL}}(P, Q) = \sum_x P(x)\log\frac{P(x)}{Q(x)}.
\end{equation}

KL-divergence is considered as a regularization term in neural network pruning.
The loss function includes both the prediction loss and KL divergence loss, which can be given by:
\begin{align}
\label{eq:defense_obj}
\mathcal{L}(f_p(\bm{x}), \bm{y}) =& \sum_i \mathcal{L}_{\mathrm{predict}}(f_p(\bm{x}_i), y_i) \nonumber\\
&+ \lambda \sum_{j, k (j\neq k)}\mathcal{L}_{\mathrm{KL}}(R(f_p(\bm{x}_j)), R(f_p(\bm{x}_k))),
\end{align}
where $\mathcal{L}_{\mathrm{KL}}$ and $\mathcal{L}_{\mathrm{predict}}$ denote the KL-divergence loss and the prediction loss (\eg, cross-entropy loss for the classification tasks), respectively. $R(\cdot)$ sorts the posteriors provided by the pruned model $f_p$ in decreasing order and $\lambda$ is a hyper-parameter to balance the two losses. It is computationally costly to calculate the KL loss for all possible pairs of data samples in the training dataset. To address this issue, we sample training pairs in each mini-batch during fine-tuning by randomly selecting two data samples as a pair without replacement. Hence, in each mini-batch with batch size $B$, KL loss consists of $B/2$ pairs of training samples.

\begin{figure*}[!tb]
\centering
\begin{subfigure}[b]{0.24\textwidth}
\includegraphics[width=\linewidth]{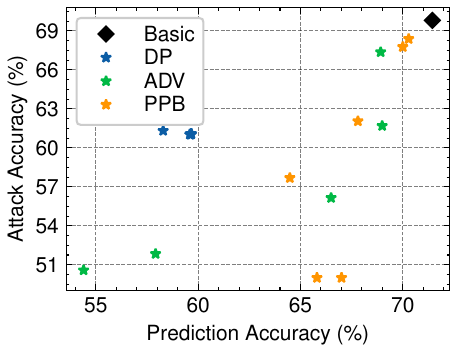}
\caption{L1 Unstructured}
\end{subfigure}
\begin{subfigure}[b]{0.24\textwidth}
\includegraphics[width=\linewidth]{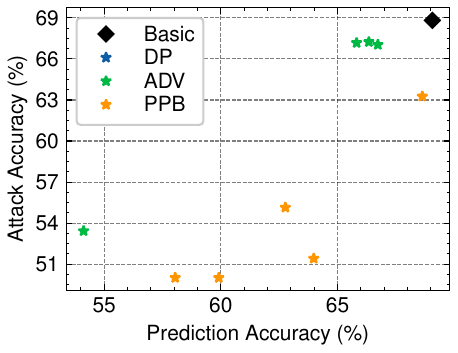}
\caption{L1 Structured}
\end{subfigure}
\begin{subfigure}[b]{0.24\textwidth}
\includegraphics[width=\linewidth]{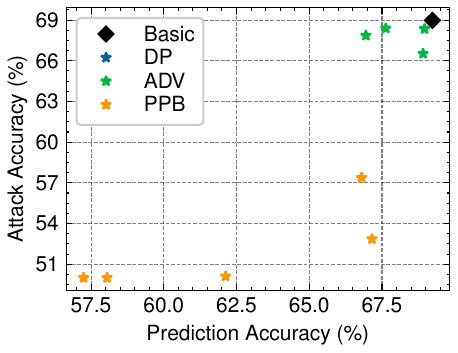}
\caption{L2 Structured}
\end{subfigure}
\begin{subfigure}[b]{0.24\textwidth}
\includegraphics[width=\linewidth]{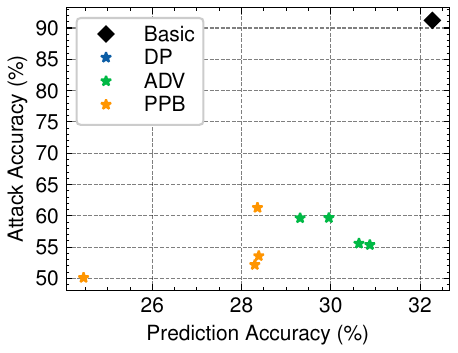}
\caption{Slimming}
\end{subfigure}
\caption{Performance of defenses for different pruning approaches (CIFAR10, ResNet18, Sparsity 0.6).}
\label{fig:defense_cifar10_resnet}
\end{figure*}

\begin{figure*}[!tb]
\centering
\begin{subfigure}[b]{0.24\textwidth}
\includegraphics[width=\linewidth]{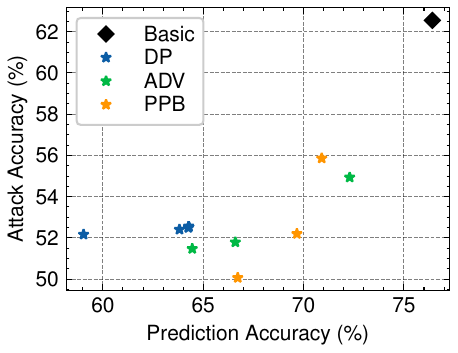}
\caption{CIFAR10, DenseNet121}
\end{subfigure}
\begin{subfigure}[b]{0.24\textwidth}
\includegraphics[width=\linewidth]{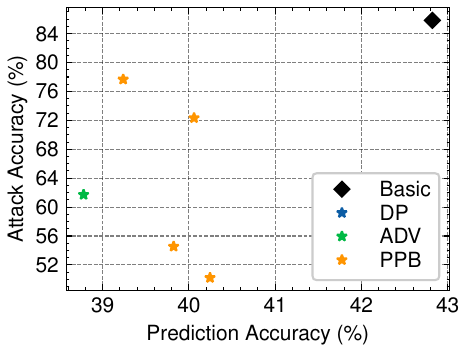}
\caption{CIFAR100, DenseNet121}
\end{subfigure}
\begin{subfigure}[b]{0.24\textwidth}
\includegraphics[width=\linewidth]{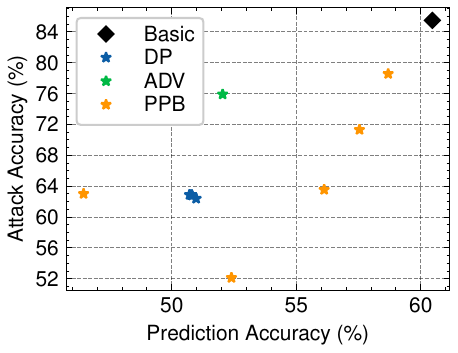}
\caption{Location, FC}
\end{subfigure}
\begin{subfigure}[b]{0.24\textwidth}
\includegraphics[width=\linewidth]{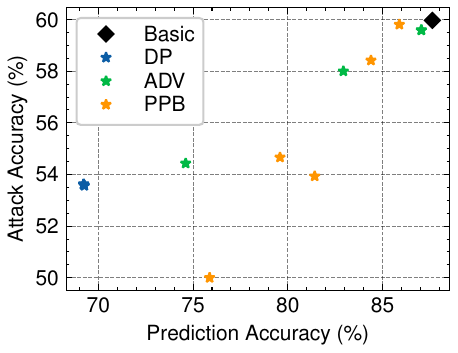}
\caption{Purchase, FC}
\end{subfigure}
\caption{Performance of defenses for different datasets (L1 Unstructured, Sparsity 0.6).}
\label{fig:defense_l1unstructured_0.6}
\end{figure*}

In addition, the PPB defense is only applied in fine-tuning of neural network pruning by using KL-divergence as a regularization term. Thus, the defense does not include the additional computational costs in the inference phase. 

After applying PPB defense, we observe the divergence between the member and non-member data is significantly reduced by comparing Figure~\ref{fig:hist_defense} with Figure~\ref{fig:histogram}. 
Such decreased divergence can be observed in different classes by comparing Figure~\ref{fig:cls_div} with Figure~\ref{fig:div_cls_defense}. 
Both changes on the distributions of the pruned model's posterior predictions indicate that the PPB defense makes the attack model fail to learn the binary classification thresholds from the prediction confidence and sensitivity.
Moreover, the PPB defense is designed to change the distribution of predictions instead of their orders. In other words, the PPB defense will not change the predicted classes of the pruned models during fine-tuning, which largely preserves the prediction accuracy of pruned models.

As shown in Figure~\ref{fig:div_cls_defense}, such decreased divergence can also be preserved in different classes. After applying PPB defense, the divergence of the pruned model is close to that of the original model (comparing Figure~\ref{fig:cls_div} with Figure~\ref{fig:div_cls_defense}), which indicates the effectiveness of the PPB defense.

\subsection{Defense Evaluation}
This section evaluates the effectiveness of PPB by comparing the performance of PPB with that of state-of-the-art defenses\footnote{The defenses are evaluated by the empirical experiments. We will investigate the strict privacy guarantee in future work.
}.

\subsubsection{State-of-the-art Defenses}
We investigate three state-of-the-art defenses against MIA attacks in neural network pruning.

\vspace{0.2em}
\noindent
\textbf{Early Stopping and L2 Regularization (Basic).} 
Early stopping and l2 regularization have been used to successfully defend membership inference attacks with competitive performance~\cite{shokri2017membership,salem2019ml,song2020systematic}. As discussed in Section~\ref{4.1}, an adversary infers the membership of a sample based on the divergence of the prediction confidences between members and non-members. Such divergence becomes more severe as the number of training epochs increases, due to the increased memorization. Hence, the early stopping mechanism with fewer training epochs and l2 regularization for penalizing the over-training can tradeoff a slight reduction in model accuracy with lower privacy risk. 
In the evaluation, we stop the training and fine-tuning when the validation loss is not decreased for five epochs using early stopping mechanism. In l2 regularization, we set the regularization factor as $0.0005$. \textit{Note that we use early stopping and l2 regularization in all the other defenses to improve the defense performance.}

\vspace{0.2em}
\noindent
\textbf{Differential Privacy (DP).}
Differential privacy is a strategy to bound the individual information exposure when running an algorithm $f$ and has been widely investigated for preventing privacy leakage against membership inference attacks~\cite{truex2019effects,rahimian2019differential,chen2020differential}.
We implement differentially private SGD (DPSGD)~\cite{abadi2016deep,rahimian2019differential}, one of the most widely-used defense techniques, to train neural networks with DP guarantees. 
Following DPSGD, we first clip the gradient, then add noise to the gradient, and use the generated noisy gradient to update the model's parameters. The noise is sampled from a Gaussian distribution $\mathcal{N}(0,\sigma)$. To achieve ($\epsilon, \delta$)-DP, the standard deviation of the Gaussian distribution, i.e., $\delta$, should be in the order of $\Omega(q\sqrt{T\log (1/\delta)/\epsilon}$, where $q$ denotes the sampling ratio and $T$ denotes the total number of iterations.
Accordingly, the privacy guarantee of the DP defense can be derived from $\delta$, which plays an important role in balancing utility and privacy.
Therefore, in the defense evaluation, we evaluate the effectiveness of DP defense and explore the impact of different privacy budgets (\ie, different values of $\delta$).

\vspace{0.2em}\noindent
\textbf{Adversarial Regularization (ADV).} 
Nasr~\etal proposed to consider the membership inference adversary in the training process~\cite{nasr2018machine}. The defender first trains a surrogate attack model to distinguish between members and non-members and then trains the target model to minimize the prediction loss while maximizing the classification loss of the surrogate attack model. A parameter $\alpha$ is used to balance the prediction performance and privacy risk. 
ADV is applied in the fine-tuning process of pruning to protect the privacy of pruned models.

\begin{figure*}[!tb]
\centering
\begin{subfigure}[b]{0.24\textwidth}
\includegraphics[width=\linewidth]{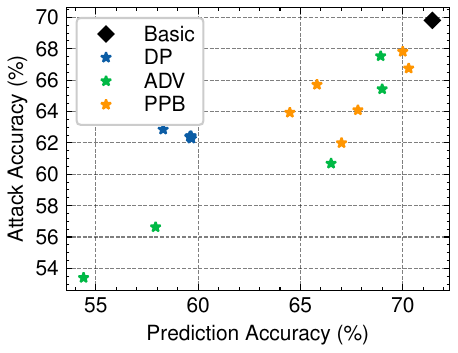}
\caption{L1 Unstructured}
\end{subfigure}
\begin{subfigure}[b]{0.24\textwidth}
\includegraphics[width=\linewidth]{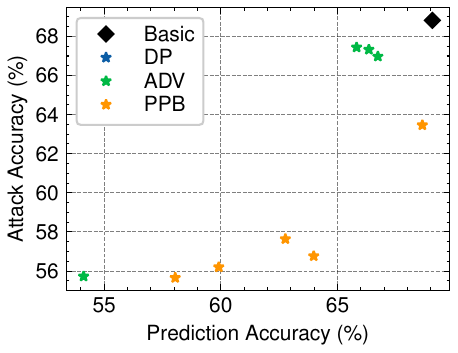}
\caption{L1 Structured}
\end{subfigure}
\begin{subfigure}[b]{0.24\textwidth}
\includegraphics[width=\linewidth]{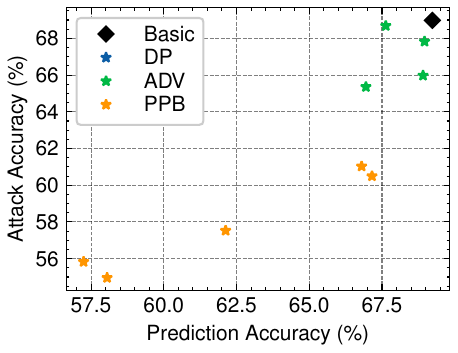}
\caption{L2 Structured}
\end{subfigure}
\begin{subfigure}[b]{0.24\textwidth}
\includegraphics[width=\linewidth]{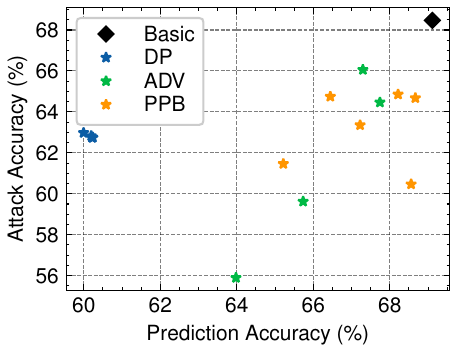}
\caption{Slimming}
\end{subfigure}
\caption{Performance of defenses against adaptive attacks for different pruning approaches (CIFAR10, ResNet18, Sparsity 0.6).}
\label{fig:defense_cifar10_resnet0.6_adp}
\end{figure*}

\begin{figure*}[!tb]
\centering
\begin{subfigure}[b]{0.24\textwidth}
\includegraphics[width=\linewidth]{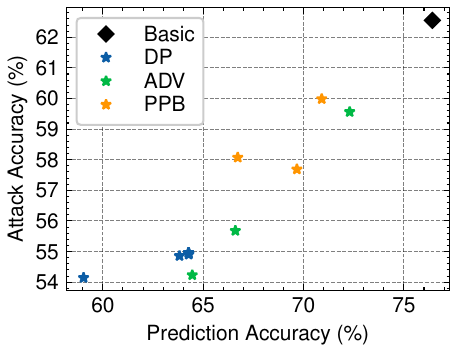}
\caption{CIFAR10, DenseNet121}
\end{subfigure}
\begin{subfigure}[b]{0.24\textwidth}
\includegraphics[width=\linewidth]{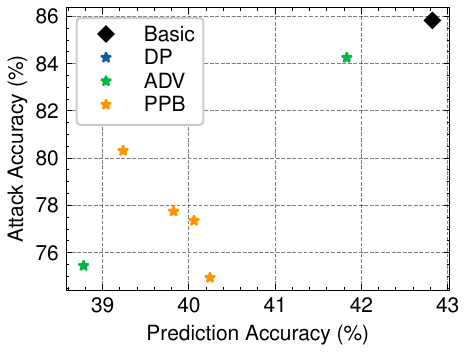}
\caption{CIFAR100, DenseNet121}
\end{subfigure}
\begin{subfigure}[b]{0.24\textwidth}
\includegraphics[width=\linewidth]{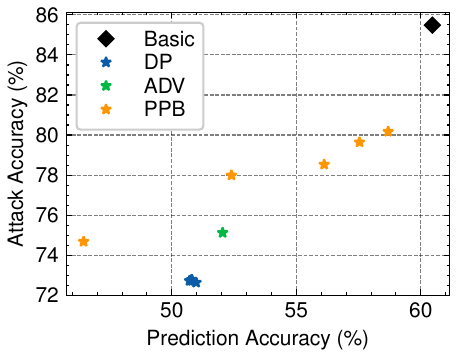}
\caption{Location, FC}
\end{subfigure}
\begin{subfigure}[b]{0.24\textwidth}
\includegraphics[width=\linewidth]{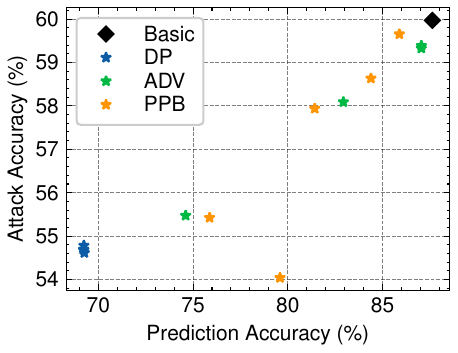}
\caption{Purchase, FC}
\end{subfigure}
\caption{Performance of defenses against adaptive attacks for different datasets (L1 Unstructured, Sparsity 0.6).}
\label{fig:defense_l1unstructured_0.6_adp}
\end{figure*}

\subsubsection{Experimental Results of Defenses}\label{sec:defense_eval}

We use the same settings of attack evaluations in Section~\ref{sec:eval} and conduct the following experiments with defenses in the process of pruning. 
Since there is always a trade-off between privacy and prediction accuracy when implementing defenses, we explore different settings of hyper-parameters in the defensive mechanisms to thoroughly evaluate the defense performance. Specifically, we set hyper-parameter $\lambda \in \{1, 2,4,8, 16\}$ in PPB, $\sigma \in \{0.01, 0.1, 1, 10, 100\}$) in the DP noise vectors, $\alpha \in \{0.5, 1, 2, 4, 8, 16\}$ in ADV, respectively. 

Figure~\ref{fig:defense_cifar10_resnet} and~\ref{fig:defense_l1unstructured_0.6} illustrate the prediction accuracy and attack accuracy with different defense mechanisms. For better illustration, we remove the results if the model with a specific hyper-parameter cannot achieve 75\% of the basic defense's prediction accuracy, \ie, poor prediction performance, or result in a higher attack accuracy, \ie, ineffective defense. 

We observe that PPB is especially effective in protecting all pruning approaches from attacks, which can reduce the attack accuracy to around 50\% (random guessing accuracy), while not degrading the prediction accuracy too much.
Hence, \textit{PPB defense provides a privacy-preserving approach with minimal degradation of prediction accuracy}. 
In addition, ADV is also effective in the L1 unstructured and Slimming pruning, but fails to achieve a good balance between prediction performance and privacy in the L1 structured and L2 structured pruning. Similar to the fact shown in recent work~\cite{jayaraman2019evaluating}, we observe DP can hardly balance this utility-privacy tradeoff.

\subsubsection{Defenses against Adaptive Attacks}
To rigorously evaluate the defense performance, we consider adaptive attacks, where the adversary knows all the details of defenses along with the pruning information. 
In adaptive attacks, the adversary trains a shadow pruned model following the same defense mechanism (\eg, Basic, DP, ADV, and proposed PPB) and pruning process. The adversary then performs the SAMIA attack based on the shadow pruned model.

As shown in Figure~\ref{fig:defense_cifar10_resnet0.6_adp} and~\ref{fig:defense_l1unstructured_0.6_adp}, we observe that PPB reduces the accuracy of adaptive attacks compared to the attacks on the pruned model without defenses and provides the best protection in L1 structured and L2 structured pruning. Besides, for the L1 unstructured and Slimming pruning, PPB and ADV are the best two defenses. 
\textit{PPB is designed towards pruned models by reducing the confidence and sensitivity gap. Therefore, in general, PPB provides good protection in all pruning approaches.
}
In addition, ADV is designed to mitigate the confidence gap, which is largely increased in L1 unstructured and slimming pruning (as discussed in Section~\ref{sec:impact_gap}). Hence, ADV is also effective in protecting pruned models using L1 unstructured and slimming pruning.

\section{Conclusion}
This paper conducted the first analysis of privacy risks in neural network pruning. We first explored the impacts of neural network pruning on prediction divergence, based on which, a new membership inference attack, \ie, self-attention membership inference attack (SAMIA), is proposed against the pruned neural network models. Through comprehensive and rigorous evaluation, we demonstrated the substantially increased privacy risks of the pruned models. We found that the privacy risks of the pruned models are tightly related to the confidence gap, sensitivity gap, and generalization gap due to pruning. Besides, even without knowing the pruning approach, the membership inference attacks can still achieve high attack accuracy against the pruned model.
Especially, the proposed SAMIA showed superiority in identifying the pruned models' prediction divergence by using finer-grained prediction metrics, which is recommended as a competitive baseline attack model for future privacy risk study of neural network pruning.

In addition, to defend the attacks, we proposed a pair-based posterior balancing named as PPB by reducing the prediction divergence of fine-tuning process during neural network pruning. We experimentally demonstrated that PPB could reduce the attack accuracy to around 50\% (random guessing accuracy) without considering adaptive attacks and achieve the best protection compared with the three existing defenses. Besides, PPB showed competitive performance even when defending adaptive attacks.

The proposed SAMIA attack will be further explored under more challenging MIA settings, such as the label-only MIA without available confidences, where the existing label-only MIA attacks using data augmentation~\cite{choquette2021label} and black-box adversary~\cite{li2021membership} can be potentially integrated for more powerful attack capability. 
We hope our work convinces the community about the importance of exploring innovative neural network pruning approaches by taking privacy-preserving into consideration.

\section*{Acknowledgement}
We would like to thank our shepherd, Yinzhi Cao, and the anonymous reviewers for their constructive suggestions. This work was supported in part by National Science Foundation (CCF-2106754).

\section*{Availability}
Our code is publicly available at \url{https://github.com/Machine-Learning-Security-Lab/mia_prune} for the purpose of reproducible research. 

\bibliographystyle{unsrt}
\bibliography{deep,compress,mia,privacy,security}

\clearpage
\appendix
\large
\onecolumn
\setcounter{page}{1}

\begin{center}
\textbf{\Large Appendix}
\end{center}
\vspace{3em}

In the Appendix, we first present the attack evaluation for the remaining experimental settings (Section A). Then we compare the defense performance of the proposed PPB with other state-of-the-art defenses for the remaining experimental settings (Section B).

\section{Extended Experiments for Attack Evaluation}
In this section, we provide the extended experiments for attack evaluation and report the prediction accuracy and the attack performance in the remaining experimental settings, \ie,  datasets (CIFAR10, CIFAR100, CHMNIST, SVHN, Location, Purchase, and Texas), model architectures (VGG16, ResNet18, DenseNet121).

\clearpage
\subsection{Prediction Accuracy of Pruned Models}
This section presents the prediction accuracy of pruned models using different pruning approaches and sparsity levels on the remaining experiment settings, including the model architectures (ResNet18, DenseNet121, VGG16, and FC) and the datasets (CIFAR10, CIFAR100, CHMNIST, SVHN, Location, Purchase, and Texas).
\begin{figure*}[!h]
\centering
\begin{subfigure}[b]{0.24\linewidth}
\centering
\includegraphics[width=\linewidth]{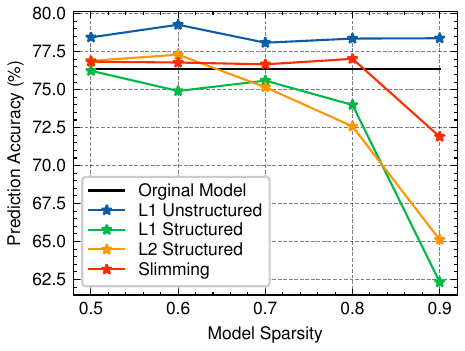}
\caption{CIFAR10, VGG16}
\label{fig:mia_cifar10_vgg16}
\end{subfigure}
\begin{subfigure}[b]{0.24\linewidth}
\centering
\includegraphics[width=\linewidth]{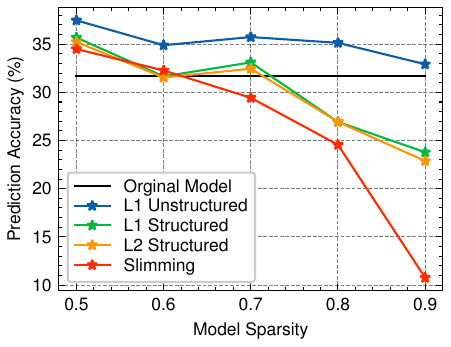}
\caption{CIFAR100, ResNet18}
\end{subfigure}
\begin{subfigure}[b]{0.24\linewidth}
\centering
\includegraphics[width=\linewidth]{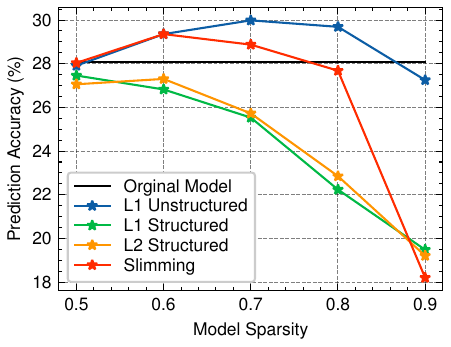}
\caption{CIFAR100, VGG16}
\end{subfigure}

\begin{subfigure}[b]{0.24\linewidth}
\centering
\includegraphics[width=\linewidth]{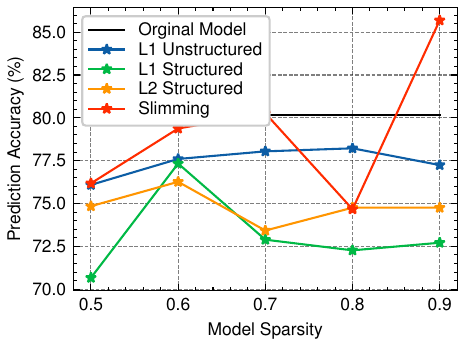}
\caption{CHMNIST, ResNet18}
\end{subfigure}
\begin{subfigure}[b]{0.24\linewidth}
\centering
\includegraphics[width=\linewidth]{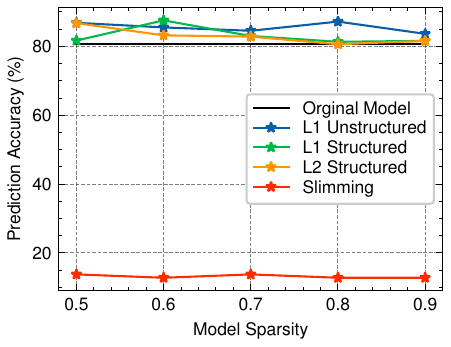}
\caption{CHMNIST, DenseNet121}
\end{subfigure}
\begin{subfigure}[b]{0.24\linewidth}
\centering
\includegraphics[width=\linewidth]{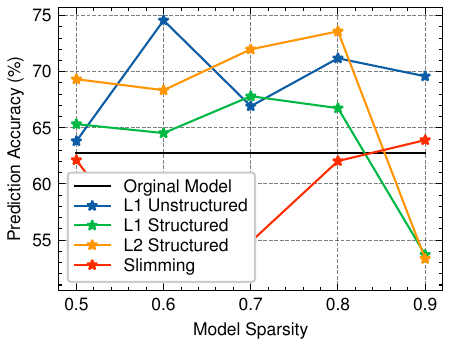}
\caption{CHMNIST, VGG16}
\end{subfigure}

\begin{subfigure}[b]{0.24\linewidth}
\centering
\includegraphics[width=\linewidth]{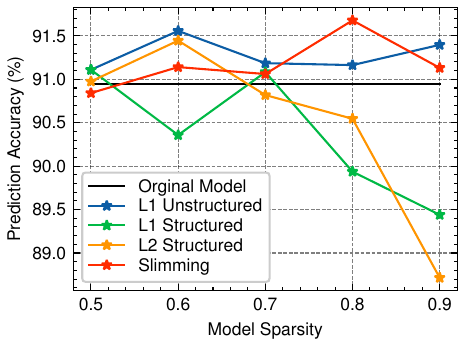}
\caption{SVHN, ResNet18}
\end{subfigure}
\begin{subfigure}[b]{0.24\linewidth}
\centering
\includegraphics[width=\linewidth]{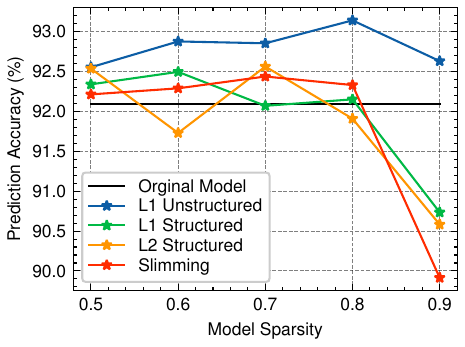}
\caption{SVHN, DenseNet121}
\end{subfigure}
\begin{subfigure}[b]{0.24\linewidth}
\centering
\includegraphics[width=\linewidth]{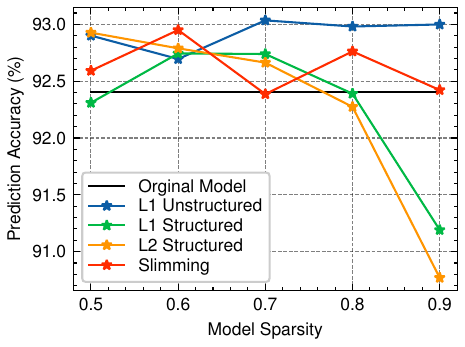}
\caption{SVHN, VGG16}
\end{subfigure}

\begin{subfigure}[b]{0.24\linewidth}
\includegraphics[width=\linewidth]{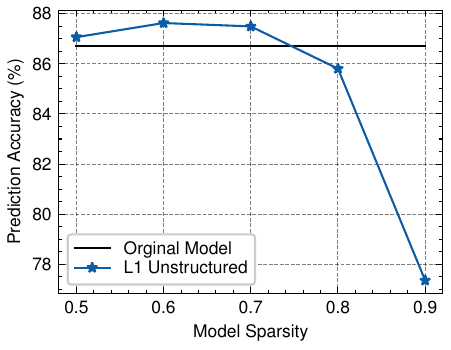}
\caption{Purchase, FC}
\end{subfigure}
\begin{subfigure}[b]{0.24\linewidth}
\includegraphics[width=\linewidth]{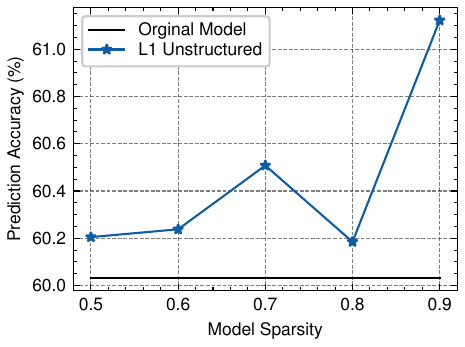}
\caption{Texas, FC}
\end{subfigure}
\caption{Prediction accuracy (test accuracy) of the pruned models using different pruning approaches and sparsity levels. Each point indicates the prediction accuracy achieved by the pruned model with a specific pruning approach and sparsity level. The black line indicates the prediction accuracy of the original models.}
\end{figure*}

\clearpage
\subsection{Privacy Risks of Pruned Models}
\label{app:attack_acc}
This section presents the attack accuracy of the remaining experiment settings. 
Similar to the observation in the main paper, we find that in most cases, the pruned models become more vulnerable to the membership inference attacks than the original models.

\begin{figure*}[!h]
    \centering
    \begin{subfigure}[b]{0.24\textwidth}
    \includegraphics[width=\linewidth]{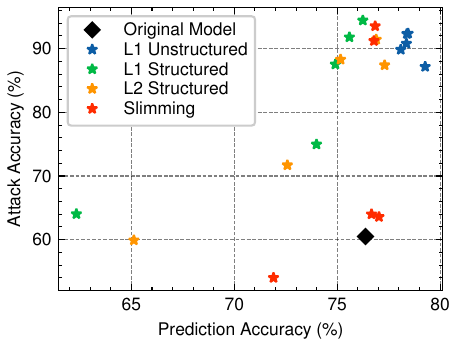}
    \caption{CIFAR10, VGG16}
    \end{subfigure}
    \begin{subfigure}[b]{0.24\textwidth}
    \includegraphics[width=\linewidth]{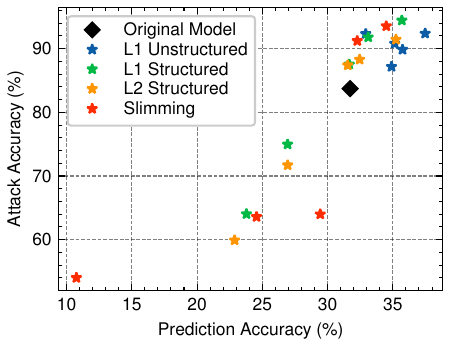}
    \caption{CIFAR100, ResNet18}
    \end{subfigure}
    \begin{subfigure}[b]{0.24\textwidth}
    \includegraphics[width=\linewidth]{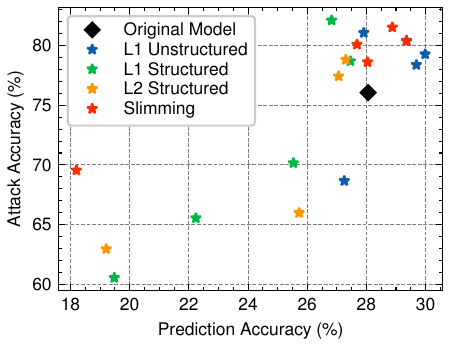}
    \caption{CIFAR100, VGG16}
    \end{subfigure}
    
    \begin{subfigure}[b]{0.24\textwidth}
    \includegraphics[width=\linewidth]{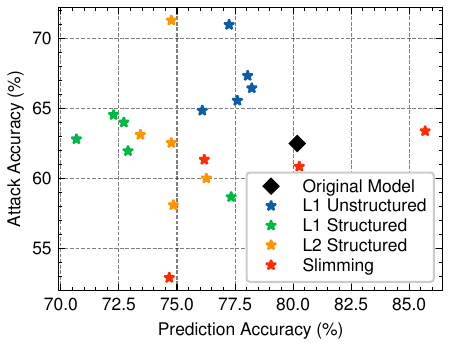}
    \caption{CHMNIST, ResNet18}
    \end{subfigure}
    \begin{subfigure}[b]{0.24\textwidth}
    \includegraphics[width=\linewidth]{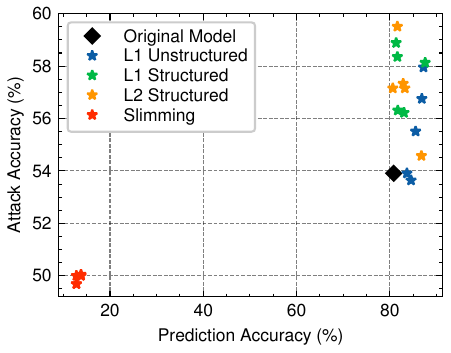}
    \caption{CHMNIST, DenseNet121}
    \end{subfigure}
    \begin{subfigure}[b]{0.24\textwidth}
    \includegraphics[width=\linewidth]{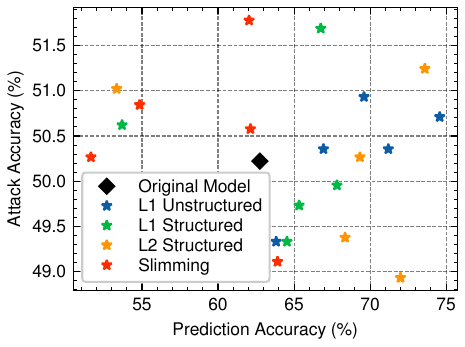}
    \caption{CHMNIST, VGG16}
    \end{subfigure}
    
    \begin{subfigure}[b]{0.24\textwidth}
    \includegraphics[width=\linewidth]{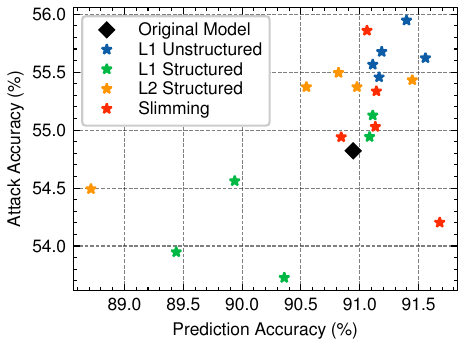}
    \caption{SVHN, ResNet18}
    \end{subfigure}
    \begin{subfigure}[b]{0.24\textwidth}
    \includegraphics[width=\linewidth]{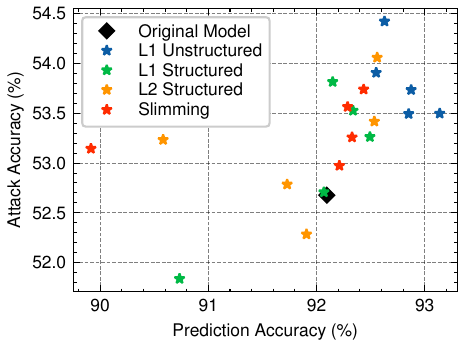}
    \caption{SVHN, DenseNet121}
    \end{subfigure}
    \begin{subfigure}[b]{0.24\textwidth}
    \includegraphics[width=\linewidth]{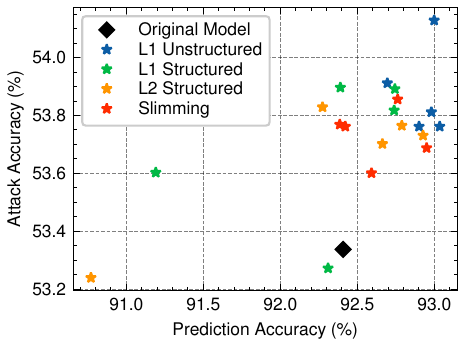}
    \caption{SVHN, VGG16}
    \end{subfigure}
    
    \begin{subfigure}[b]{0.24\textwidth}
    \includegraphics[width=\linewidth]{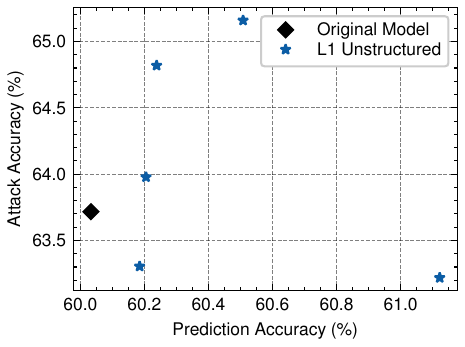}
    \caption{Texas, FC}
    \end{subfigure}
    \begin{subfigure}[b]{0.24\textwidth}
    \includegraphics[width=\linewidth]{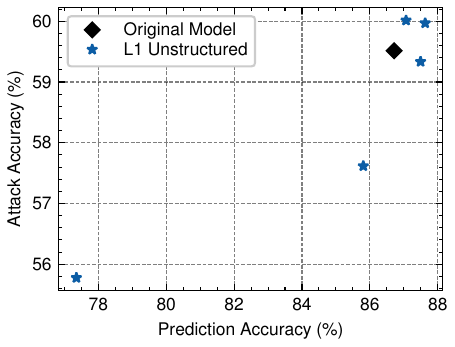}
    \caption{Purchase, FC}
    \end{subfigure}
    \caption{Privacy Risks of Neural Network Pruning (w.r.t. prediction accuracy). Most pruning approaches results in a higher attack accuracy when considering a similar prediction accuracy, compared with the original models. We present the attack accuracy of SAMIA for pruned models and the attack accuracy of Conf attack for the original models.}
\end{figure*}

\newpage

\subsection{Effectiveness of SAMIA}
We investigate the effectiveness of SAMIA.
Although most investigated MIAs are effective in attacking pruned models, SAMIA results in higher attack accuracy than other attacks in most cases.
SAMIA does not perform among the best attacks mainly in two cases: 1) the attack accuracy is close to 50\% (\ie, random guessing); 2) the sparsity level is low (\eg, $0.5$). When the sparsity level is low, the divergence of confidence and sensitivity is reduced. In this case, the features of confidence and sensitivity cannot be fully utilized for SAMIA, which makes the SAMIA's attack performance close to the existing attacks.

\begin{figure*}[!h]
\centering
\begin{subfigure}[b]{0.24\linewidth}
\includegraphics[width=0.95\linewidth]{figs/exp/mias/mias_cifar10_densenet121_level.pdf}
\caption{L1 unstructured}
\end{subfigure}
\begin{subfigure}[b]{0.24\linewidth}
\includegraphics[width=0.95\linewidth]{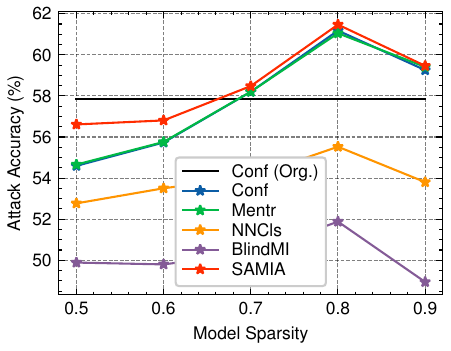}
\caption{L1 structured}
\end{subfigure}
\begin{subfigure}[b]{0.24\linewidth}
\includegraphics[width=0.95\linewidth]{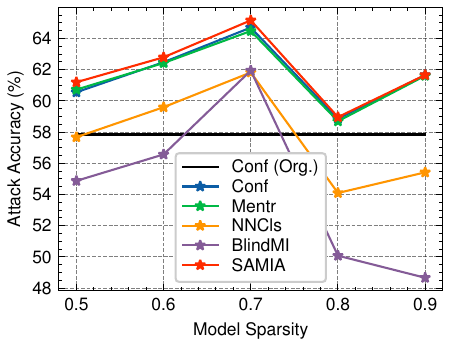}
\caption{L2 structured}
\end{subfigure}
\begin{subfigure}[b]{0.24\linewidth}
\includegraphics[width=0.95\linewidth]{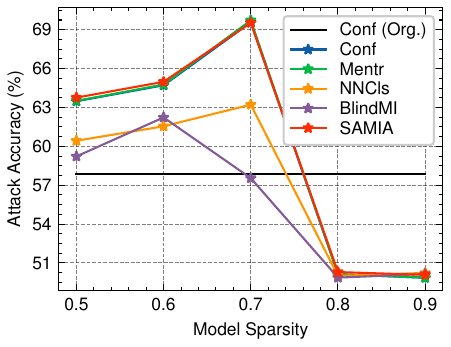}
\caption{Slimming}
\end{subfigure}
\begin{subfigure}[b]{0.24\linewidth}
\includegraphics[width=0.95\linewidth]{figs/exp/mias/mias_cifar10_densenet121_level_2.pdf}
\caption{L1 unstructured}
\end{subfigure}
\begin{subfigure}[b]{0.24\linewidth}
\includegraphics[width=0.95\linewidth]{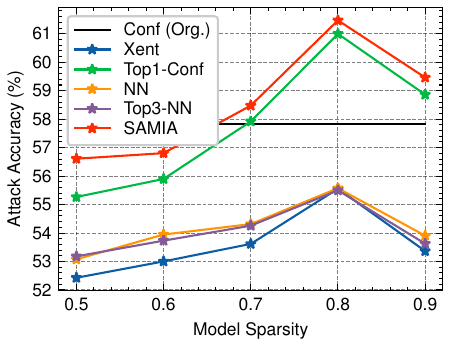}
\caption{L1 structured}
\end{subfigure}
\begin{subfigure}[b]{0.24\linewidth}
\includegraphics[width=0.95\linewidth]{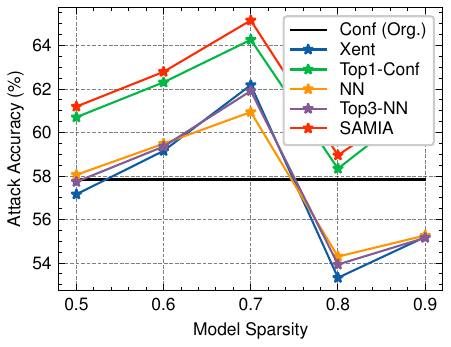}
\caption{L2 structured}
\end{subfigure}
\begin{subfigure}[b]{0.24\linewidth}
\includegraphics[width=0.95\linewidth]{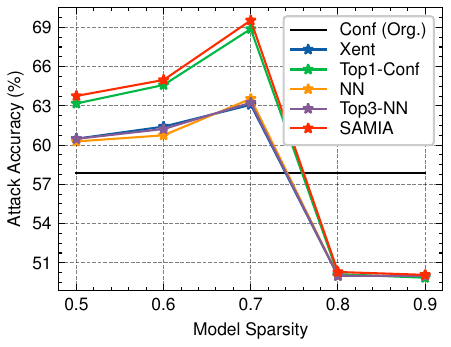}
\caption{Slimming}
\end{subfigure}
\caption{Attack performance comparison of MIAs (CIFAR10, DenseNet121).}
\end{figure*}

\begin{figure*}[!h]
\centering
\begin{subfigure}[b]{0.24\linewidth}
\includegraphics[width=0.95\linewidth]{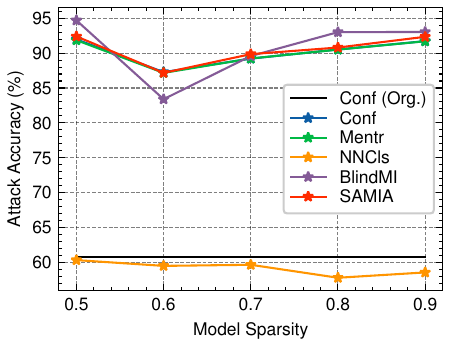}
\caption{L1 unstructured}
\end{subfigure}
\begin{subfigure}[b]{0.24\linewidth}
\includegraphics[width=0.95\linewidth]{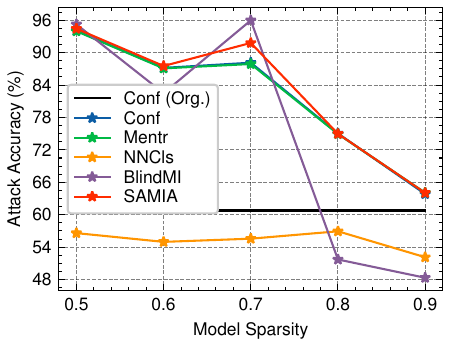}
\caption{L1 structured}
\end{subfigure}
\begin{subfigure}[b]{0.24\linewidth}
\includegraphics[width=0.95\linewidth]{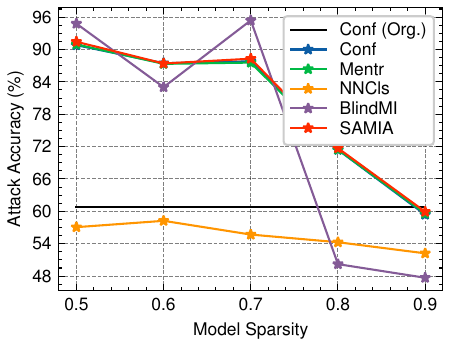}
\caption{L2 structured}
\end{subfigure}
\begin{subfigure}[b]{0.24\linewidth}
\includegraphics[width=0.95\linewidth]{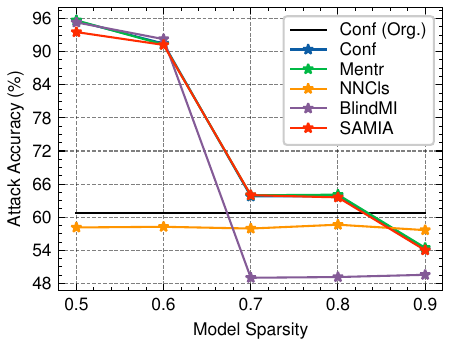}
\caption{Slimming}
\end{subfigure}
\begin{subfigure}[b]{0.24\linewidth}
\includegraphics[width=0.95\linewidth]{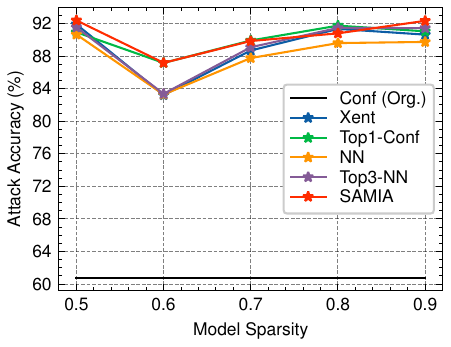}
\caption{L1 unstructured}
\end{subfigure}
\begin{subfigure}[b]{0.24\linewidth}
\includegraphics[width=0.95\linewidth]{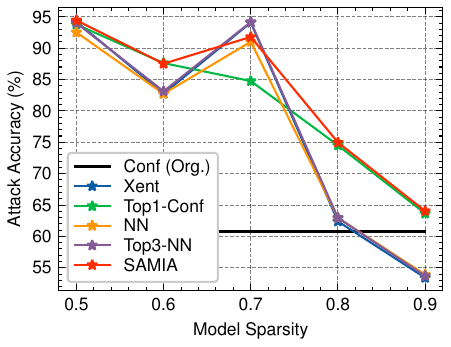}
\caption{L1 structured}
\end{subfigure}
\begin{subfigure}[b]{0.24\linewidth}
\includegraphics[width=0.95\linewidth]{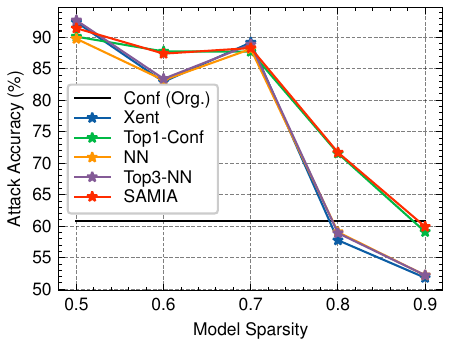}
\caption{L2 structured}
\end{subfigure}
\begin{subfigure}[b]{0.24\linewidth}
\includegraphics[width=0.95\linewidth]{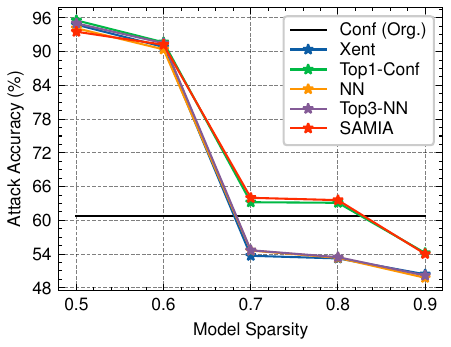}
\caption{Slimming}
\end{subfigure}
\caption{Attack performance comparison of MIAs (CIFAR10, VGG16).}
\end{figure*}

\begin{figure*}[!h]
\centering
\begin{subfigure}[b]{0.24\linewidth}
\includegraphics[width=0.95\linewidth]{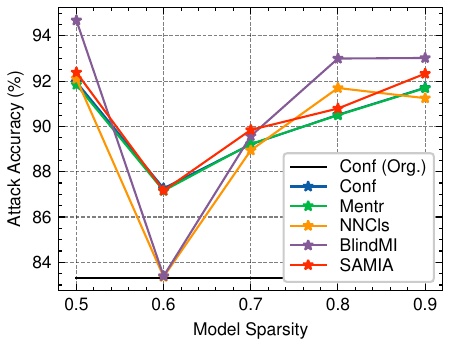}
\caption{L1 unstructured}
\end{subfigure}
\begin{subfigure}[b]{0.24\linewidth}
\includegraphics[width=0.95\linewidth]{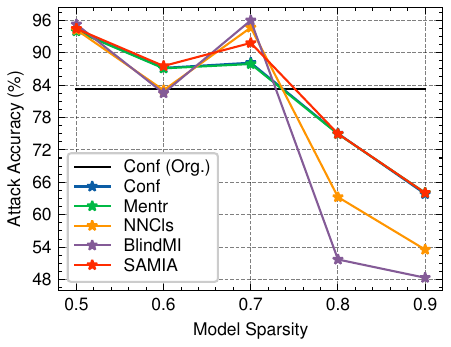}
\caption{L1 structured}
\end{subfigure}
\begin{subfigure}[b]{0.24\linewidth}
\includegraphics[width=0.95\linewidth]{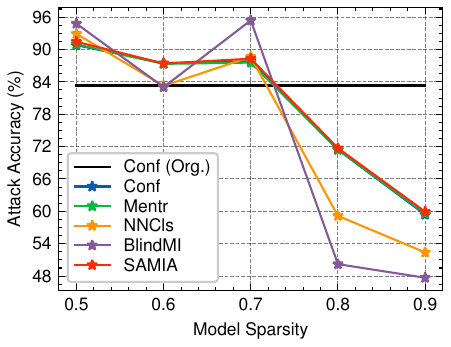}
\caption{L2 structured}
\end{subfigure}
\begin{subfigure}[b]{0.24\linewidth}
\includegraphics[width=0.95\linewidth]{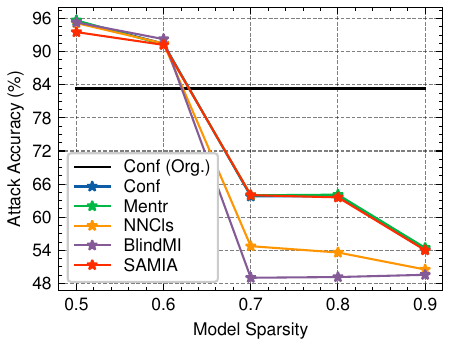}
\caption{Slimming}
\end{subfigure}
\begin{subfigure}[b]{0.24\linewidth}
\includegraphics[width=0.95\linewidth]{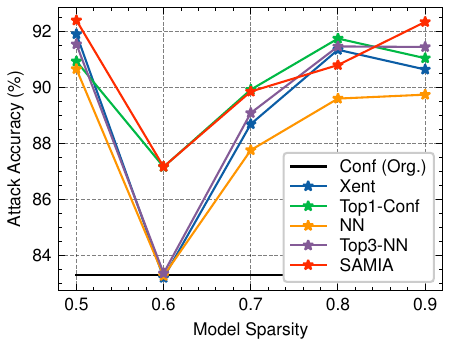}
\caption{L1 unstructured}
\end{subfigure}
\begin{subfigure}[b]{0.24\linewidth}
\includegraphics[width=0.95\linewidth]{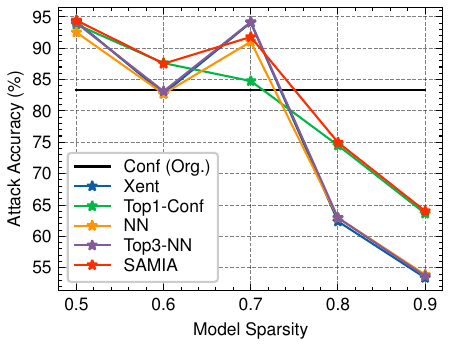}
\caption{L1 structured}
\end{subfigure}
\begin{subfigure}[b]{0.24\linewidth}
\includegraphics[width=0.95\linewidth]{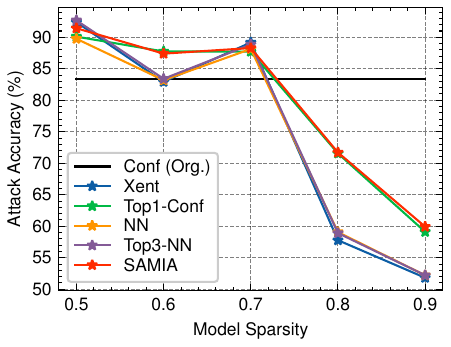}
\caption{L2 structured}
\end{subfigure}
\begin{subfigure}[b]{0.24\linewidth}
\includegraphics[width=0.95\linewidth]{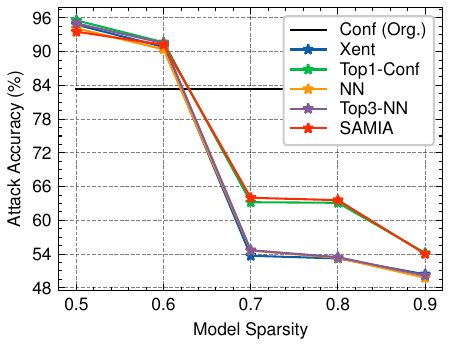}
\caption{Slimming}
\end{subfigure}
\caption{Attack performance comparison of MIAs (CIFAR100, ResNet18). }
\end{figure*}

\begin{figure*}[!h]
\centering
\begin{subfigure}[b]{0.24\linewidth}
\includegraphics[width=0.95\linewidth]{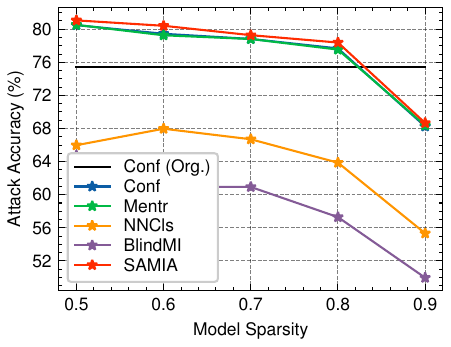}
\caption{L1 unstructured}
\end{subfigure}
\begin{subfigure}[b]{0.24\linewidth}
\includegraphics[width=0.95\linewidth]{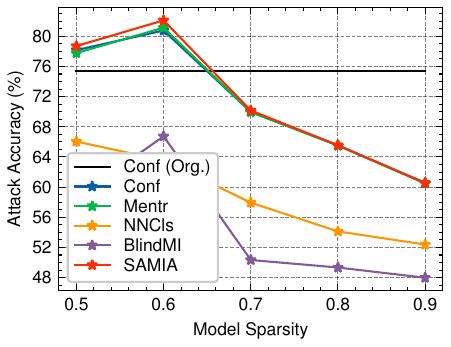}
\caption{L1 structured}
\end{subfigure}
\begin{subfigure}[b]{0.24\linewidth}
\includegraphics[width=0.95\linewidth]{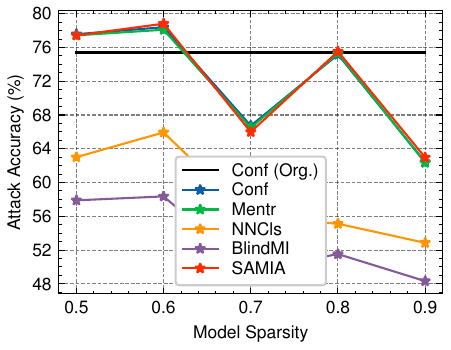}
\caption{L2 structured}
\end{subfigure}
\begin{subfigure}[b]{0.24\linewidth}
\includegraphics[width=0.95\linewidth]{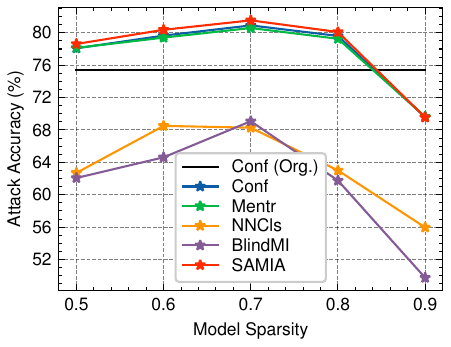}
\caption{Slimming}
\end{subfigure}
\begin{subfigure}[b]{0.24\linewidth}
\includegraphics[width=0.95\linewidth]{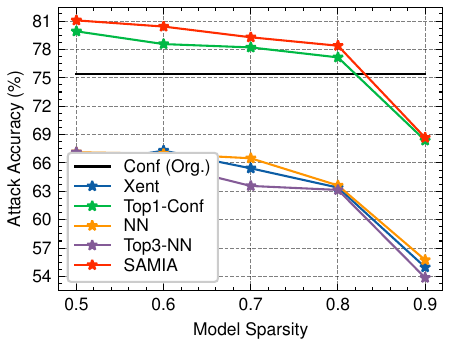}
\caption{L1 unstructured}
\end{subfigure}
\begin{subfigure}[b]{0.24\linewidth}
\includegraphics[width=0.95\linewidth]{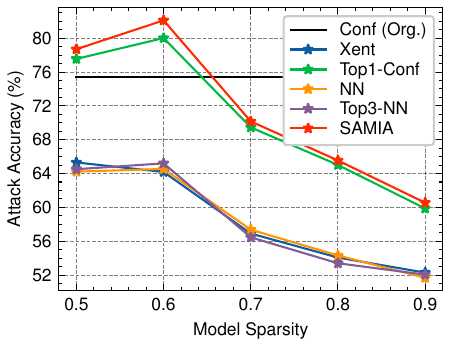}
\caption{L1 structured}
\end{subfigure}
\begin{subfigure}[b]{0.24\linewidth}
\includegraphics[width=0.95\linewidth]{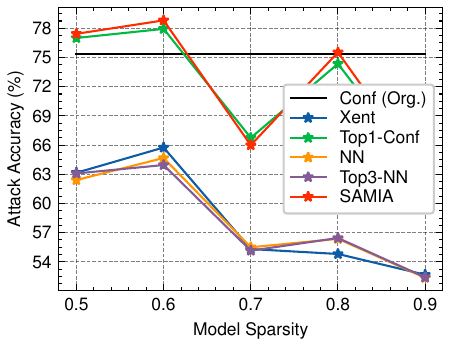}
\caption{L2 structured}
\end{subfigure}
\begin{subfigure}[b]{0.24\linewidth}
\includegraphics[width=0.95\linewidth]{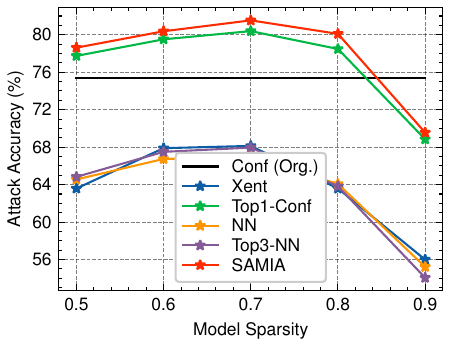}
\caption{Slimming}
\end{subfigure}
\caption{Attack performance comparison of MIAs (CIFAR100, VGG16). }

\end{figure*}

\begin{figure*}[!h]
\centering
\begin{subfigure}[b]{0.24\linewidth}
\includegraphics[width=0.95\linewidth]{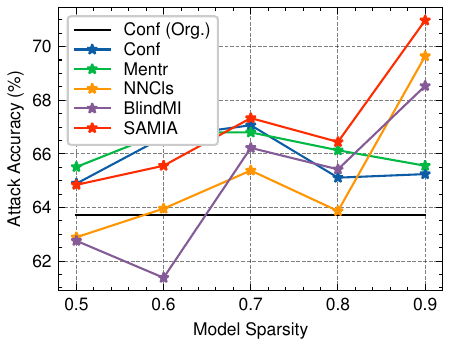}
\caption{L1 unstructured}
\end{subfigure}
\begin{subfigure}[b]{0.24\linewidth}
\includegraphics[width=0.95\linewidth]{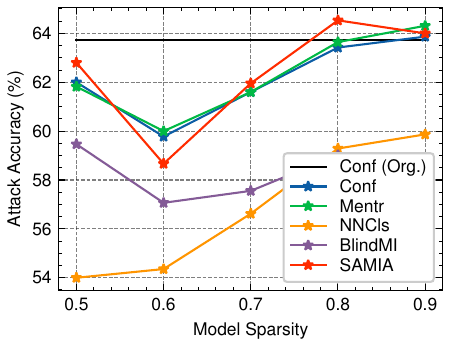}
\caption{L1 structured}
\end{subfigure}
\begin{subfigure}[b]{0.24\linewidth}
\includegraphics[width=0.95\linewidth]{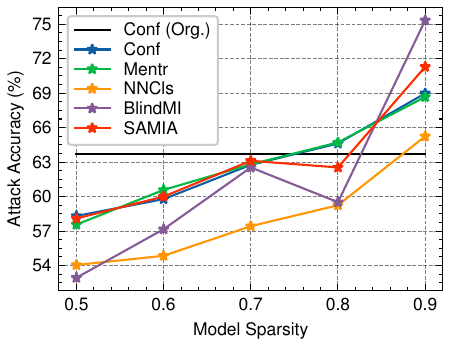}
\caption{L2 structured}
\end{subfigure}
\begin{subfigure}[b]{0.24\linewidth}
\includegraphics[width=0.95\linewidth]{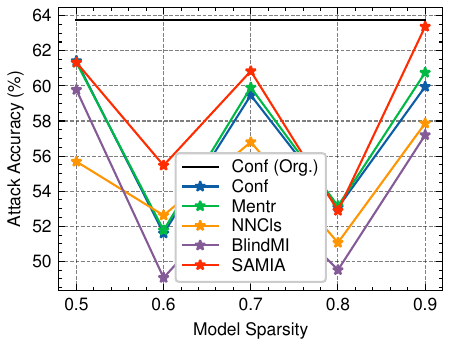}
\caption{Slimming}
\end{subfigure}
\begin{subfigure}[b]{0.24\linewidth}
\includegraphics[width=0.95\linewidth]{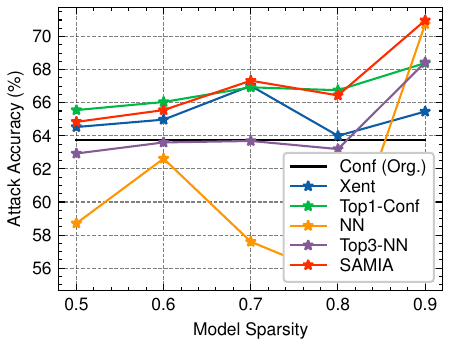}
\caption{L1 unstructured}
\end{subfigure}
\begin{subfigure}[b]{0.24\linewidth}
\includegraphics[width=0.95\linewidth]{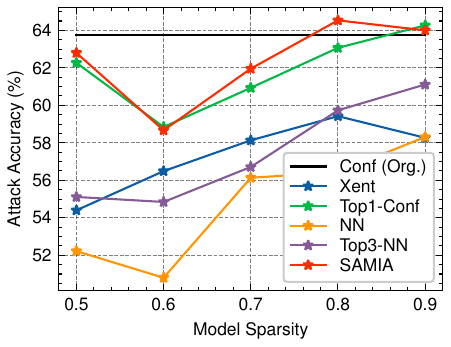}
\caption{L1 structured}
\end{subfigure}
\begin{subfigure}[b]{0.24\linewidth}
\includegraphics[width=0.95\linewidth]{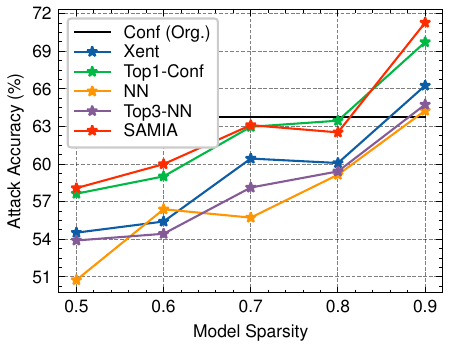}
\caption{L2 structured}
\end{subfigure}
\begin{subfigure}[b]{0.24\linewidth}
\includegraphics[width=0.95\linewidth]{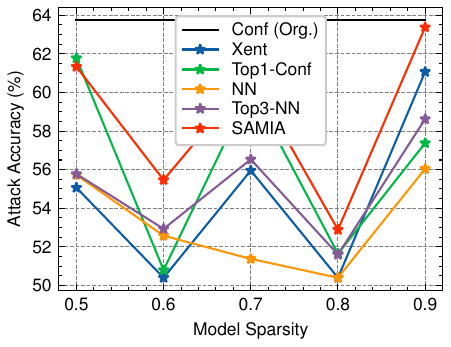}
\caption{Slimming}
\end{subfigure}
\caption{Attack performance comparison of MIAs (CHMNIST, ResNet18). }

\end{figure*}

\begin{figure*}[!h]
\centering
\begin{subfigure}[b]{0.24\linewidth}
\includegraphics[width=0.95\linewidth]{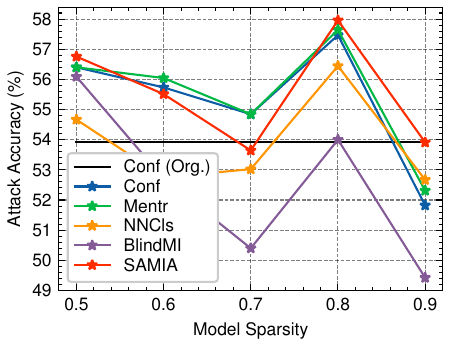}
\caption{L1 unstructured}
\end{subfigure}
\begin{subfigure}[b]{0.24\linewidth}
\includegraphics[width=0.95\linewidth]{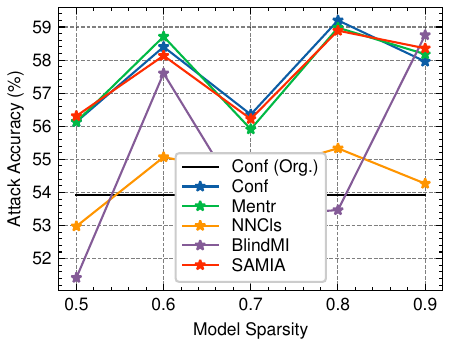}
\caption{L1 structured}
\end{subfigure}
\begin{subfigure}[b]{0.24\linewidth}
\includegraphics[width=0.95\linewidth]{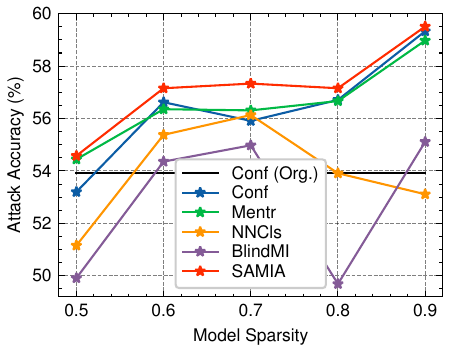}
\caption{L2 structured}
\end{subfigure}
\begin{subfigure}[b]{0.24\linewidth}
\includegraphics[width=0.95\linewidth]{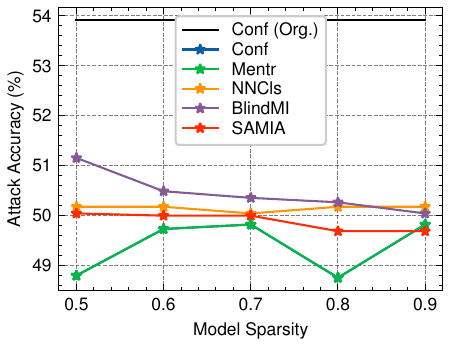}
\caption{Slimming}
\end{subfigure}
\begin{subfigure}[b]{0.24\linewidth}
\includegraphics[width=0.95\linewidth]{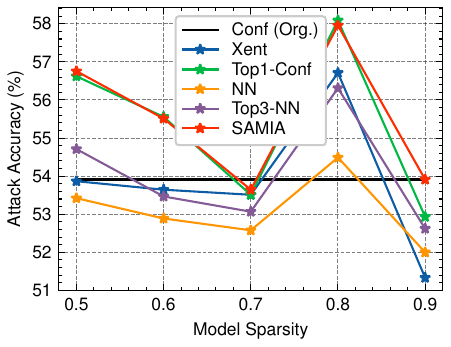}
\caption{L1 unstructured}
\end{subfigure}
\begin{subfigure}[b]{0.24\linewidth}
\includegraphics[width=0.95\linewidth]{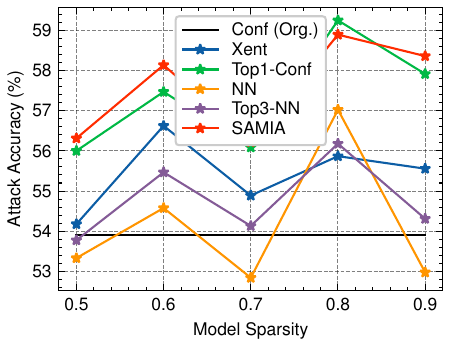}
\caption{L1 structured}
\end{subfigure}
\begin{subfigure}[b]{0.24\linewidth}
\includegraphics[width=0.95\linewidth]{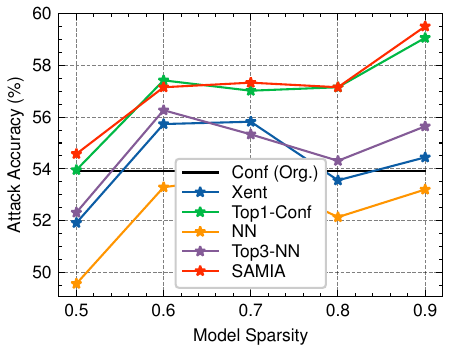}
\caption{L2 structured}
\end{subfigure}
\begin{subfigure}[b]{0.24\linewidth}
\includegraphics[width=0.95\linewidth]{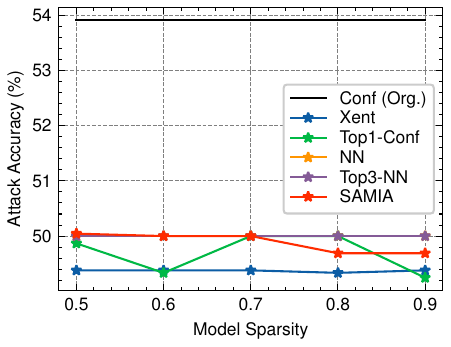}
\caption{Slimming}
\end{subfigure}
\caption{Attack performance comparison of MIAs (CHMNIST, DenseNet121). }
\end{figure*}

\begin{figure*}[!h]
\centering
\begin{subfigure}[b]{0.24\linewidth}
\includegraphics[width=0.95\linewidth]{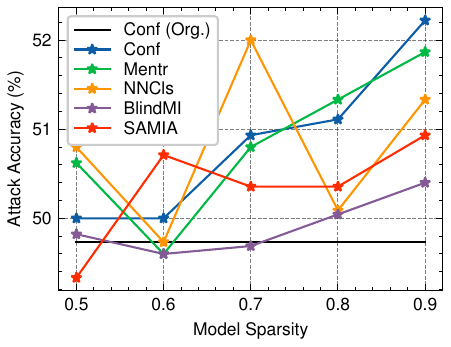}
\caption{L1 unstructured}
\end{subfigure}
\begin{subfigure}[b]{0.24\linewidth}
\includegraphics[width=0.95\linewidth]{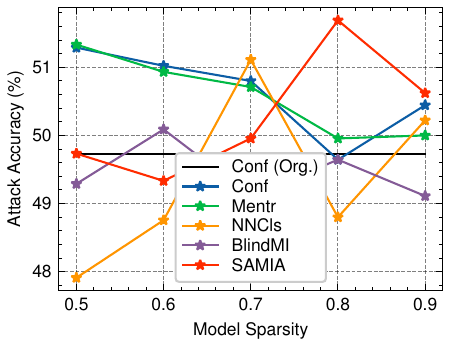}
\caption{L1 structured}
\end{subfigure}
\begin{subfigure}[b]{0.24\linewidth}
\includegraphics[width=0.95\linewidth]{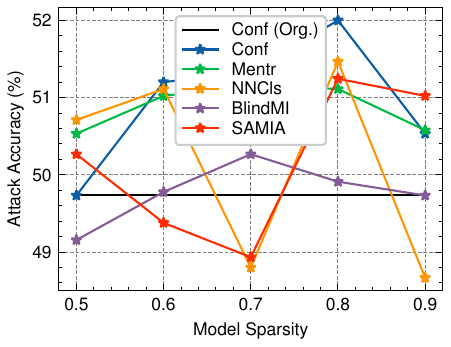}
\caption{L2 structured}
\end{subfigure}
\begin{subfigure}[b]{0.24\linewidth}
\includegraphics[width=0.95\linewidth]{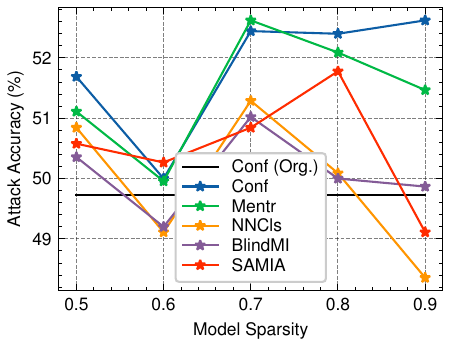}
\caption{Slimming}
\end{subfigure}
\begin{subfigure}[b]{0.24\linewidth}
\includegraphics[width=0.95\linewidth]{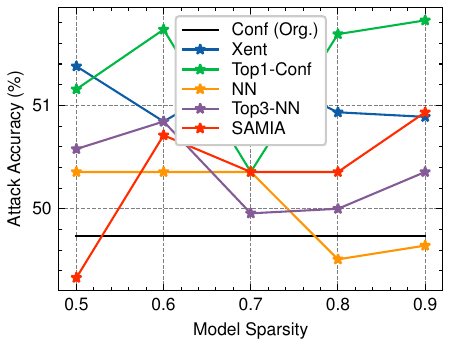}
\caption{L1 unstructured}
\end{subfigure}
\begin{subfigure}[b]{0.24\linewidth}
\includegraphics[width=0.95\linewidth]{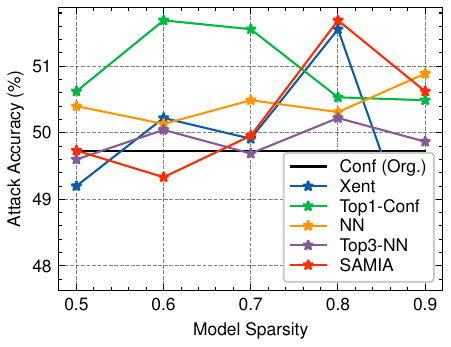}
\caption{L1 structured}
\end{subfigure}
\begin{subfigure}[b]{0.24\linewidth}
\includegraphics[width=0.95\linewidth]{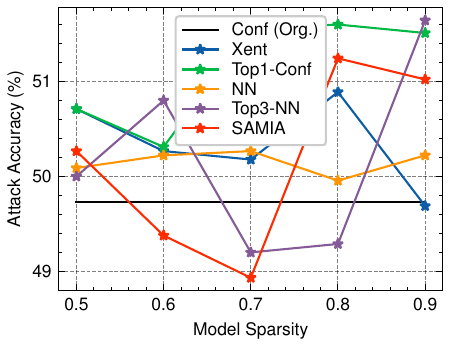}
\caption{L2 structured}
\end{subfigure}
\begin{subfigure}[b]{0.24\linewidth}
\includegraphics[width=0.95\linewidth]{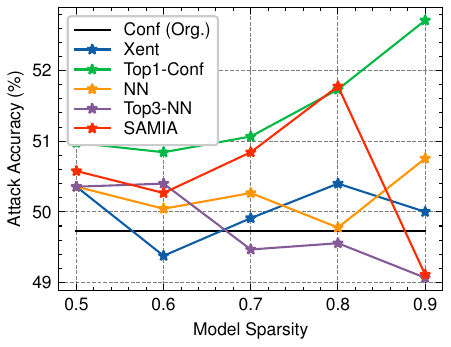}
\caption{Slimming}
\end{subfigure}
\caption{Attack performance comparison of MIAs (CHMNIST, VGG16). }
\end{figure*}

\begin{figure*}[!h]
\centering
\begin{subfigure}[b]{0.24\linewidth}
\includegraphics[width=0.95\linewidth]{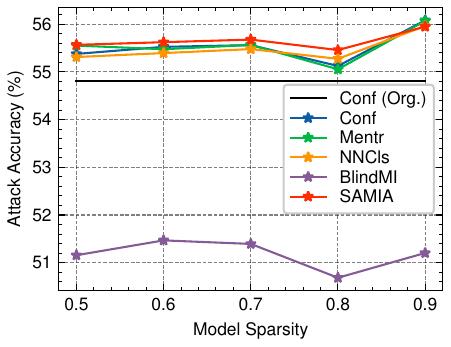}
\caption{L1 unstructured}
\end{subfigure}
\begin{subfigure}[b]{0.24\linewidth}
\includegraphics[width=0.95\linewidth]{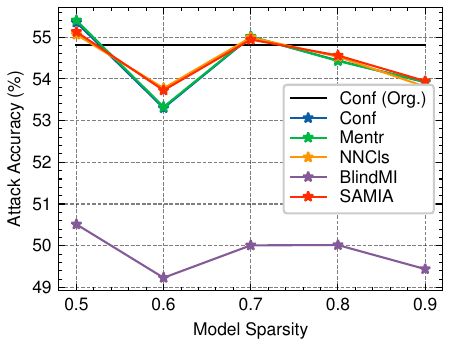}
\caption{L1 structured}
\end{subfigure}
\begin{subfigure}[b]{0.24\linewidth}
\includegraphics[width=0.95\linewidth]{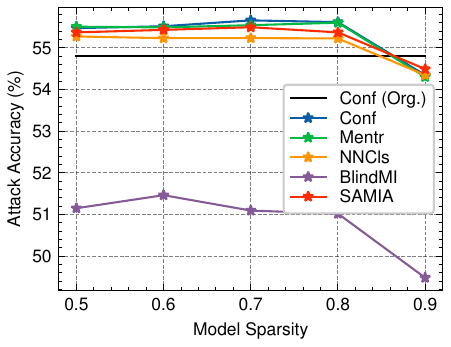}
\caption{L2 structured}
\end{subfigure}
\begin{subfigure}[b]{0.24\linewidth}
\includegraphics[width=0.95\linewidth]{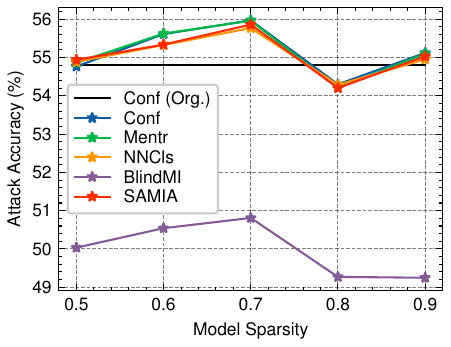}
\caption{Slimming}
\end{subfigure}
\begin{subfigure}[b]{0.24\linewidth}
\includegraphics[width=0.95\linewidth]{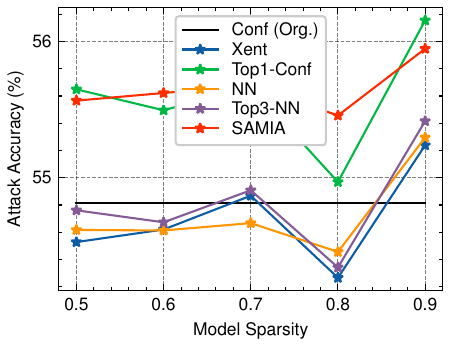}
\caption{L1 unstructured}
\end{subfigure}
\begin{subfigure}[b]{0.24\linewidth}
\includegraphics[width=0.95\linewidth]{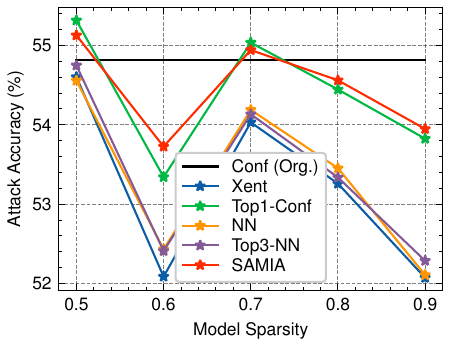}
\caption{L1 structured}
\end{subfigure}
\begin{subfigure}[b]{0.24\linewidth}
\includegraphics[width=0.95\linewidth]{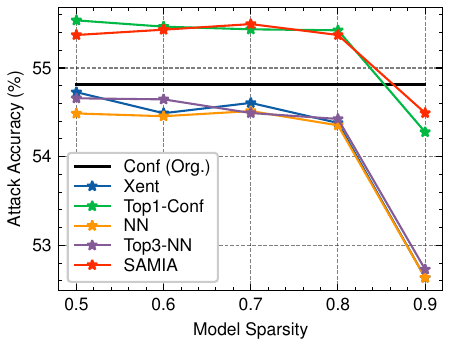}
\caption{L2 structured}
\end{subfigure}
\begin{subfigure}[b]{0.24\linewidth}
\includegraphics[width=0.95\linewidth]{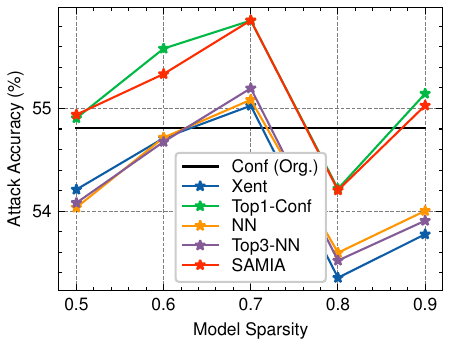}
\caption{Slimming}
\end{subfigure}
\caption{Attack performance comparison of MIAs (SVHN, ResNet18). }
\label{fig:mia_svhn_resnet}
\end{figure*}

\begin{figure*}[!h]
\centering
\begin{subfigure}[b]{0.24\linewidth}
\includegraphics[width=0.95\linewidth]{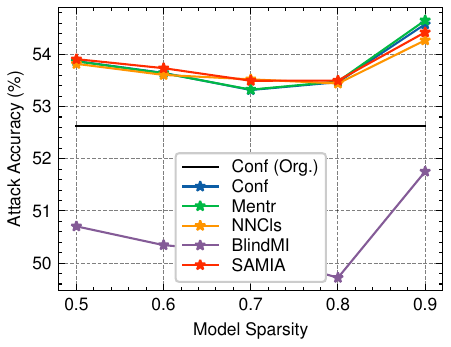}
\caption{L1 unstructured}
\end{subfigure}
\begin{subfigure}[b]{0.24\linewidth}
\includegraphics[width=0.95\linewidth]{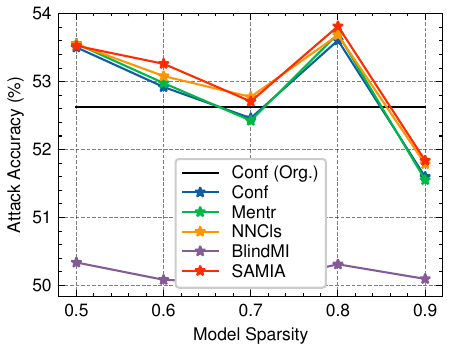}
\caption{L1 structured}
\end{subfigure}
\begin{subfigure}[b]{0.24\linewidth}
\includegraphics[width=0.95\linewidth]{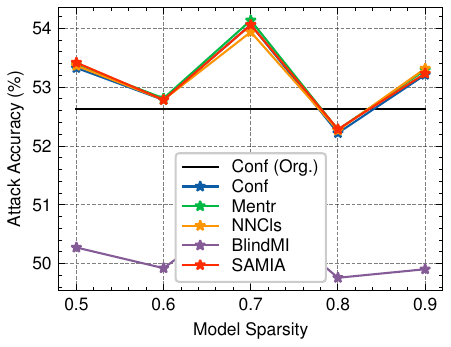}
\caption{L2 structured}
\end{subfigure}
\begin{subfigure}[b]{0.24\linewidth}
\includegraphics[width=0.95\linewidth]{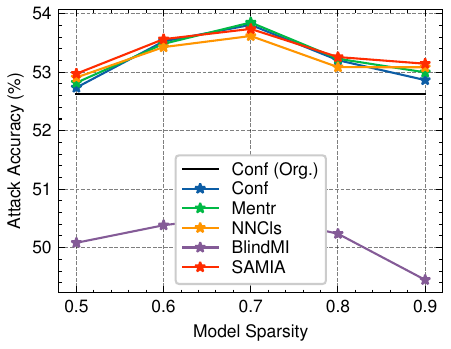}
\caption{Slimming}
\end{subfigure}
\begin{subfigure}[b]{0.24\linewidth}
\includegraphics[width=0.95\linewidth]{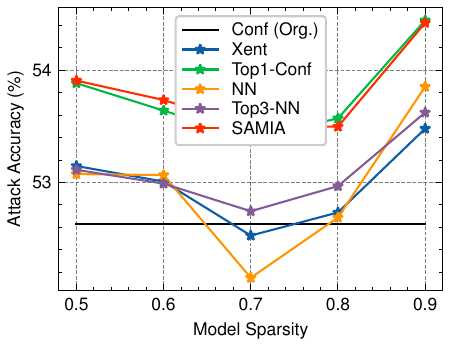}
\caption{L1 unstructured}
\end{subfigure}
\begin{subfigure}[b]{0.24\linewidth}
\includegraphics[width=0.95\linewidth]{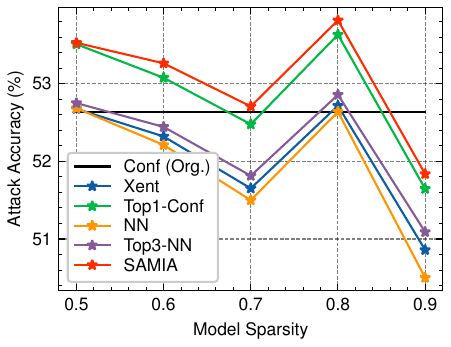}
\caption{L1 structured}
\end{subfigure}
\begin{subfigure}[b]{0.24\linewidth}
\includegraphics[width=0.95\linewidth]{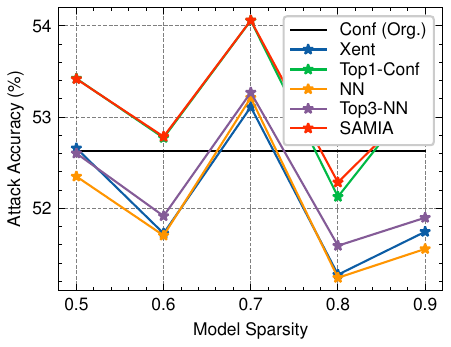}
\caption{L2 structured}
\end{subfigure}
\begin{subfigure}[b]{0.24\linewidth}
\includegraphics[width=0.95\linewidth]{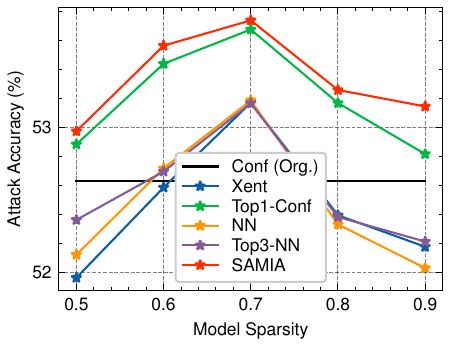}
\caption{Slimming}
\end{subfigure}
\caption{Attack performance comparison of MIAs (SVHN, DenseNet121). }
\end{figure*}

\begin{figure*}[!h]
\centering
\begin{subfigure}[b]{0.24\linewidth}
\includegraphics[width=0.95\linewidth]{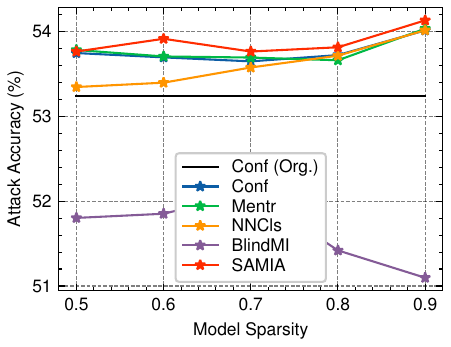}
\caption{L1 unstructured}
\end{subfigure}
\begin{subfigure}[b]{0.24\linewidth}
\includegraphics[width=0.95\linewidth]{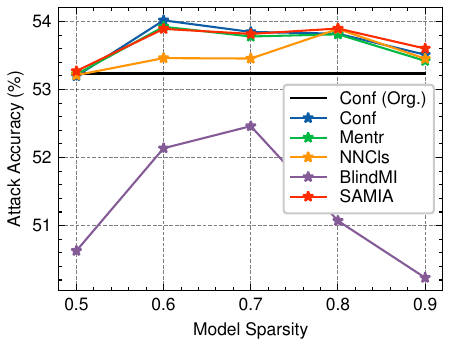}
\caption{L1 structured}
\end{subfigure}
\begin{subfigure}[b]{0.24\linewidth}
\includegraphics[width=0.95\linewidth]{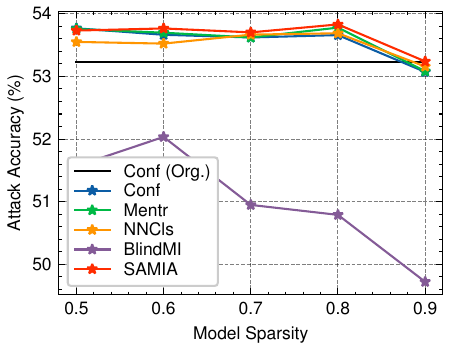}
\caption{L2 structured}
\end{subfigure}
\begin{subfigure}[b]{0.24\linewidth}
\includegraphics[width=0.95\linewidth]{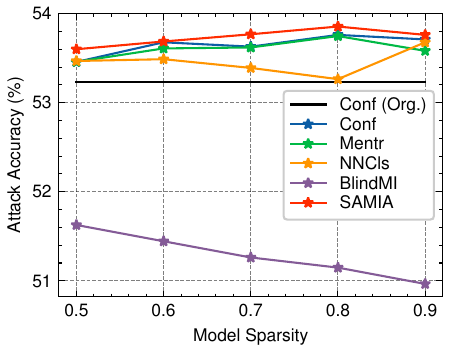}
\caption{Slimming}
\end{subfigure}
\begin{subfigure}[b]{0.24\linewidth}
\includegraphics[width=0.95\linewidth]{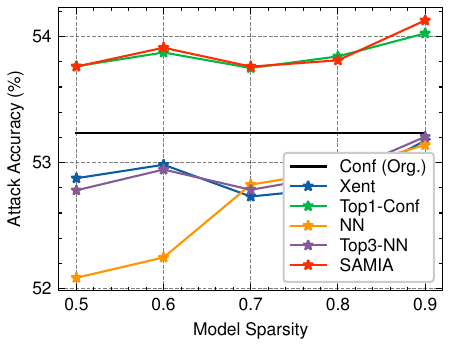}
\caption{L1 unstructured}
\end{subfigure}
\begin{subfigure}[b]{0.24\linewidth}
\includegraphics[width=0.95\linewidth]{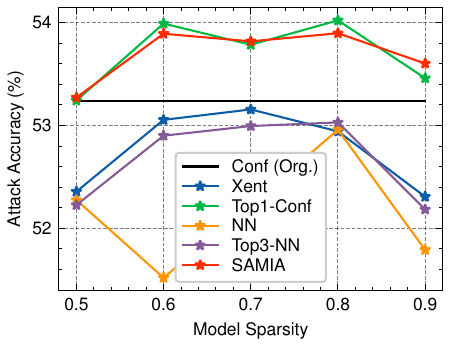}
\caption{L1 structured}
\end{subfigure}
\begin{subfigure}[b]{0.24\linewidth}
\includegraphics[width=0.95\linewidth]{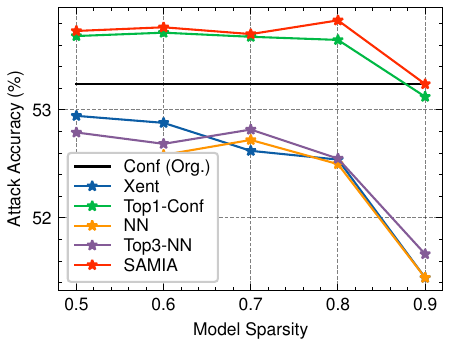}
\caption{L2 structured}
\end{subfigure}
\begin{subfigure}[b]{0.24\linewidth}
\includegraphics[width=0.95\linewidth]{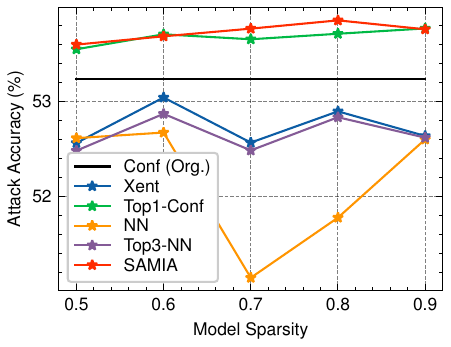}
\caption{Slimming}
\end{subfigure}
\caption{Attack performance comparison of MIAs (SVHN, VGG16). }
\end{figure*}

\begin{figure*}[!h]
\centering
\begin{subfigure}[b]{0.24\linewidth}
\includegraphics[width=0.95\linewidth]{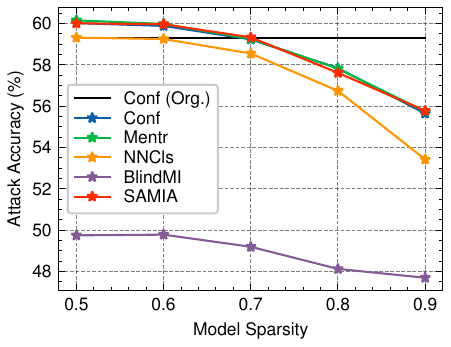}
\caption{Purchase, FC}
\end{subfigure}
\begin{subfigure}[b]{0.24\linewidth}
\includegraphics[width=0.95\linewidth]{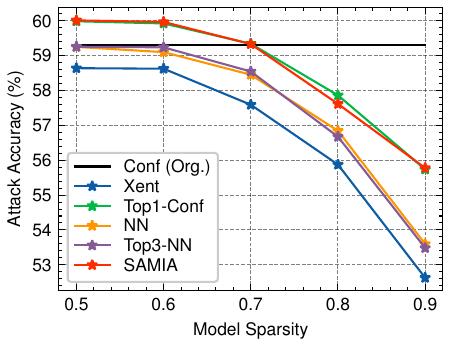}
\caption{Purchase, FC}
\end{subfigure}
\begin{subfigure}[b]{0.24\linewidth}
\includegraphics[width=0.95\linewidth]{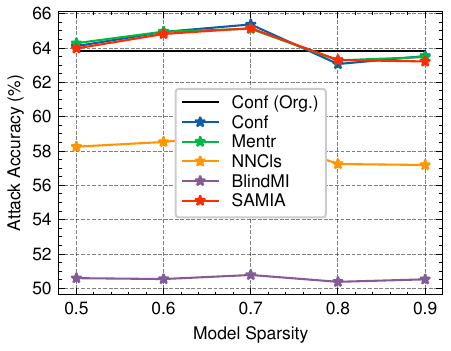}
\caption{Texas, FC}
\end{subfigure}
\begin{subfigure}[b]{0.24\linewidth}
\includegraphics[width=0.95\linewidth]{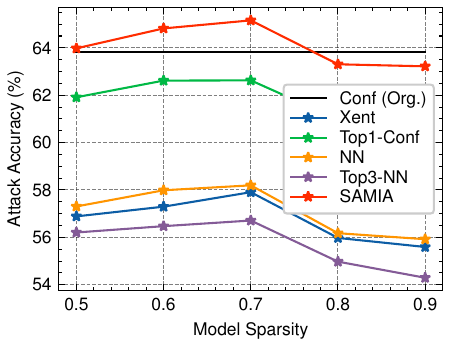}
\caption{Texas, FC}
\end{subfigure}
\caption{Attack performance comparison of MIAs (L1 Unstructured).}
\end{figure*}

\clearpage
\subsection{Impact of Confidence Gap, Sensitivity Gap, and Generalization Gap}
This section reports the impact of confidence gap, sensitivity gap, and generalization gap in the remaining experimental settings. We observe the strong correlation between gaps (\ie, confidence gap, sensitivity gap, and generalization gap) and attack accuracy.

\begin{figure*}[!h]
\centering
\begin{subfigure}[b]{0.24\textwidth}
\includegraphics[width=\linewidth]{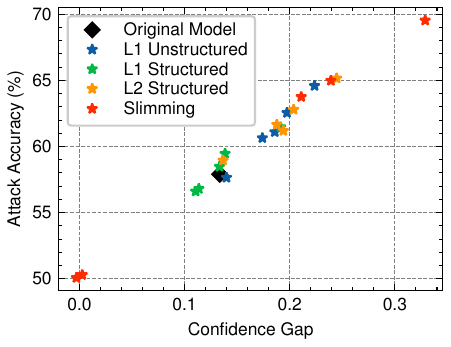}
\caption{Confidence gap}
\end{subfigure}
\begin{subfigure}[b]{0.24\textwidth}
\includegraphics[width=\linewidth]{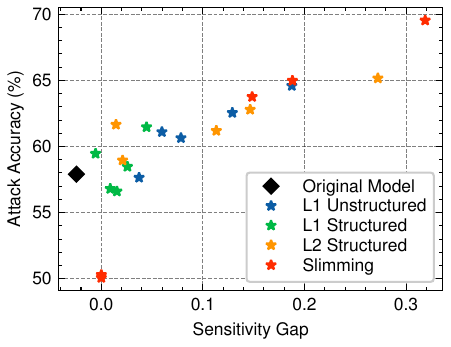}
\caption{Sensitivity gap}
\end{subfigure}
\begin{subfigure}[b]{0.24\textwidth}
\includegraphics[width=\linewidth]{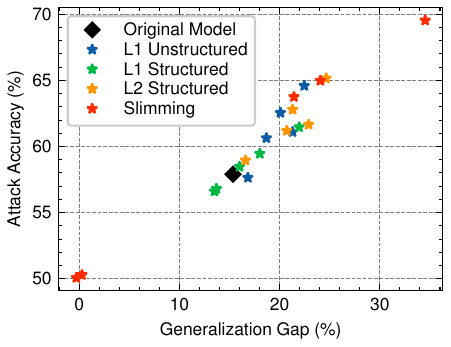}
\caption{Generalization gap}
\end{subfigure}
\caption{Impact of confidence gap, sensitivity gap, and generalization gap (CIFAR10, DenseNet121).}
\end{figure*}

\begin{figure*}[!h]
\centering
\begin{subfigure}[b]{0.24\textwidth}
\includegraphics[width=\linewidth]{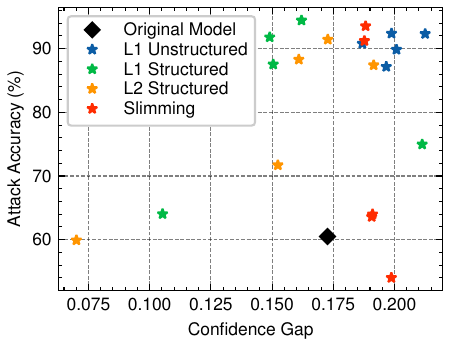}
\caption{Confidence gap}
\end{subfigure}
\begin{subfigure}[b]{0.24\textwidth}
\includegraphics[width=\linewidth]{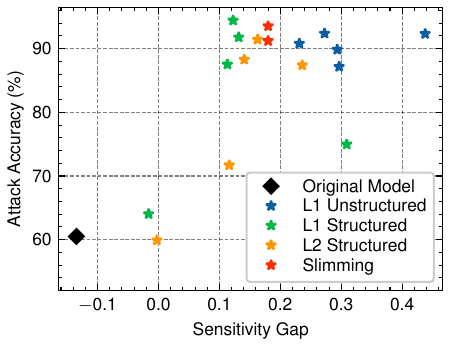}
\caption{Sensitivity gap}
\end{subfigure}
\begin{subfigure}[b]{0.24\textwidth}
\includegraphics[width=\linewidth]{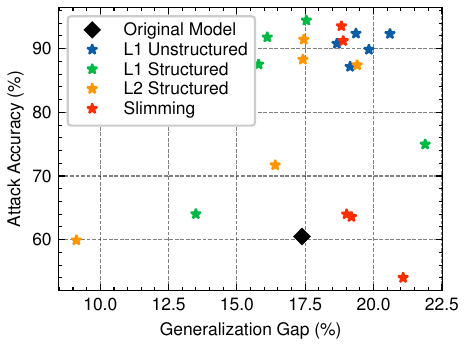}
\caption{Generalization gap}
\end{subfigure}
\caption{Impact of confidence gap, sensitivity gap, and generalization gap (CIFAR10, VGG16).}
\end{figure*}

\begin{figure*}[!h]
\centering
\begin{subfigure}[b]{0.24\textwidth}
\includegraphics[width=\linewidth]{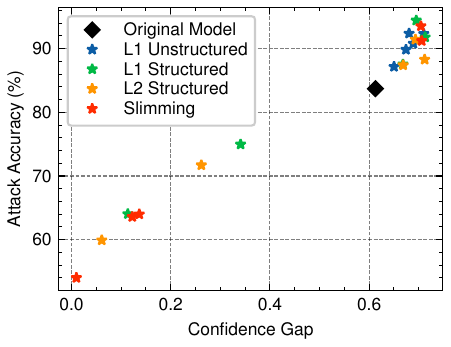}
\caption{Confidence gap}
\end{subfigure}
\begin{subfigure}[b]{0.24\textwidth}
\includegraphics[width=\linewidth]{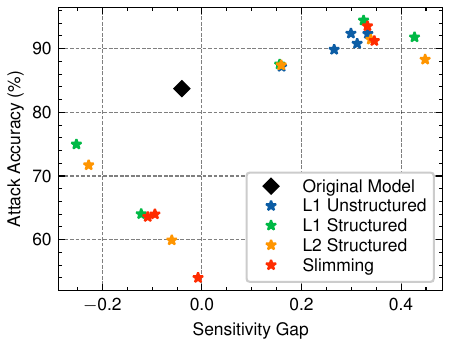}
\caption{Sensitivity gap}
\end{subfigure}
\begin{subfigure}[b]{0.24\textwidth}
\includegraphics[width=\linewidth]{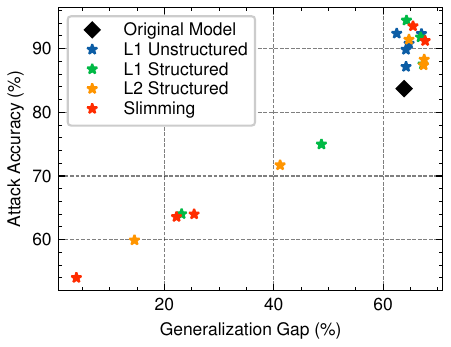}
\caption{Generalization gap}
\end{subfigure}
\caption{Impact of confidence gap, sensitivity gap, and generalization gap (CIFAR100, ResNet18).}

\end{figure*}

\begin{figure*}[!h]
\centering
\begin{subfigure}[b]{0.24\textwidth}
\includegraphics[width=\linewidth]{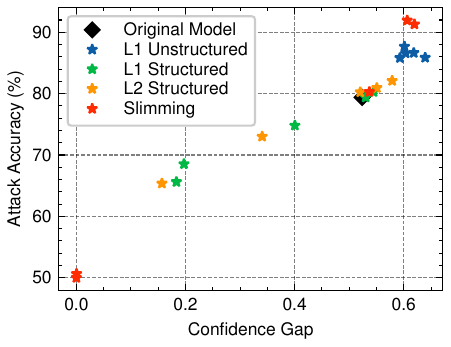}
\caption{Confidence gap}
\end{subfigure}
\begin{subfigure}[b]{0.24\textwidth}
\includegraphics[width=\linewidth]{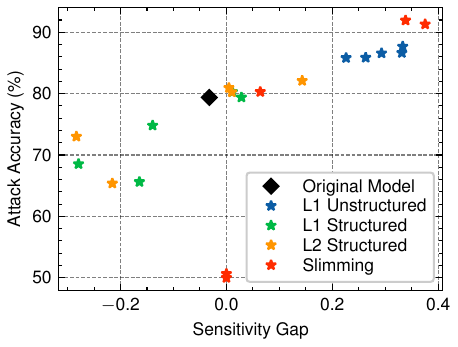}
\caption{Sensitivity gap}
\end{subfigure}
\begin{subfigure}[b]{0.24\textwidth}
\includegraphics[width=\linewidth]{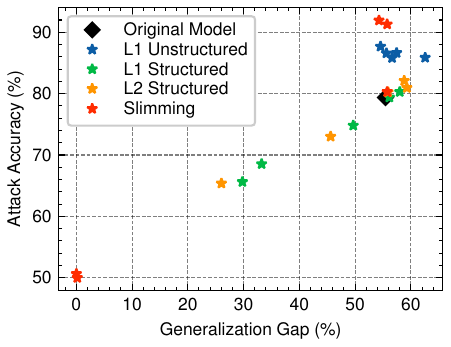}
\caption{Generalization gap}
\label{fig:gen_gap_cifar100_densenet}
\end{subfigure}
\caption{Impact of confidence gap, sensitivity gap, and generalization gap (CIFAR100, DenseNet121).}
\end{figure*}

\begin{figure*}[!h]
\centering
\begin{subfigure}[b]{0.24\textwidth}
\includegraphics[width=\linewidth]{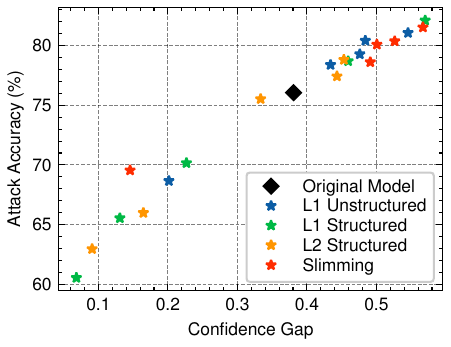}
\caption{Confidence gap}
\end{subfigure}
\begin{subfigure}[b]{0.24\textwidth}
\includegraphics[width=\linewidth]{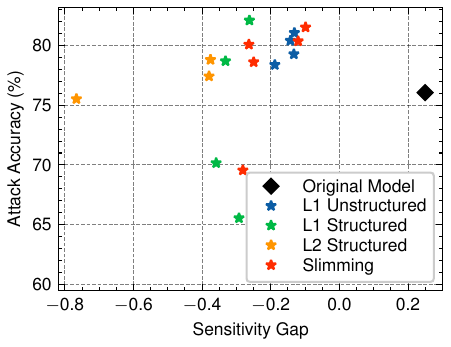}
\caption{Sensitivity gap}
\end{subfigure}
\begin{subfigure}[b]{0.24\textwidth}
\includegraphics[width=\linewidth]{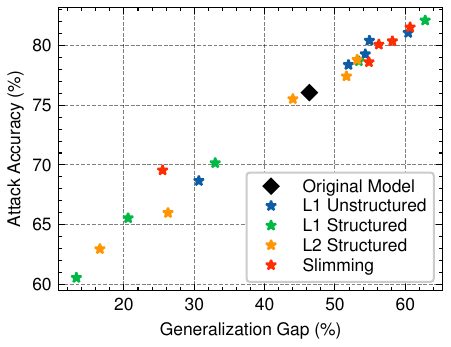}
\caption{Generalization gap}
\end{subfigure}
\caption{Impact of confidence gap, sensitivity gap, and generalization gap (CIFAR100, VGG16).}
\end{figure*}

\begin{figure*}[!h]
\centering
\begin{subfigure}[b]{0.24\textwidth}
\includegraphics[width=\linewidth]{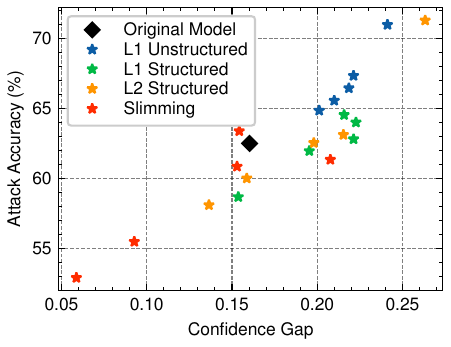}
\caption{Confidence gap}
\end{subfigure}
\begin{subfigure}[b]{0.24\textwidth}
\includegraphics[width=\linewidth]{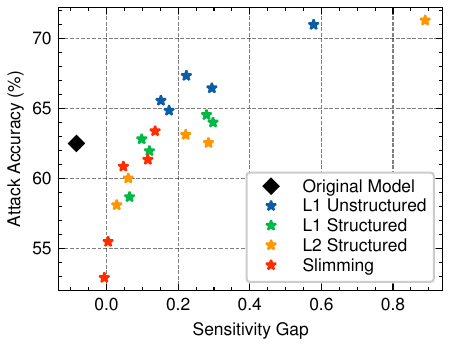}
\caption{Sensitivity gap}
\end{subfigure}
\begin{subfigure}[b]{0.24\textwidth}
\includegraphics[width=\linewidth]{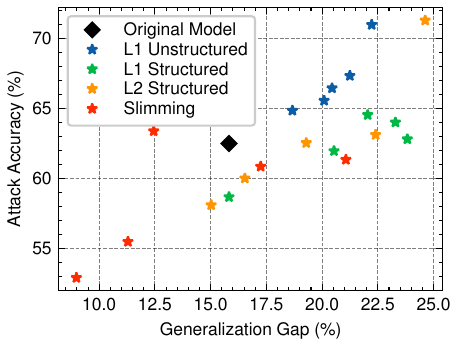}
\caption{Generalization gap}
\end{subfigure}
\caption{Impact of confidence gap, sensitivity gap, and generalization gap (CHMNIST, ResNet18).}

\end{figure*}

\begin{figure*}[!h]
\centering
\begin{subfigure}[b]{0.24\textwidth}
\includegraphics[width=\linewidth]{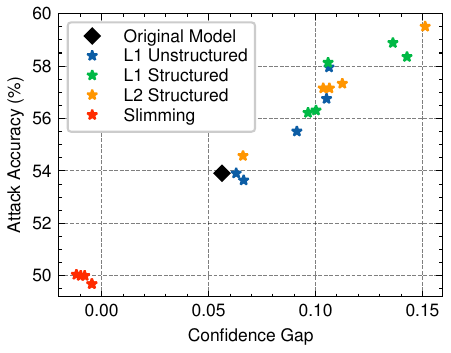}
\caption{Confidence gap}
\end{subfigure}
\begin{subfigure}[b]{0.24\textwidth}
\includegraphics[width=\linewidth]{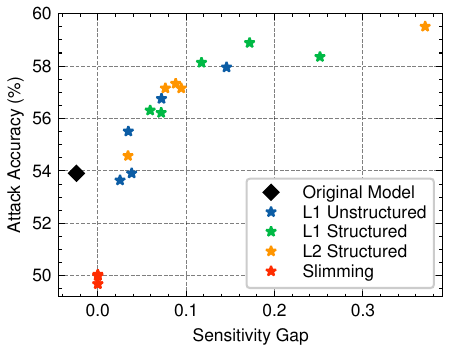}
\caption{Sensitivity gap}
\end{subfigure}
\begin{subfigure}[b]{0.24\textwidth}
\includegraphics[width=\linewidth]{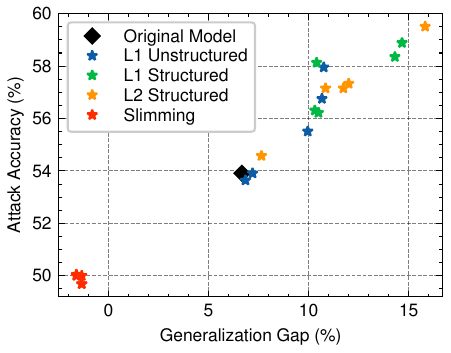}
\caption{Generalization gap}
\end{subfigure}

\caption{Impact of confidence gap, sensitivity gap, and generalization gap (CHMNIST, DenseNet121).}

\end{figure*}

\begin{figure*}[!h]
\centering
\begin{subfigure}[b]{0.24\textwidth}
\includegraphics[width=\linewidth]{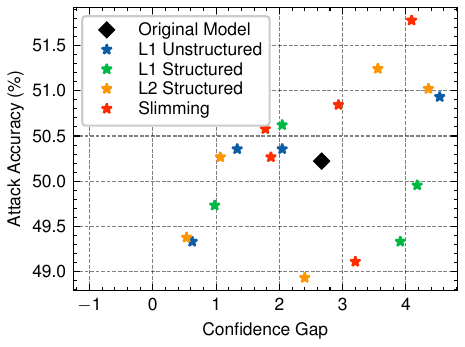}
\caption{Confidence gap}
\end{subfigure}
\begin{subfigure}[b]{0.24\textwidth}
\includegraphics[width=\linewidth]{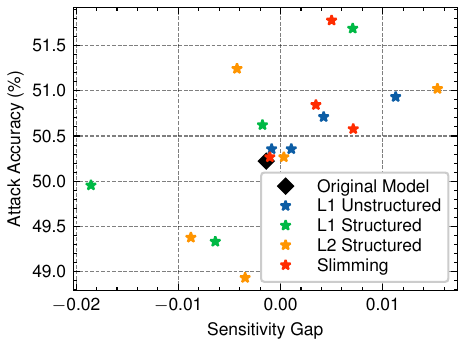}
\caption{Sensitivity gap}
\end{subfigure}
\begin{subfigure}[b]{0.24\textwidth}
\includegraphics[width=\linewidth]{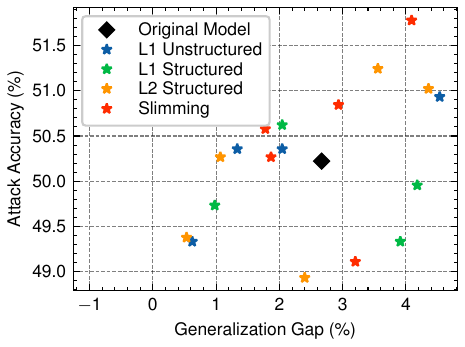}
\caption{Generalization gap}
\end{subfigure}
\caption{Impact of confidence gap, sensitivity gap, and generalization gap (CHMNIST, VGG16).}

\end{figure*}

\begin{figure*}[!h]
\centering
\begin{subfigure}[b]{0.24\textwidth}
\includegraphics[width=\linewidth]{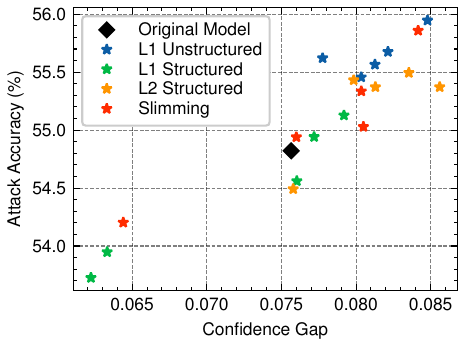}
\caption{Confidence gap}
\end{subfigure}
\begin{subfigure}[b]{0.24\textwidth}
\includegraphics[width=\linewidth]{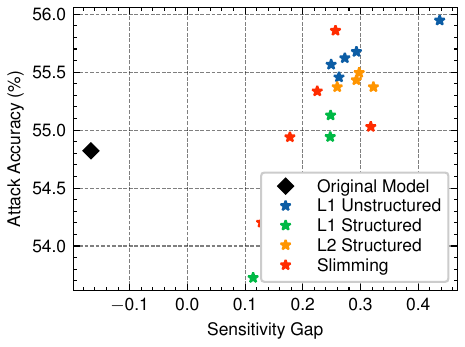}
\caption{Sensitivity gap}
\end{subfigure}
\begin{subfigure}[b]{0.24\textwidth}
\includegraphics[width=\linewidth]{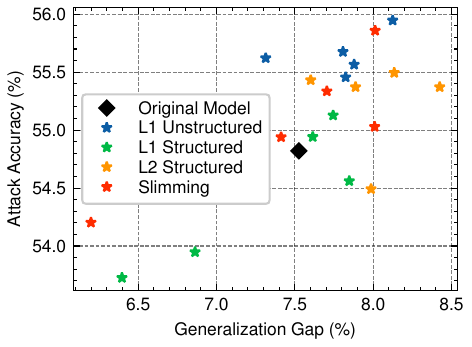}
\caption{Generalization gap}
\end{subfigure}
\caption{Impact of confidence gap, sensitivity gap, and generalization gap (SVHN, ResNet18).}

\end{figure*}

\begin{figure*}[!h]
\centering
\begin{subfigure}[b]{0.24\textwidth}
\includegraphics[width=\linewidth]{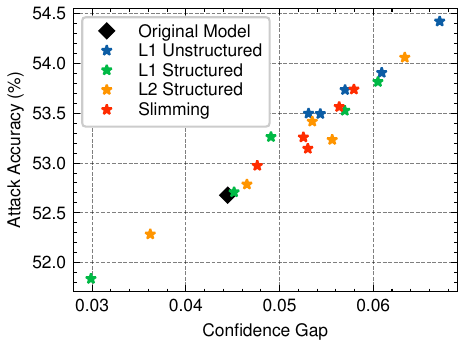}
\caption{Confidence gap}
\end{subfigure}
\begin{subfigure}[b]{0.24\textwidth}
\includegraphics[width=\linewidth]{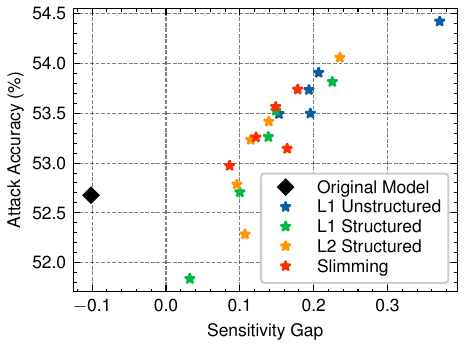}
\caption{Sensitivity gap}
\end{subfigure}
\begin{subfigure}[b]{0.24\textwidth}
\includegraphics[width=\linewidth]{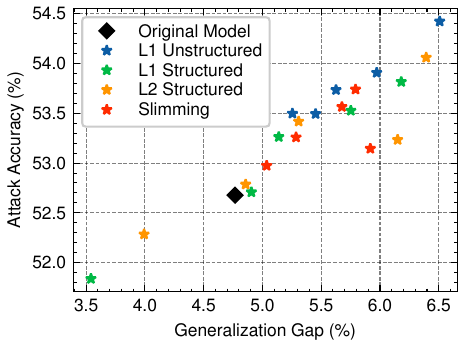}
\caption{Generalization gap}
\end{subfigure}
\caption{Impact of confidence gap, sensitivity gap, and generalization gap (SVHN, DenseNet121).}
\end{figure*}

\begin{figure*}[!h]
\centering
\begin{subfigure}[b]{0.24\textwidth}
\includegraphics[width=\linewidth]{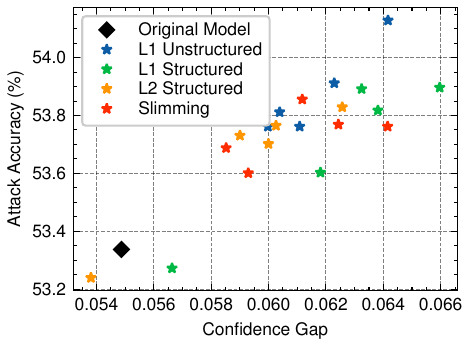}
\caption{Confidence gap}
\end{subfigure}
\begin{subfigure}[b]{0.24\textwidth}
\includegraphics[width=\linewidth]{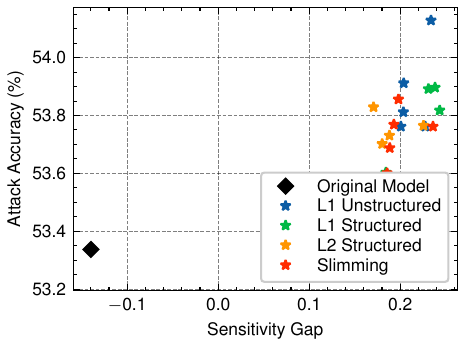}
\caption{Sensitivity gap}
\end{subfigure}
\begin{subfigure}[b]{0.24\textwidth}
\includegraphics[width=\linewidth]{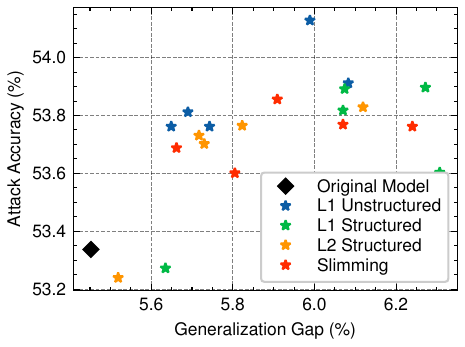}
\caption{Generalization gap}
\end{subfigure}
\caption{Impact of confidence gap, sensitivity gap, and generalization gap (SVHN, VGG16).}

\end{figure*}

\begin{figure*}[!h]
\centering
\begin{subfigure}[b]{0.24\textwidth}
\includegraphics[width=\linewidth]{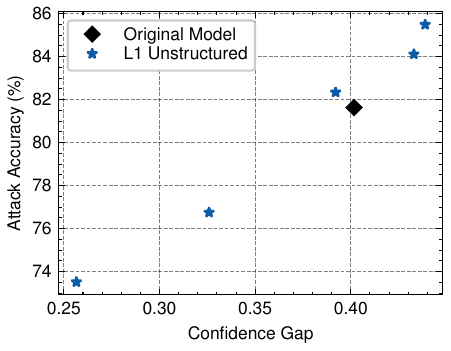}
\caption{Confidence gap}
\end{subfigure}
\begin{subfigure}[b]{0.24\textwidth}
\includegraphics[width=\linewidth]{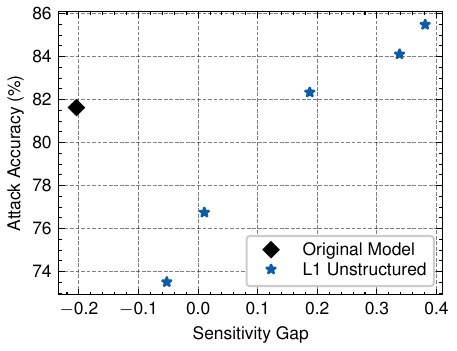}
\caption{Sensitivity gap}
\label{fig:sens_gap_location}
\end{subfigure}
\begin{subfigure}[b]{0.24\textwidth}
\includegraphics[width=\linewidth]{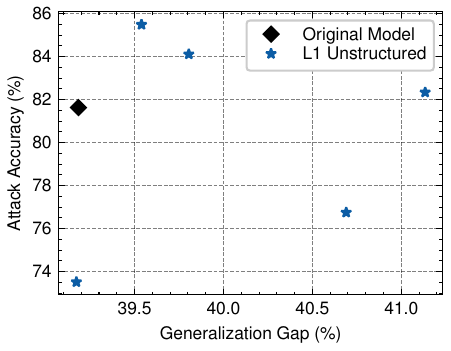}
\caption{Generalization gap}
\end{subfigure}
\vspace{-0.5em}
\caption{Impact of confidence gap, sensitivity gap, and generalization gap (Location, FC).}
\label{fig:gap_attack_location_fc}
\end{figure*}

\begin{figure*}[!h]
\centering
\begin{subfigure}[b]{0.24\textwidth}
\includegraphics[width=\linewidth]{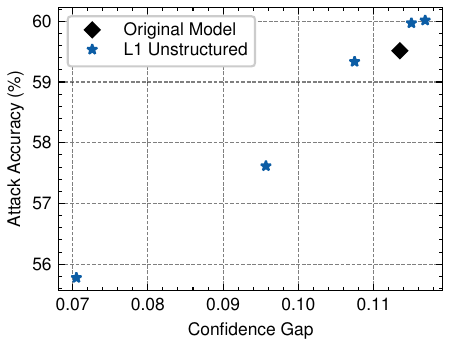}
\caption{Confidence gap}
\end{subfigure}
\begin{subfigure}[b]{0.24\textwidth}
\includegraphics[width=\linewidth]{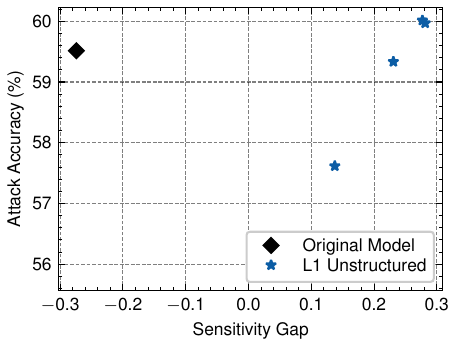}
\caption{Sensitivity gap}
\end{subfigure}
\begin{subfigure}[b]{0.24\textwidth}
\includegraphics[width=\linewidth]{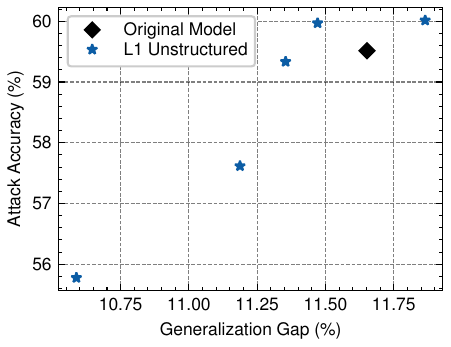}
\caption{Generalization gap}
\end{subfigure}
\caption{Impact of confidence gap, sensitivity gap, and generalization gap (Purchase, FC).}
\end{figure*}

\begin{figure*}[!h]
\centering
\begin{subfigure}[b]{0.24\textwidth}
\includegraphics[width=\linewidth]{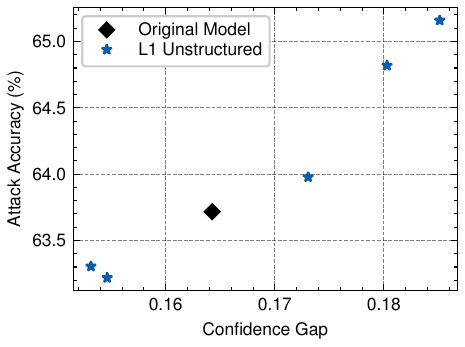}
\caption{Confidence gap}
\end{subfigure}
\begin{subfigure}[b]{0.24\textwidth}
\includegraphics[width=\linewidth]{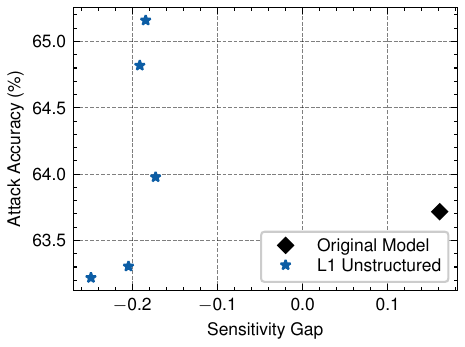}
\caption{Sensitivity gap}
\end{subfigure}
\begin{subfigure}[b]{0.24\textwidth}
\includegraphics[width=\linewidth]{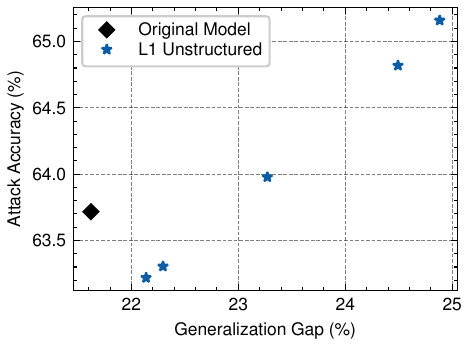}
\caption{Generalization gap}
\end{subfigure}
\caption{Impact of confidence gap, sensitivity gap, and generalization gap (Texas, FC).}
\end{figure*}

\clearpage
\subsection{Impact of Dataset Split Strategies}
\label{sec:dataset_split}
This section reports the attack performance when the target model is trained on an unevenly split dataset. 
In the main paper, we split datasets into training, test, and validation in the ratio 45\%, 45\%, 10\%, so that we get the same number of training and test data and the random guessing attack accuracy is 50\%. Here we evaluate the attack performance using different and uneven ratios. Given the uneven dataset, we use Area Under the Curve (AUC) to report the attack performance. As shown in Table~\ref{tab:split}, the uneven split may cause a slight decrease of attack AUC compared with an even split, but the privacy risk is still high (over 70\% attack AUC).
\begin{table*}[!h]
\centering
\begin{tabular}{@{}cc@{}}
\toprule
Dataset Split Ratio              & \multirow{2}{*}{Attack AUC} \\
Train(\%)/Test(\%)/Dev(\%) &                      \\ \midrule
45/45/10                  & 75.5                 \\
50/40/10                   & 72.7                 \\
60/30/10                   & 71.6                 \\ 
70/20/10                & 72.7 \\\bottomrule
\end{tabular}%
\caption{Impact of different dataset split strategies (CIFAR10, ResNet18, L1 Unstructured, Sparsity 0.7).}
\label{tab:split}
\end{table*}

\clearpage
\subsection{Attack Accuracy With Unknown Sparsity Levels}
\label{app:attack_unknown}
This section reports the attack accuracy with unknown sparsity levels in the CIFAR10, CIFAR100, Purchase, and Texas datasets. Without knowing the sparsity level (\ie different sparsity levels used in the target model and the shadow model), the adversary can still achieve similar (even higher in some cases) attack accuracy. 

\begin{figure*}[!h]
\centering
\begin{subfigure}[b]{0.24\linewidth}
\includegraphics[width=\linewidth]{figs/exp/unknown/unknown_cifar10_densenet121_level.pdf}
\caption{L1 Unstructured}
\end{subfigure}
\begin{subfigure}[b]{0.24\linewidth}
\includegraphics[width=\linewidth]{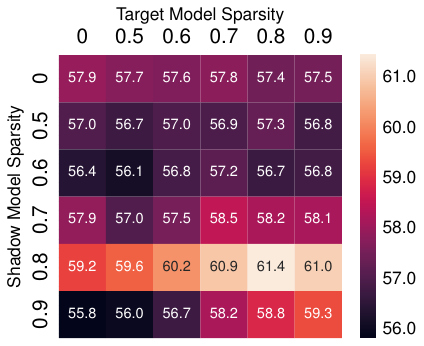}
\caption{L1 Structured}
\end{subfigure}
\begin{subfigure}[b]{0.24\linewidth}
\includegraphics[width=\linewidth]{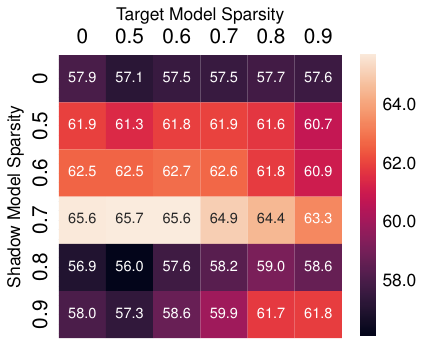}
\caption{L2 Structured}
\end{subfigure}
\begin{subfigure}[b]{0.24\linewidth}
\includegraphics[width=\linewidth]{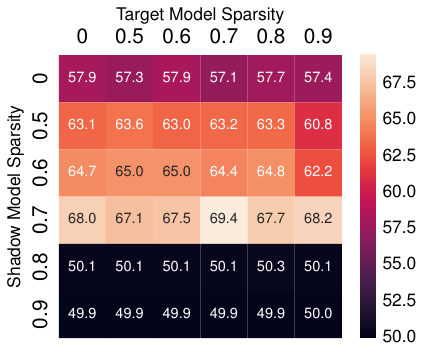}
\caption{Slimming}
\end{subfigure}
\caption{Attack accuracy with unknown sparsity levels (CIFAR10, DenseNet121).}
\end{figure*}

\begin{figure*}[!h]
\centering
\begin{subfigure}[b]{0.24\linewidth}
\includegraphics[width=\linewidth]{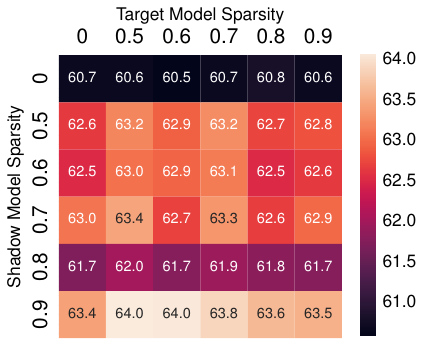}
\caption{L1 Unstructured}
\end{subfigure}
\begin{subfigure}[b]{0.24\linewidth}
\includegraphics[width=\linewidth]{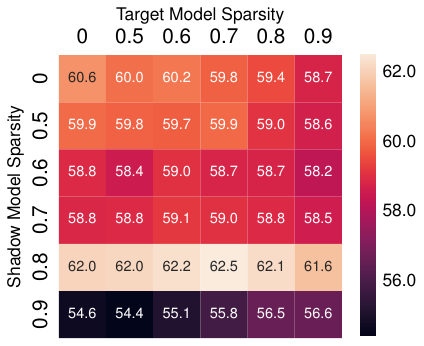}
\caption{L1 Structured}
\end{subfigure}
\begin{subfigure}[b]{0.24\linewidth}
\includegraphics[width=\linewidth]{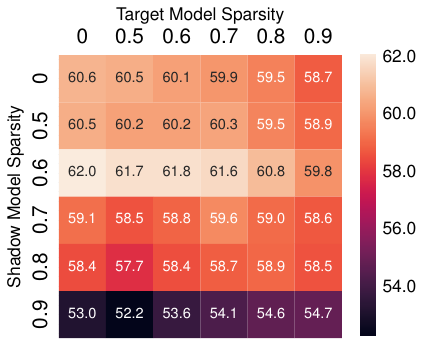}
\caption{L2 Structured}
\end{subfigure}
\begin{subfigure}[b]{0.24\linewidth}
\includegraphics[width=\linewidth]{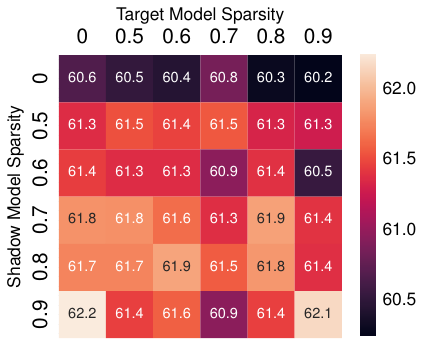}
\caption{Slimming}
\end{subfigure}
\caption{Attack accuracy with unknown sparsity levels (CIFAR10, VGG16).}
\end{figure*}

\begin{figure*}[!h]
\centering
\begin{subfigure}[b]{0.24\linewidth}
\includegraphics[width=\linewidth]{figs/exp/unknown/unknown_cifar100_densenet121_level.pdf}
\caption{L1 Unstructured}
\end{subfigure}
\begin{subfigure}[b]{0.24\linewidth}
\includegraphics[width=\linewidth]{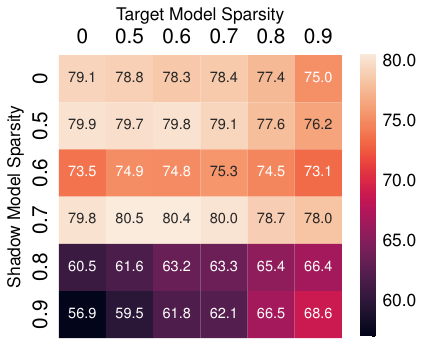}
\caption{L1 Structured}
\end{subfigure}
\begin{subfigure}[b]{0.24\linewidth}
\includegraphics[width=\linewidth]{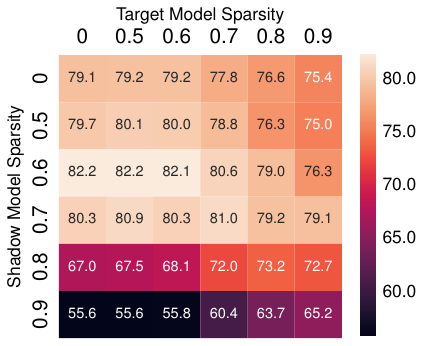}
\caption{L2 Structured}
\end{subfigure}
\begin{subfigure}[b]{0.24\linewidth}
\includegraphics[width=\linewidth]{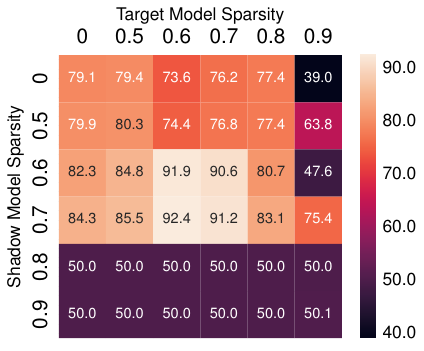}
\caption{Slimming}
\end{subfigure}

\caption{Attack accuracy with unknown sparsity levels (CIFAR100, DenseNet121).}

\end{figure*}

\begin{figure*}[!h]
\centering
\begin{subfigure}[b]{0.24\linewidth}
\includegraphics[width=\linewidth]{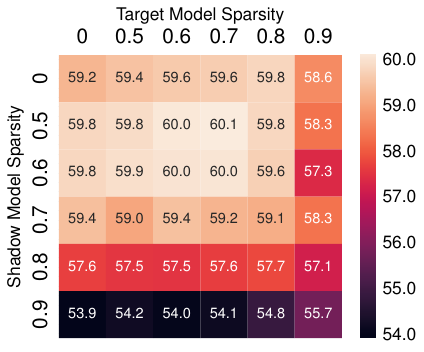}
\caption{Purchase, FC}
\end{subfigure}
\begin{subfigure}[b]{0.24\linewidth}
\includegraphics[width=\linewidth]{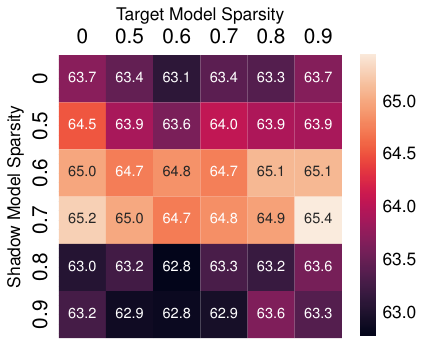}
\caption{Texas, FC}
\end{subfigure}
\caption{Attack accuracy with unknown sparsity levels.}
\end{figure*}

\clearpage
\section{Defense Evaluation}
This section reports the defense performance of our proposed PPD and three existing defensive mechanisms (Basic, DP, and ADV). We consider the defense performance with and without the adaptive attacks.
Due to the large computation for exploring proper hyper-parameters in all the defenses, we only report the defense performance on a subset of settings.

\subsection{Defense Performance Against Non-adaptive Attacks}
The proposed PPB defense can significantly reduce the attack accuracy to around 50\% in most settings, which demonstrates the effectiveness of PPB. Comparing with the existing defenses, PPB achieves a better trade-off between utility and privacy (\ie, high prediction accuracy and low attack accuracy).

\begin{figure*}[!h]
\centering
\begin{subfigure}[b]{0.24\textwidth}
\includegraphics[width=\linewidth]{figs/exp/defend/defend_cifar10_densenet121_level_0.6.pdf}
\caption{L1 Unstructured}
\end{subfigure}
\begin{subfigure}[b]{0.24\textwidth}
\includegraphics[width=\linewidth]{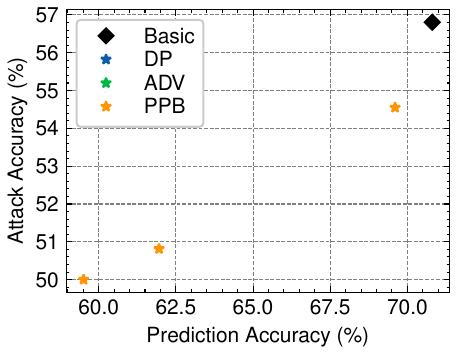}
\caption{L1 Structured}
\end{subfigure}
\begin{subfigure}[b]{0.24\textwidth}
\includegraphics[width=\linewidth]{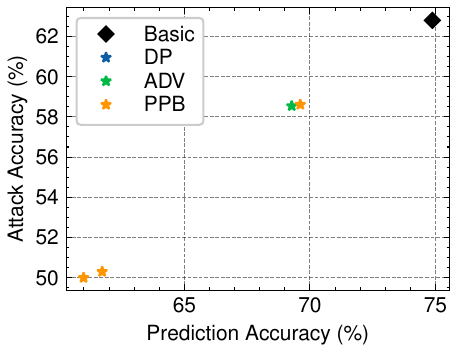}
\caption{L2 Structured}
\end{subfigure}
\begin{subfigure}[b]{0.24\textwidth}
\includegraphics[width=\linewidth]{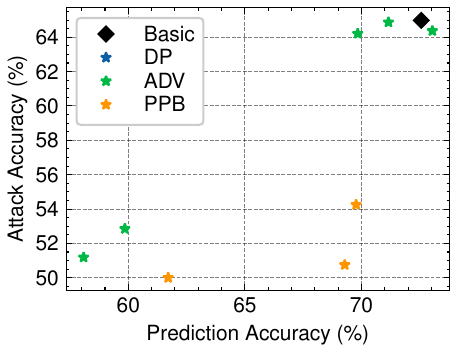}
\caption{Slimming}
\end{subfigure}
\caption{Performance of defenses (CIFAR10, DenseNet121, Sparsity 0.6).}
\end{figure*}

\begin{figure*}[!h]
\centering
\begin{subfigure}[b]{0.24\textwidth}
\includegraphics[width=\linewidth]{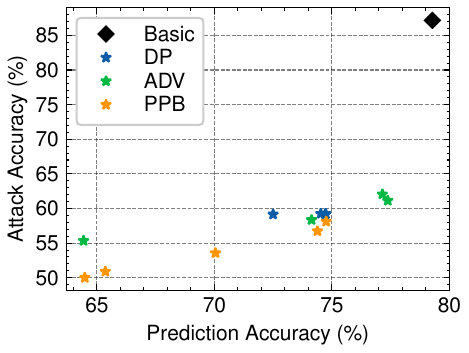}
\caption{L1 Unstructured}
\end{subfigure}
\begin{subfigure}[b]{0.24\textwidth}
\includegraphics[width=\linewidth]{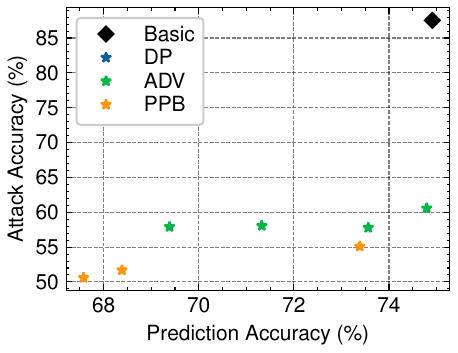}
\caption{L1 Structured}
\end{subfigure}
\begin{subfigure}[b]{0.24\textwidth}
\includegraphics[width=\linewidth]{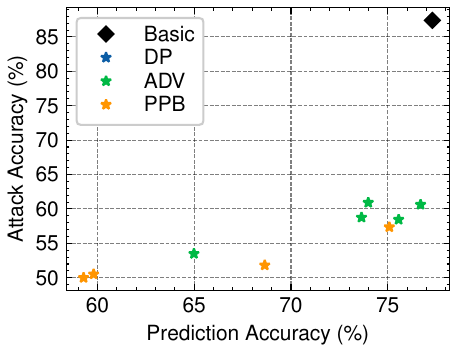}
\caption{L2 Structured}
\end{subfigure}
\begin{subfigure}[b]{0.24\textwidth}
\includegraphics[width=\linewidth]{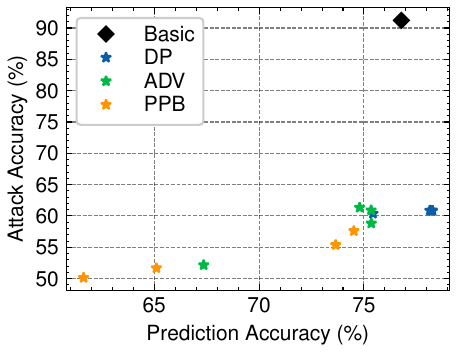}
\caption{Slimming}
\end{subfigure}
\caption{Performance of defenses (CIFAR10, VGG16, Sparsity 0.6).}
\end{figure*}

\begin{figure*}[!h]
\centering
\begin{subfigure}[b]{0.24\textwidth}
\includegraphics[width=\linewidth]{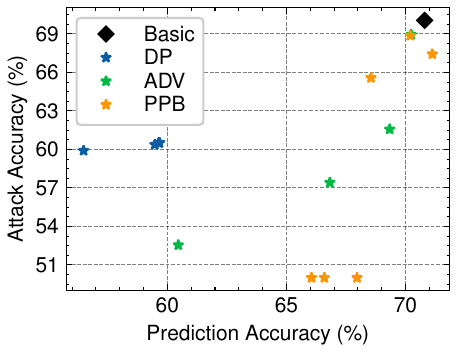}
\caption{L1 Unstructured}
\end{subfigure}
\begin{subfigure}[b]{0.24\textwidth}
\includegraphics[width=\linewidth]{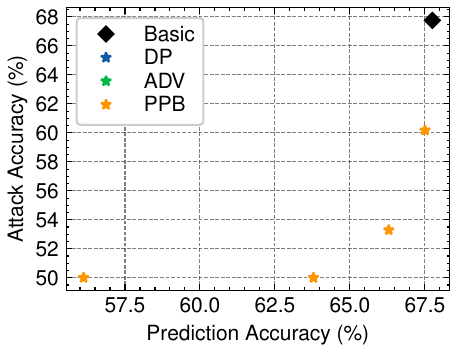}
\caption{L1 Structured}
\end{subfigure}
\begin{subfigure}[b]{0.24\textwidth}
\includegraphics[width=\linewidth]{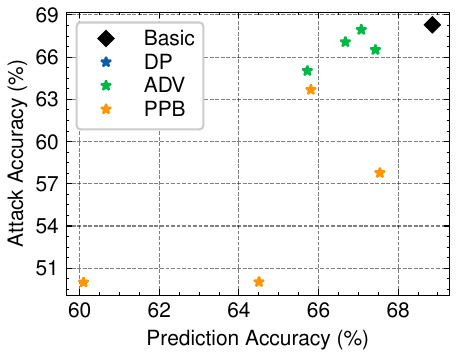}
\caption{L2 Structured}
\end{subfigure}
\begin{subfigure}[b]{0.24\textwidth}
\includegraphics[width=\linewidth]{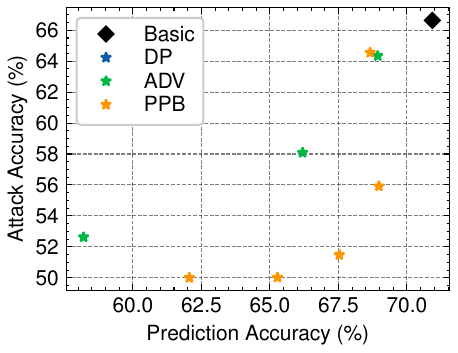}
\caption{Slimming}
\end{subfigure}
\caption{Performance of defenses (CIFAR10, ResNet18, Sparsity 0.7).}
\end{figure*}

\begin{figure*}[!h]
\centering
\begin{subfigure}[b]{0.24\textwidth}
\includegraphics[width=\linewidth]{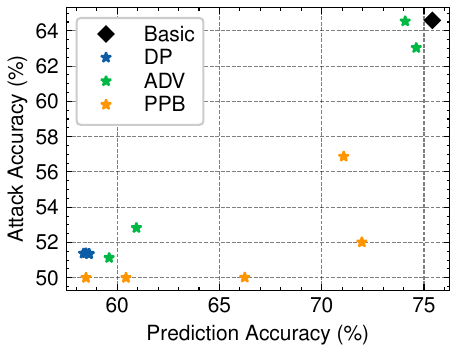}
\caption{L1 Unstructured}
\end{subfigure}
\begin{subfigure}[b]{0.24\textwidth}
\includegraphics[width=\linewidth]{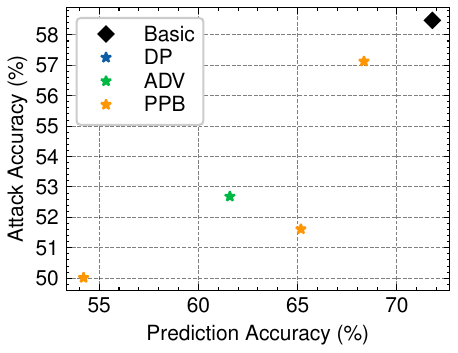}
\caption{L1 Structured}
\end{subfigure}
\begin{subfigure}[b]{0.24\textwidth}
\includegraphics[width=\linewidth]{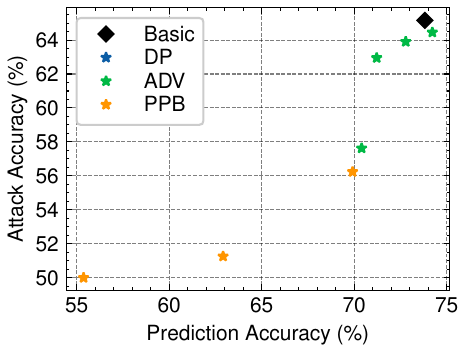}
\caption{L2 Structured}
\end{subfigure}
\begin{subfigure}[b]{0.24\textwidth}
\includegraphics[width=\linewidth]{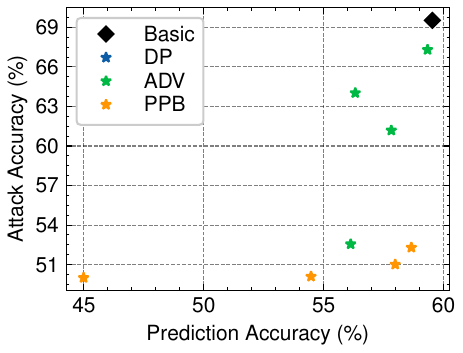}
\caption{Slimming}
\end{subfigure}
\caption{Performance of defenses (CIFAR10, DenseNet121, Sparsity 0.7).}
\end{figure*}

\begin{figure*}[!h]
\centering
\begin{subfigure}[b]{0.24\textwidth}
\includegraphics[width=\linewidth]{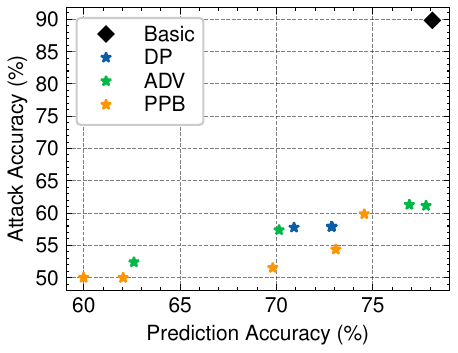}
\caption{L1 Unstructured}
\end{subfigure}
\begin{subfigure}[b]{0.24\textwidth}
\includegraphics[width=\linewidth]{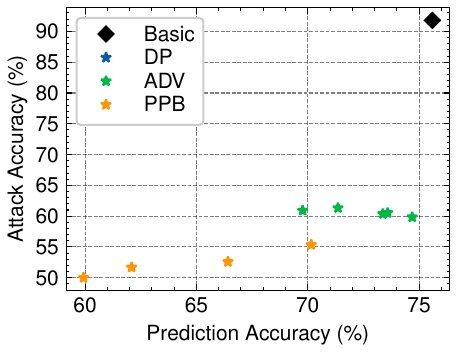}
\caption{L1 Structured}
\end{subfigure}
\begin{subfigure}[b]{0.24\textwidth}
\includegraphics[width=\linewidth]{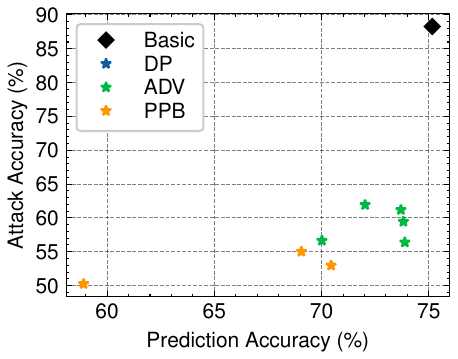}
\caption{L2 Structured}
\end{subfigure}
\begin{subfigure}[b]{0.24\textwidth}
\includegraphics[width=\linewidth]{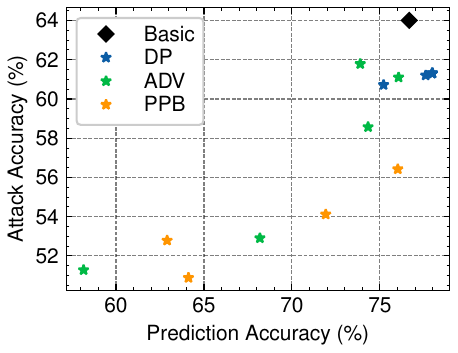}
\caption{Slimming}
\end{subfigure}
\caption{Performance of defenses (CIFAR10, VGG16, Sparsity 0.7).}
\end{figure*}

\begin{figure*}[!h]
\centering
\begin{subfigure}[b]{0.24\textwidth}
\includegraphics[width=\linewidth]{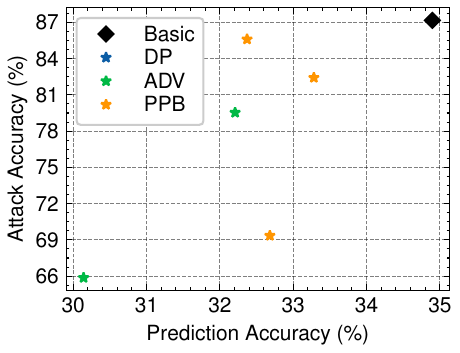}
\caption{L1 Unstructured}
\end{subfigure}
\begin{subfigure}[b]{0.24\textwidth}
\includegraphics[width=\linewidth]{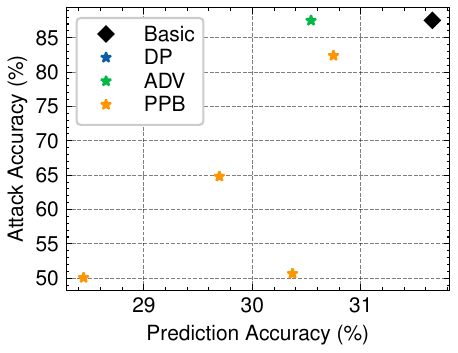}
\caption{L1 Structured}
\end{subfigure}
\begin{subfigure}[b]{0.24\textwidth}
\includegraphics[width=\linewidth]{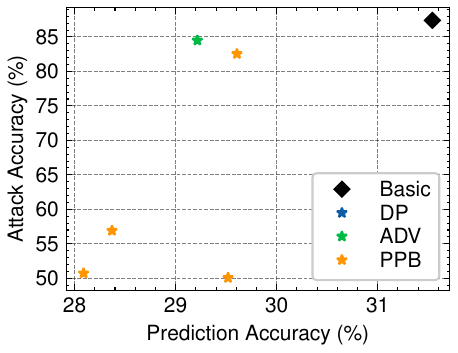}
\caption{L2 Structured}
\end{subfigure}
\begin{subfigure}[b]{0.24\textwidth}
\includegraphics[width=\linewidth]{figs/exp/defend/defend_cifar100_resnet18_slim_0.6.pdf}
\caption{Slimming}
\end{subfigure}
\caption{Performance of defenses (CIFAR100, ResNet18, Sparsity 0.6).}
\end{figure*}

\begin{figure*}[!h]
\centering
\begin{subfigure}[b]{0.24\textwidth}
\includegraphics[width=\linewidth]{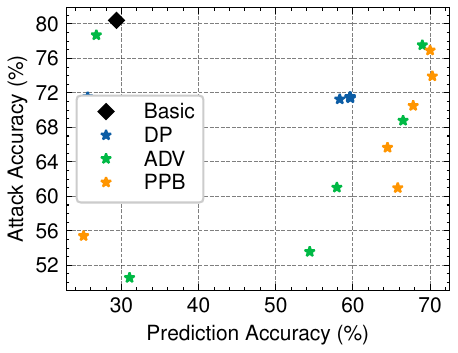}
\caption{L1 Unstructured}
\end{subfigure}
\begin{subfigure}[b]{0.24\textwidth}
\includegraphics[width=\linewidth]{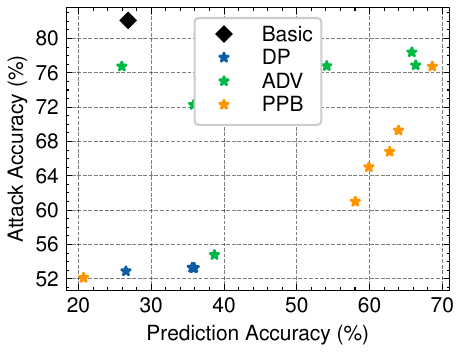}
\caption{L1 Structured}
\end{subfigure}
\begin{subfigure}[b]{0.24\textwidth}
\includegraphics[width=\linewidth]{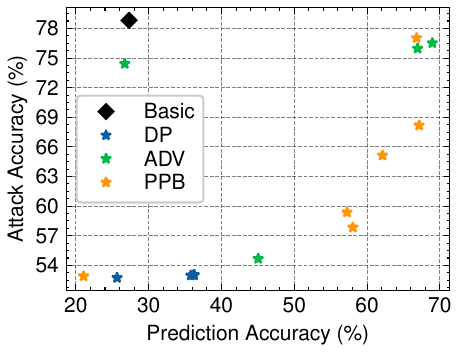}
\caption{L2 Structured}
\end{subfigure}
\begin{subfigure}[b]{0.24\textwidth}
\includegraphics[width=\linewidth]{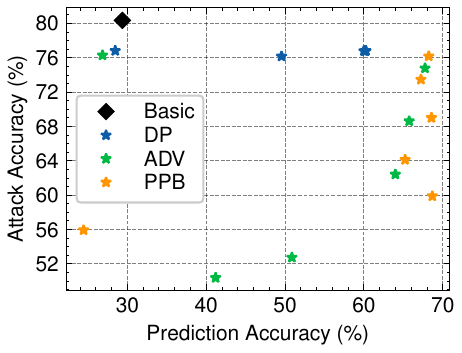}
\caption{Slimming}
\end{subfigure}
\caption{Performance of defenses (CIFAR100, VGG16, Sparsity 0.6).}

\end{figure*}

\begin{figure*}[!h]
\centering
\begin{subfigure}[b]{0.24\textwidth}
\includegraphics[width=\linewidth]{figs/exp/defend/defend_cifar100_densenet121_level_0.6.pdf}
\caption{L1 Unstructured}
\end{subfigure}
\begin{subfigure}[b]{0.24\textwidth}
\includegraphics[width=\linewidth]{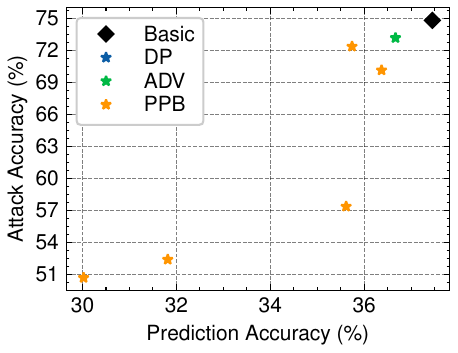}
\caption{L1 Structured}
\end{subfigure}
\begin{subfigure}[b]{0.24\textwidth}
\includegraphics[width=\linewidth]{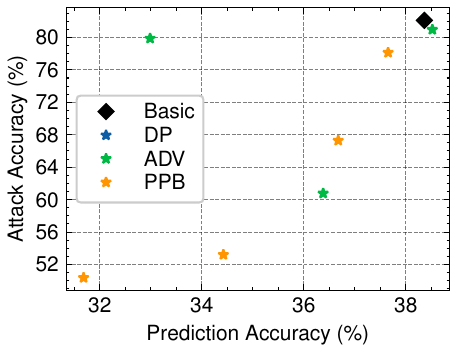}
\caption{L2 Structured}
\end{subfigure}
\begin{subfigure}[b]{0.24\textwidth}
\includegraphics[width=\linewidth]{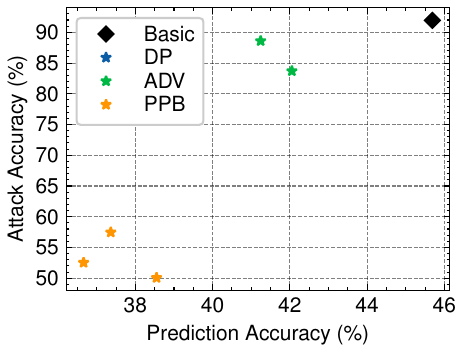}
\caption{Slimming}
\end{subfigure}
\caption{Performance of defenses (CIFAR100, DenseNet121, Sparsity 0.6).}
\end{figure*}

\begin{figure*}[!h]
\centering
\begin{subfigure}[b]{0.24\textwidth}
\includegraphics[width=\linewidth]{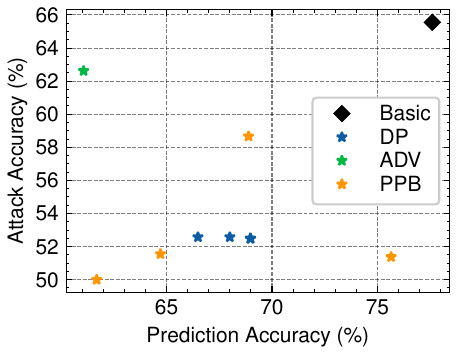}
\caption{L1 Unstructured}
\end{subfigure}
\begin{subfigure}[b]{0.24\textwidth}
\includegraphics[width=\linewidth]{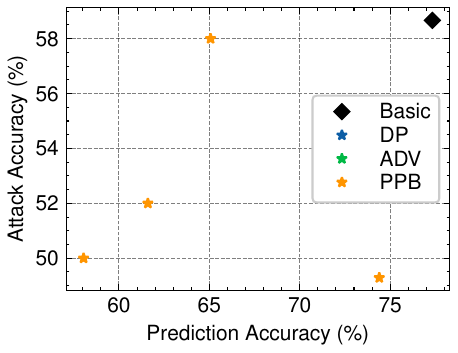}
\caption{L1 Structured}
\end{subfigure}
\begin{subfigure}[b]{0.24\textwidth}
\includegraphics[width=\linewidth]{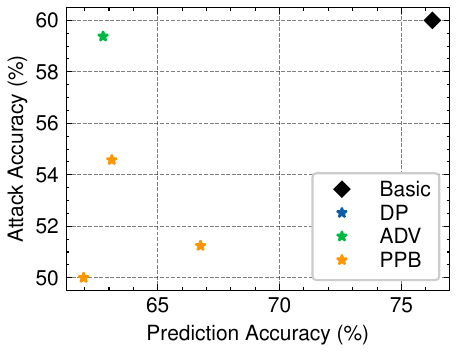}
\caption{L2 Structured}
\end{subfigure}
\begin{subfigure}[b]{0.24\textwidth}
\includegraphics[width=\linewidth]{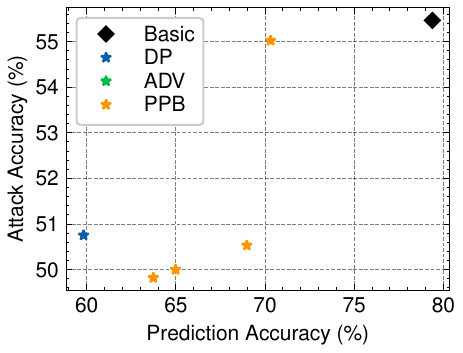}
\caption{Slimming}
\end{subfigure}
\caption{Performance of defenses (CHMNIST, ResNet18, Sparsity 0.6).}
\end{figure*}

\begin{figure*}[!h]
\centering
\begin{subfigure}[b]{0.24\textwidth}
\includegraphics[width=\linewidth]{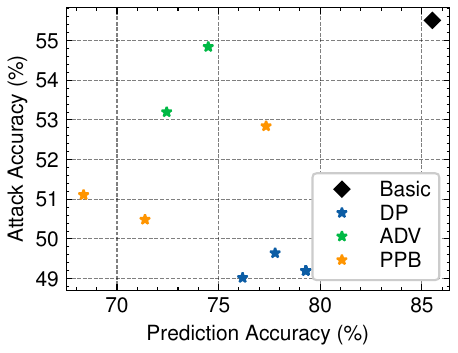}
\caption{L1 Unstructured}
\end{subfigure}
\begin{subfigure}[b]{0.24\textwidth}
\includegraphics[width=\linewidth]{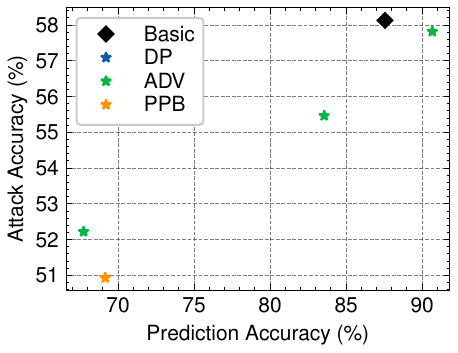}
\caption{L1 Structured}
\end{subfigure}
\begin{subfigure}[b]{0.24\textwidth}
\includegraphics[width=\linewidth]{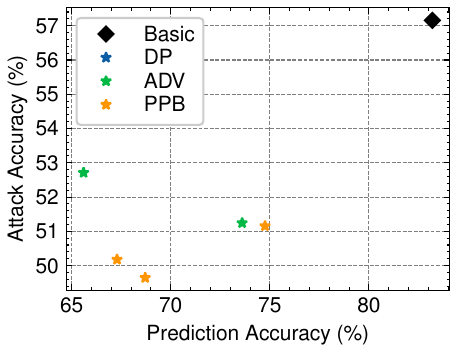}
\caption{L2 Structured}
\end{subfigure}
\begin{subfigure}[b]{0.24\textwidth}
\includegraphics[width=\linewidth]{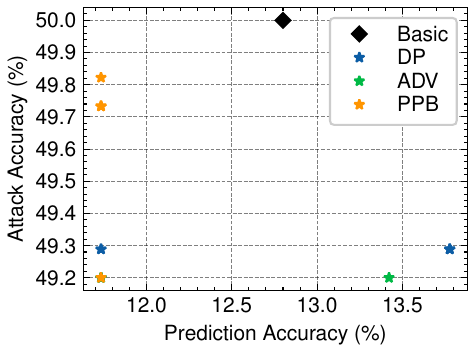}
\caption{Slimming}
\end{subfigure}
\caption{Performance of defenses (CHMNIST, DenseNet121, Sparsity 0.6).}
\end{figure*}

\begin{figure*}[!h]
\centering
\begin{subfigure}[b]{0.24\textwidth}
\includegraphics[width=\linewidth]{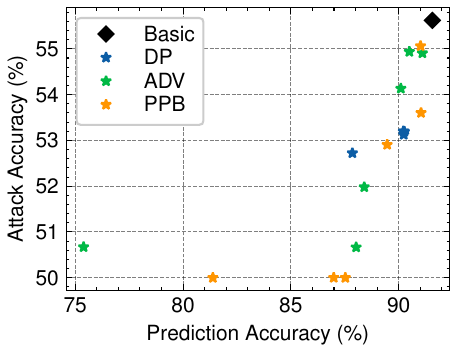}
\caption{L1 Unstructured}
\end{subfigure}
\begin{subfigure}[b]{0.24\textwidth}
\includegraphics[width=\linewidth]{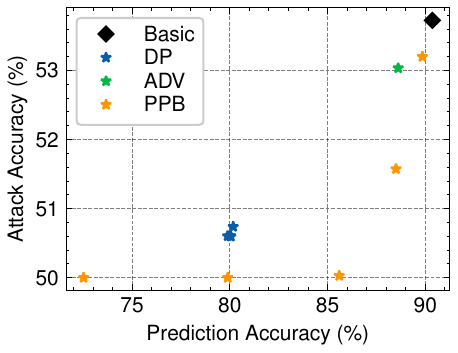}
\caption{L1 Structured}
\end{subfigure}
\begin{subfigure}[b]{0.24\textwidth}
\includegraphics[width=\linewidth]{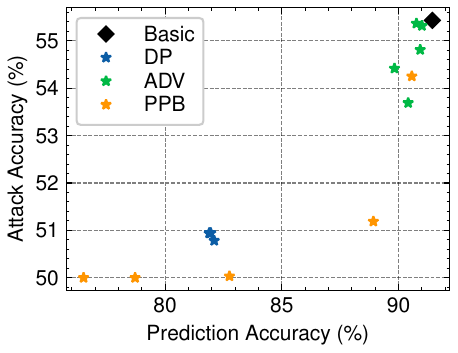}
\caption{L2 Structured}
\end{subfigure}
\begin{subfigure}[b]{0.24\textwidth}
\includegraphics[width=\linewidth]{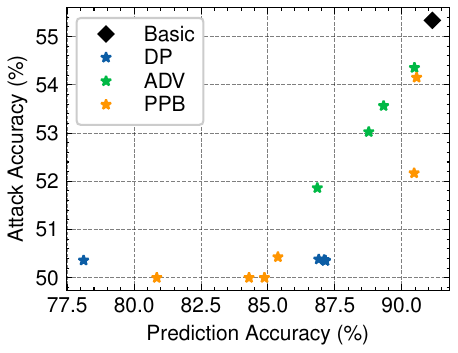}
\caption{Slimming}
\end{subfigure}
\caption{Performance of defenses (SVHN, ResNet18, Sparsity 0.6).}

\end{figure*}

\begin{figure*}[!h]
\centering
\begin{subfigure}[b]{0.24\textwidth}
\includegraphics[width=\linewidth]{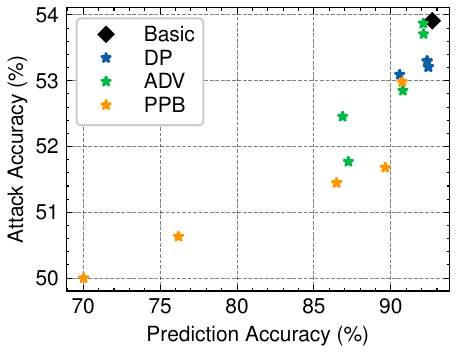}
\caption{L1 Unstructured}
\end{subfigure}
\begin{subfigure}[b]{0.24\textwidth}
\includegraphics[width=\linewidth]{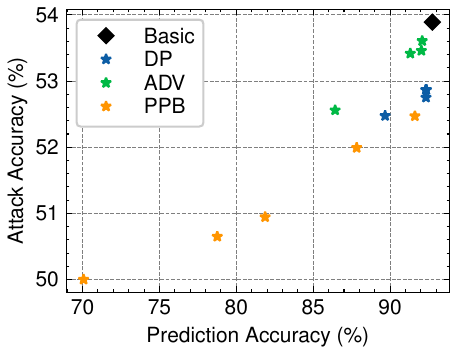}
\caption{L1 Structured}
\end{subfigure}
\begin{subfigure}[b]{0.24\textwidth}
\includegraphics[width=\linewidth]{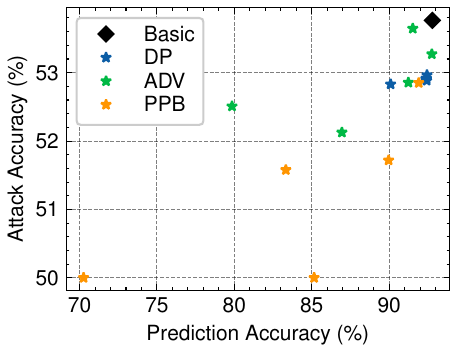}
\caption{L2 Structured}
\end{subfigure}
\begin{subfigure}[b]{0.24\textwidth}
\includegraphics[width=\linewidth]{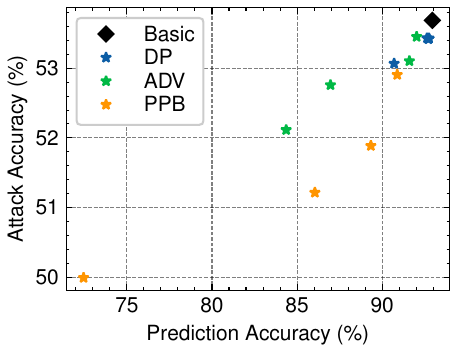}
\caption{Slimming}
\end{subfigure}
\caption{Performance of defenses (SVHN, VGG16, Sparsity 0.6).}

\end{figure*}

\begin{figure*}[!h]
\centering
\begin{subfigure}[b]{0.24\textwidth}
\includegraphics[width=\linewidth]{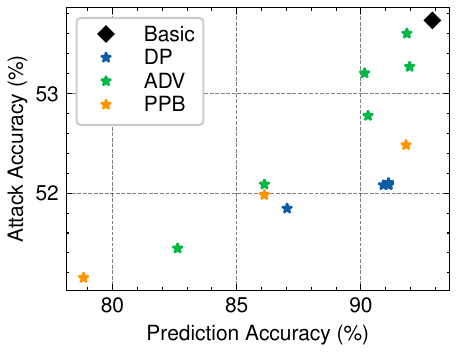}
\caption{L1 Unstructured}
\end{subfigure}
\begin{subfigure}[b]{0.24\textwidth}
\includegraphics[width=\linewidth]{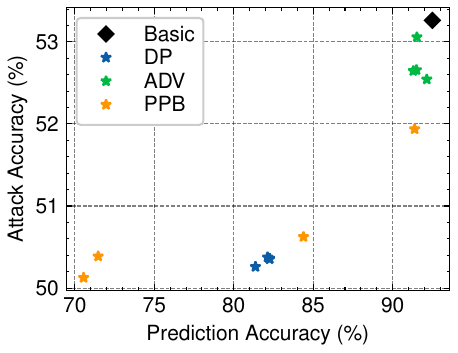}
\caption{L1 Structured}
\end{subfigure}
\begin{subfigure}[b]{0.24\textwidth}
\includegraphics[width=\linewidth]{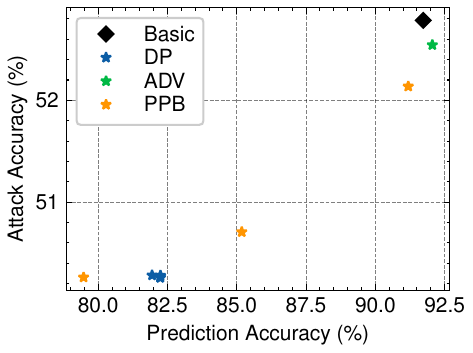}
\caption{L2 Structured}
\end{subfigure}
\begin{subfigure}[b]{0.24\textwidth}
\includegraphics[width=\linewidth]{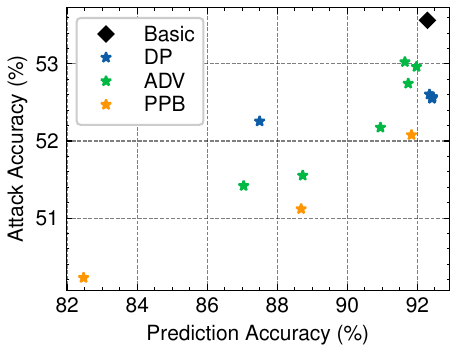}
\caption{Slimming}
\end{subfigure}
\caption{Performance of defenses (SVHN, DenseNet121, Sparsity 0.6).}
\end{figure*}

\clearpage
\begin{figure*}[!h]
\centering
\begin{subfigure}[b]{0.24\textwidth}
\includegraphics[width=\linewidth]{figs/exp/defend/defend_location_column_level_0.6.pdf}
\caption{Sparsity 0.6}
\end{subfigure}
\begin{subfigure}[b]{0.24\textwidth}
\includegraphics[width=\linewidth]{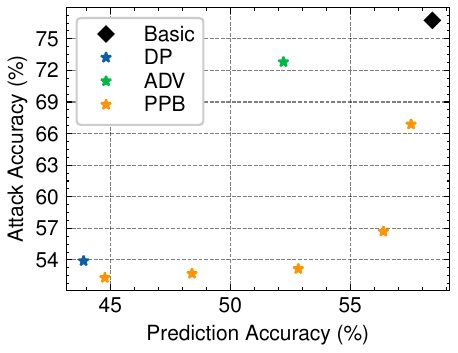}
\caption{Sparsity 0.7}
\end{subfigure}
\caption{Performance of defenses (Location, FC, L1 Unstructured).}
\end{figure*}

\begin{figure*}[!h]
\centering
\begin{subfigure}[b]{0.24\textwidth}
\includegraphics[width=\linewidth]{figs/exp/defend/defend_purchase_column_level_0.6.pdf}
\caption{Sparsity 0.6}
\end{subfigure}
\begin{subfigure}[b]{0.24\textwidth}
\includegraphics[width=\linewidth]{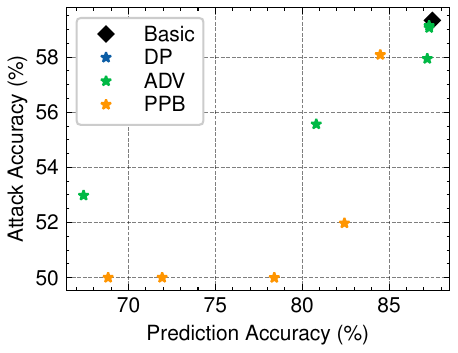}
\caption{Sparsity 0.7}
\end{subfigure}
\caption{Performance of defenses (Purchase, FC, L1 Unstructured).}
\end{figure*}

\begin{figure*}[!h]
\centering
\begin{subfigure}[b]{0.24\textwidth}
\includegraphics[width=\linewidth]{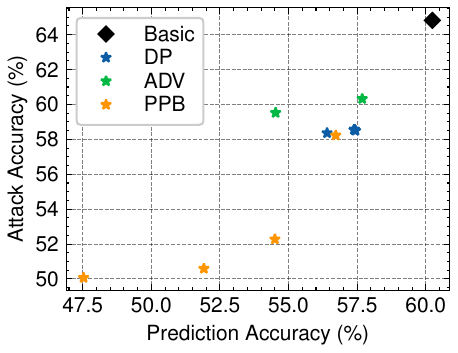}
\caption{Sparsity 0.6}
\end{subfigure}
\begin{subfigure}[b]{0.24\textwidth}
\includegraphics[width=\linewidth]{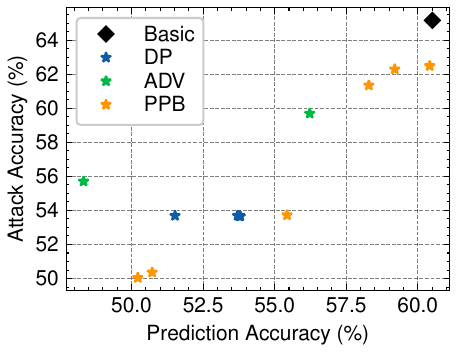}
\caption{Sparsity 0.7}
\end{subfigure}
\caption{Performance of defenses (Location, FC, L1 Unstructured).}
\end{figure*}

\clearpage
\subsection{Defense Performance Against Adaptive Attacks}
The proposed PPB defense can still reduce the attack accuracy by a significant degree when considering adaptive attacks. PPB and ADV are the best two defenses for L1 Unstructured and Slimming pruning. PPB performs the best for L1 Structured and L2 Structured pruning in most cases.

\begin{figure*}[!h]
\centering
\begin{subfigure}[b]{0.24\textwidth}
\includegraphics[width=\linewidth]{figs/exp/defend/defend_cifar10_densenet121_level_0.6_adp.pdf}
\caption{L1 Unstructured}
\end{subfigure}
\begin{subfigure}[b]{0.24\textwidth}
\includegraphics[width=\linewidth]{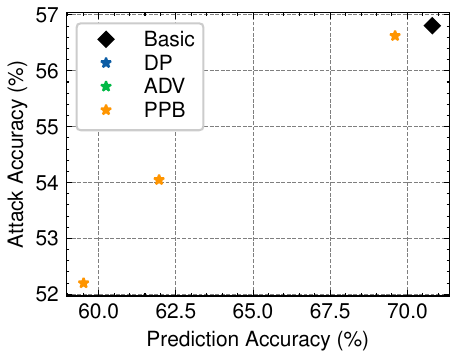}
\caption{L1 Structured}
\end{subfigure}
\begin{subfigure}[b]{0.24\textwidth}
\includegraphics[width=\linewidth]{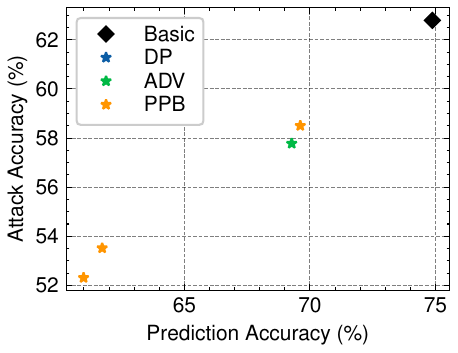}
\caption{L2 Structured}
\end{subfigure}
\begin{subfigure}[b]{0.24\textwidth}
\includegraphics[width=\linewidth]{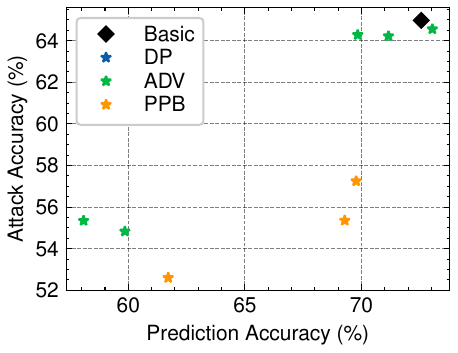}
\caption{Slimming}
\end{subfigure}
\caption{Performance of defenses against adaptive attacks (CIFAR10, DenseNet121, Sparsity 0.6).}
\end{figure*}

\begin{figure*}[!h]
\centering
\begin{subfigure}[b]{0.24\textwidth}
\includegraphics[width=\linewidth]{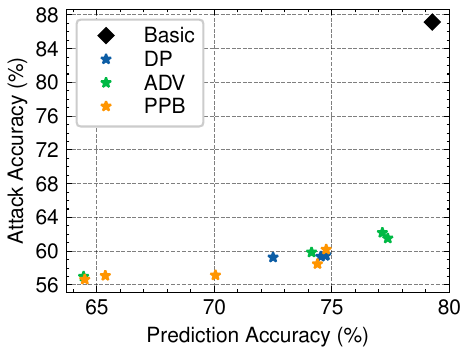}
\caption{L1 Unstructured}
\end{subfigure}
\begin{subfigure}[b]{0.24\textwidth}
\includegraphics[width=\linewidth]{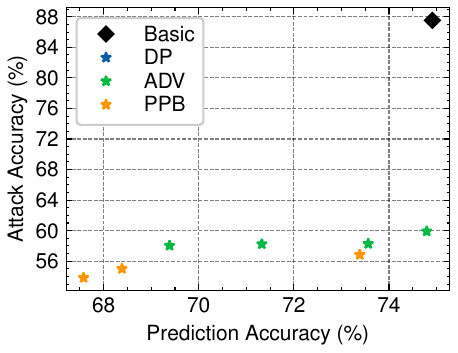}
\caption{L1 Structured}
\end{subfigure}
\begin{subfigure}[b]{0.24\textwidth}
\includegraphics[width=\linewidth]{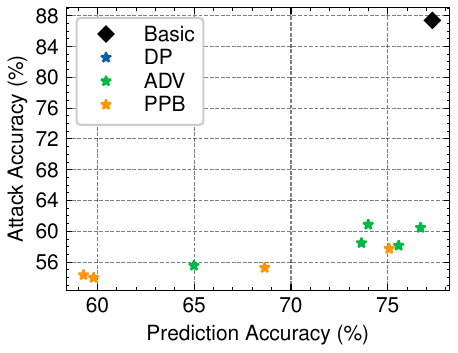}
\caption{L2 Structured}
\end{subfigure}
\begin{subfigure}[b]{0.24\textwidth}
\includegraphics[width=\linewidth]{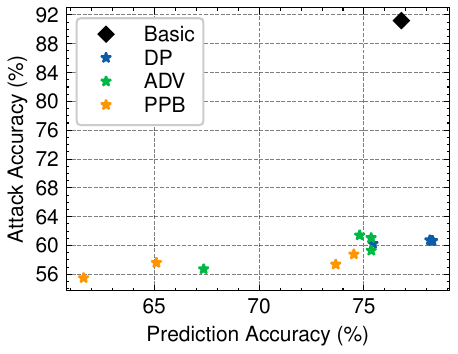}
\caption{Slimming}
\end{subfigure}
\caption{Performance of defenses against adaptive attacks (CIFAR10, VGG16, Sparsity 0.6).}
\end{figure*}

\begin{figure*}[!h]
\centering
\begin{subfigure}[b]{0.24\textwidth}
\includegraphics[width=\linewidth]{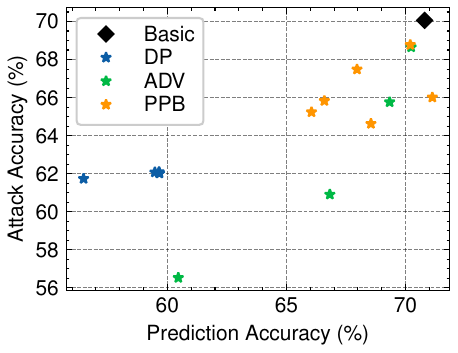}
\caption{L1 Unstructured}
\end{subfigure}
\begin{subfigure}[b]{0.24\textwidth}
\includegraphics[width=\linewidth]{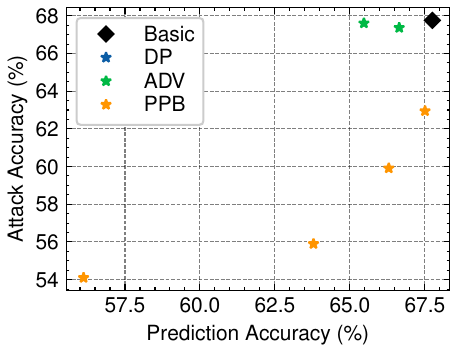}
\caption{L1 Structured}
\end{subfigure}
\begin{subfigure}[b]{0.24\textwidth}
\includegraphics[width=\linewidth]{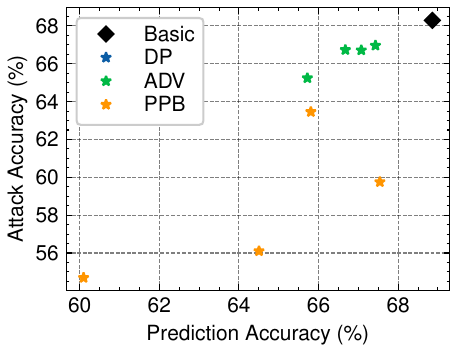}
\caption{L2 Structured}
\end{subfigure}
\begin{subfigure}[b]{0.24\textwidth}
\includegraphics[width=\linewidth]{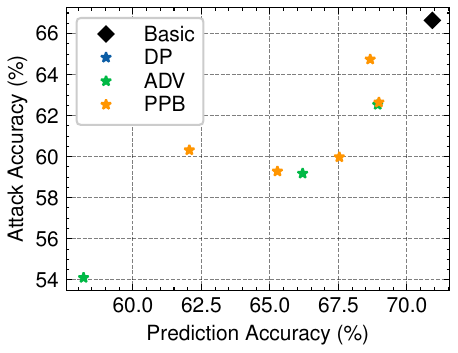}
\caption{Slimming}
\end{subfigure}
\caption{Performance of defenses against adaptive attacks (CIFAR10, ResNet18, Sparsity 0.7).}
\end{figure*}

\begin{figure*}[!h]
\centering
\begin{subfigure}[b]{0.24\textwidth}
\includegraphics[width=\linewidth]{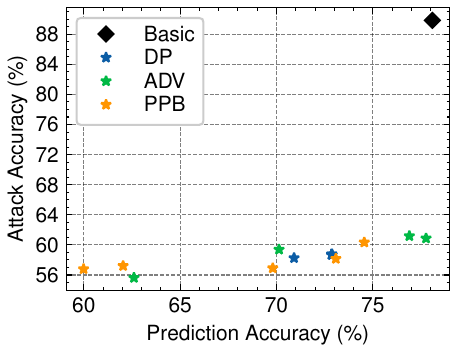}
\caption{L1 Unstructured}
\end{subfigure}
\begin{subfigure}[b]{0.24\textwidth}
\includegraphics[width=\linewidth]{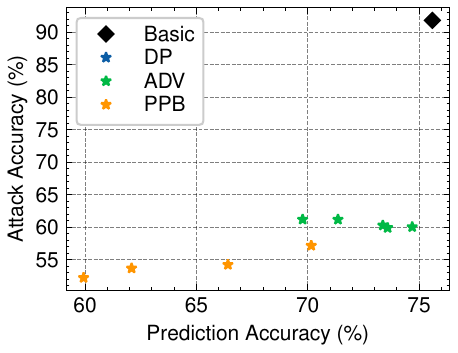}
\caption{L1 Structured}
\end{subfigure}
\begin{subfigure}[b]{0.24\textwidth}
\includegraphics[width=\linewidth]{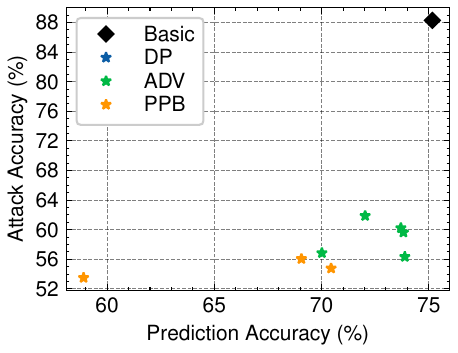}
\caption{L2 Structured}
\end{subfigure}
\begin{subfigure}[b]{0.24\textwidth}
\includegraphics[width=\linewidth]{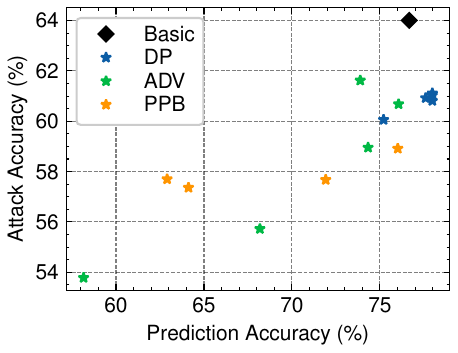}
\caption{Slimming}
\end{subfigure}
\caption{Performance of defenses against adaptive attacks (CIFAR10, VGG16, Sparsity 0.7).}
\end{figure*}

\begin{figure*}[!h]
\centering
\begin{subfigure}[b]{0.24\textwidth}
\includegraphics[width=\linewidth]{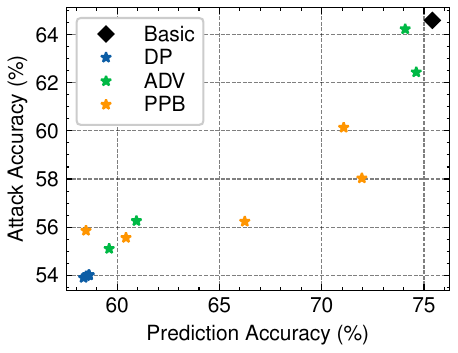}
\caption{L1 Unstructured}
\end{subfigure}
\begin{subfigure}[b]{0.24\textwidth}
\includegraphics[width=\linewidth]{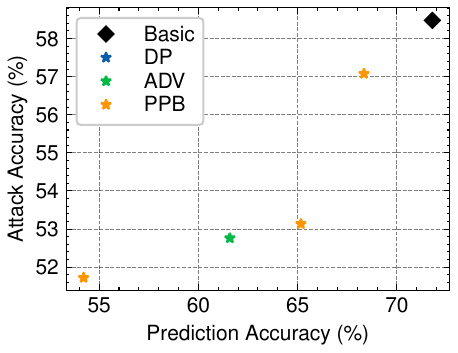}
\caption{L1 Structured}
\end{subfigure}
\begin{subfigure}[b]{0.24\textwidth}
\includegraphics[width=\linewidth]{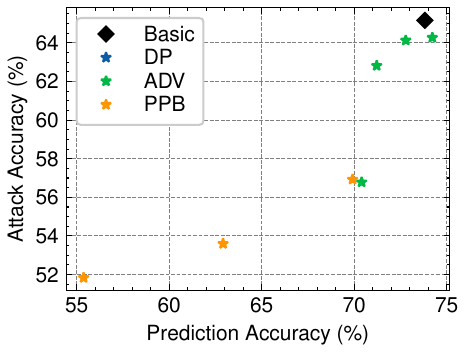}
\caption{L2 Structured}
\end{subfigure}
\begin{subfigure}[b]{0.24\textwidth}
\includegraphics[width=\linewidth]{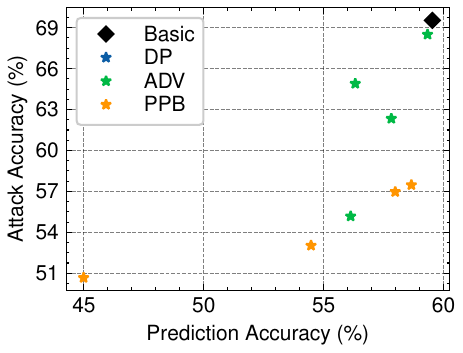}
\caption{Slimming}
\end{subfigure}
\caption{Performance of defenses against adaptive attacks (CIFAR10, DenseNet121, Sparsity 0.7).}
\end{figure*}

\begin{figure*}[!h]
\centering
\begin{subfigure}[b]{0.24\textwidth}
\includegraphics[width=\linewidth]{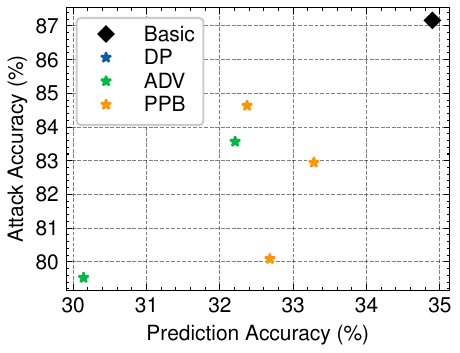}
\caption{L1 Unstructured}
\end{subfigure}
\begin{subfigure}[b]{0.24\textwidth}
\includegraphics[width=\linewidth]{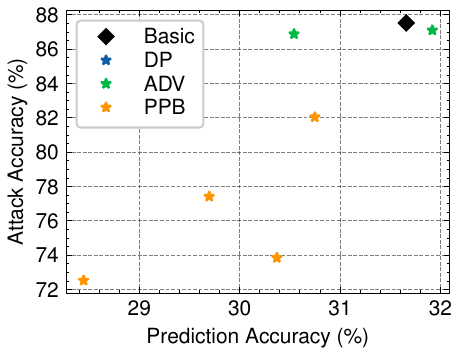}
\caption{L1 Structured}
\end{subfigure}
\begin{subfigure}[b]{0.24\textwidth}
\includegraphics[width=\linewidth]{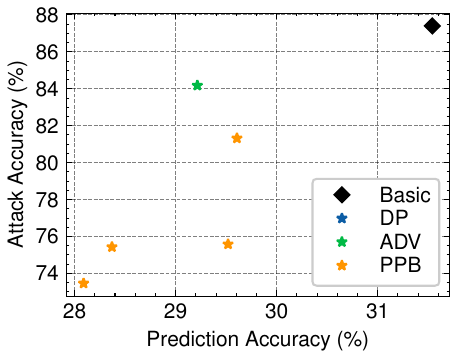}
\caption{L2 Structured}
\end{subfigure}
\begin{subfigure}[b]{0.24\textwidth}
\includegraphics[width=\linewidth]{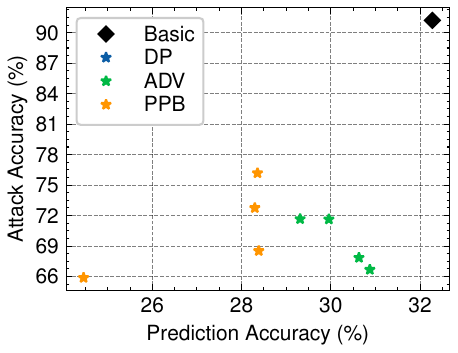}
\caption{Slimming}
\end{subfigure}
\caption{Performance of defenses against adaptive attacks (CIFAR100, ResNet18, Sparsity 0.6).}

\end{figure*}

\begin{figure*}[!h]
\centering
\begin{subfigure}[b]{0.24\textwidth}
\includegraphics[width=\linewidth]{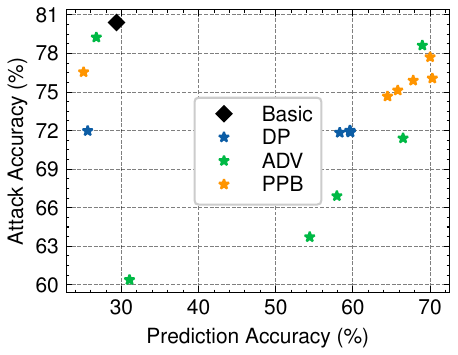}
\caption{L1 Unstructured}
\end{subfigure}
\begin{subfigure}[b]{0.24\textwidth}
\includegraphics[width=\linewidth]{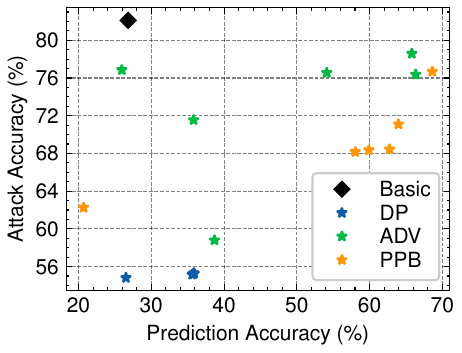}
\caption{L1 Structured}
\end{subfigure}
\begin{subfigure}[b]{0.24\textwidth}
\includegraphics[width=\linewidth]{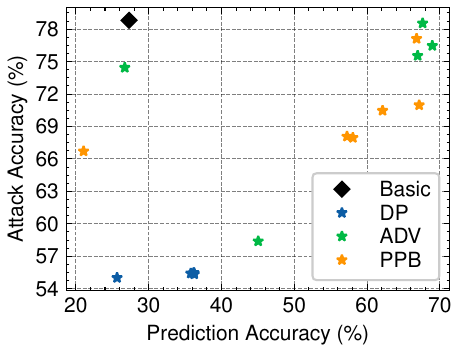}
\caption{L2 Structured}
\end{subfigure}
\begin{subfigure}[b]{0.24\textwidth}
\includegraphics[width=\linewidth]{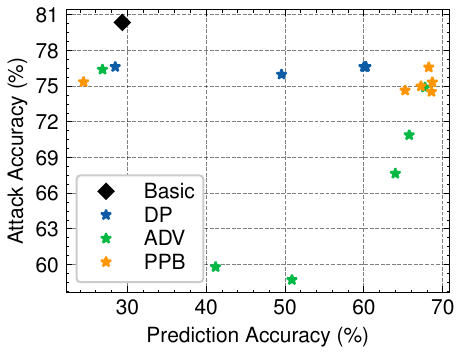}
\caption{Slimming}
\end{subfigure}
\caption{Performance of defenses against adaptive attacks (CIFAR100, VGG16, Sparsity 0.6).}

\end{figure*}

\begin{figure*}[!h]
\centering
\begin{subfigure}[b]{0.24\textwidth}
\includegraphics[width=\linewidth]{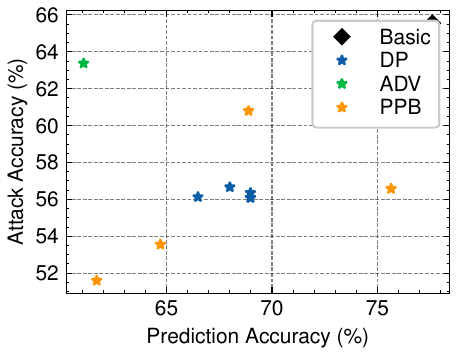}
\caption{L1 Unstructured}
\end{subfigure}
\begin{subfigure}[b]{0.24\textwidth}
\includegraphics[width=\linewidth]{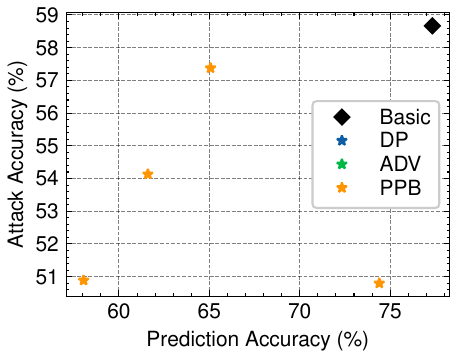}
\caption{L1 Structured}
\end{subfigure}
\begin{subfigure}[b]{0.24\textwidth}
\includegraphics[width=\linewidth]{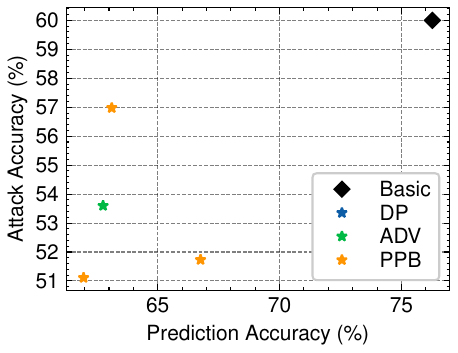}
\caption{L2 Structured}
\end{subfigure}
\begin{subfigure}[b]{0.24\textwidth}
\includegraphics[width=\linewidth]{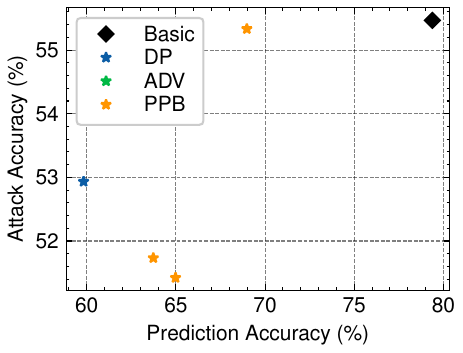}
\caption{Slimming}
\end{subfigure}
\caption{Performance of defenses against adaptive attacks (CHMNIST, ResNet18, Sparsity 0.6).}
\end{figure*}

\begin{figure*}[!h]
\centering
\begin{subfigure}[b]{0.24\textwidth}
\includegraphics[width=\linewidth]{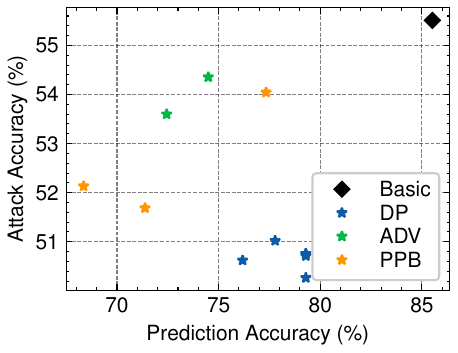}
\caption{L1 Unstructured}
\end{subfigure}
\begin{subfigure}[b]{0.24\textwidth}
\includegraphics[width=\linewidth]{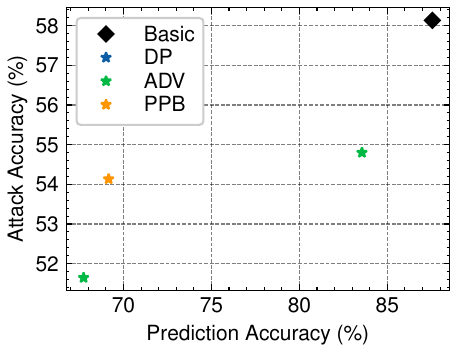}
\caption{L1 Structured}
\end{subfigure}
\begin{subfigure}[b]{0.24\textwidth}
\includegraphics[width=\linewidth]{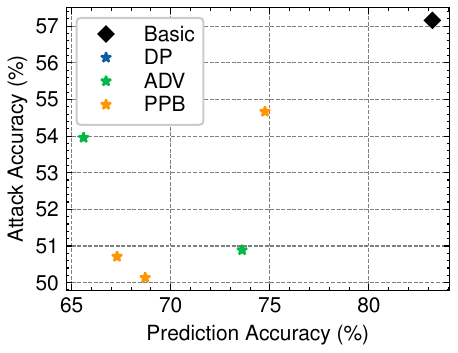}
\caption{L2 Structured}
\end{subfigure}
\begin{subfigure}[b]{0.24\textwidth}
\includegraphics[width=\linewidth]{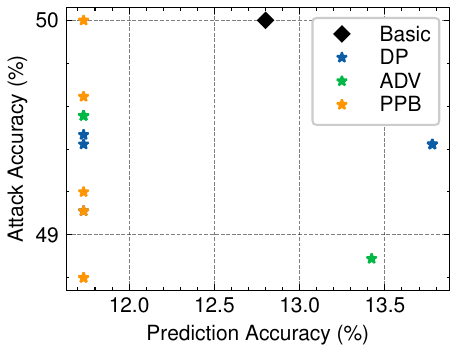}
\caption{Slimming}
\end{subfigure}
\caption{Performance of defenses against adaptive attacks (CHMNIST, DenseNet121, Sparsity 0.6).}
\end{figure*}

\begin{figure*}[!h]
\centering
\begin{subfigure}[b]{0.24\textwidth}
\includegraphics[width=\linewidth]{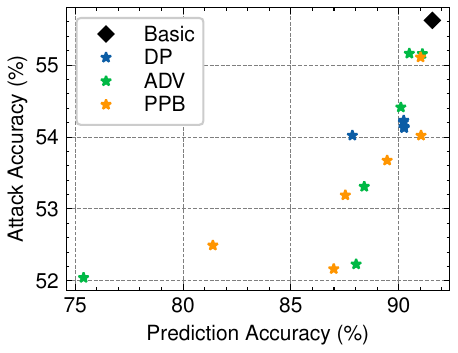}
\caption{L1 Unstructured}
\end{subfigure}
\begin{subfigure}[b]{0.24\textwidth}
\includegraphics[width=\linewidth]{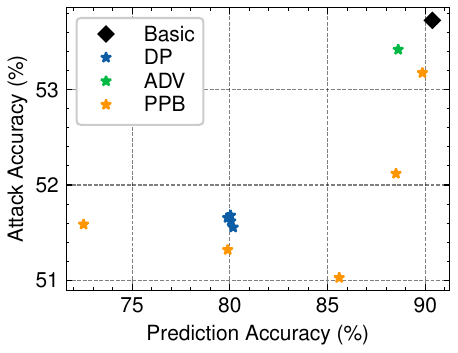}
\caption{L1 Structured}
\end{subfigure}
\begin{subfigure}[b]{0.24\textwidth}
\includegraphics[width=\linewidth]{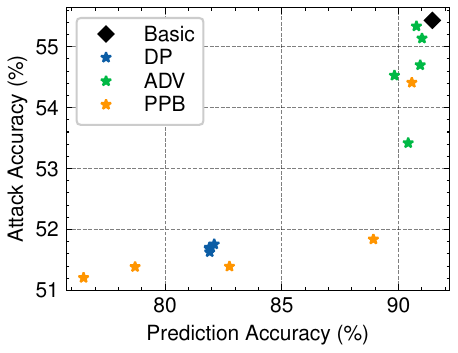}
\caption{L2 Structured}
\end{subfigure}
\begin{subfigure}[b]{0.24\textwidth}
\includegraphics[width=\linewidth]{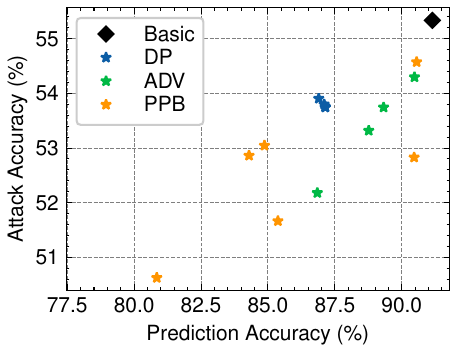}
\caption{Slimming}
\end{subfigure}
\caption{Performance of defenses against adaptive attacks (SVHN, ResNet18, Sparsity 0.6).}

\end{figure*}

\begin{figure*}[!h]
\centering
\begin{subfigure}[b]{0.24\textwidth}
\includegraphics[width=\linewidth]{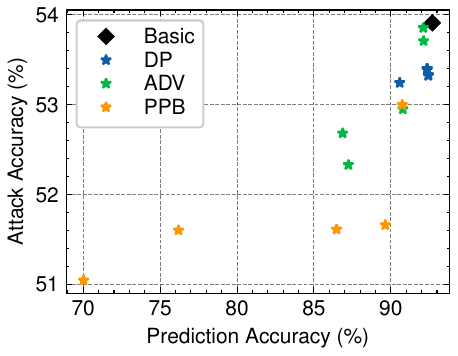}
\caption{L1 Unstructured}
\end{subfigure}
\begin{subfigure}[b]{0.24\textwidth}
\includegraphics[width=\linewidth]{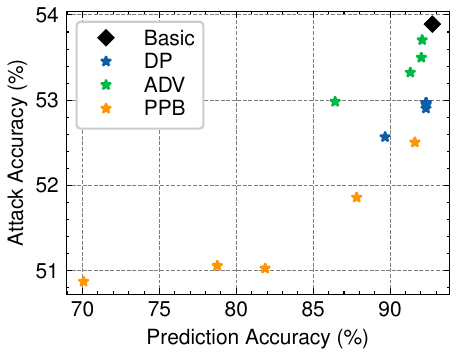}
\caption{L1 Structured}
\end{subfigure}
\begin{subfigure}[b]{0.24\textwidth}
\includegraphics[width=\linewidth]{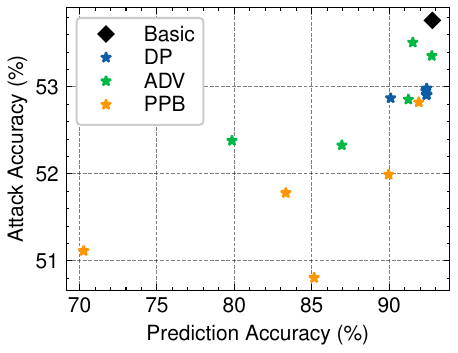}
\caption{L2 Structured}
\end{subfigure}
\begin{subfigure}[b]{0.24\textwidth}
\includegraphics[width=\linewidth]{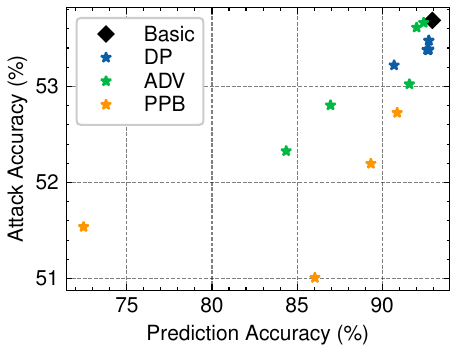}
\caption{Slimming}
\end{subfigure}
\caption{Performance of defenses against adaptive attacks (SVHN, VGG16, Sparsity 0.6).}

\end{figure*}

\begin{figure*}[!h]
\centering
\begin{subfigure}[b]{0.24\textwidth}
\includegraphics[width=\linewidth]{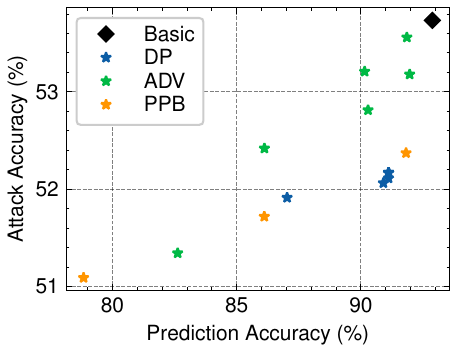}
\caption{L1 Unstructured}
\end{subfigure}
\begin{subfigure}[b]{0.24\textwidth}
\includegraphics[width=\linewidth]{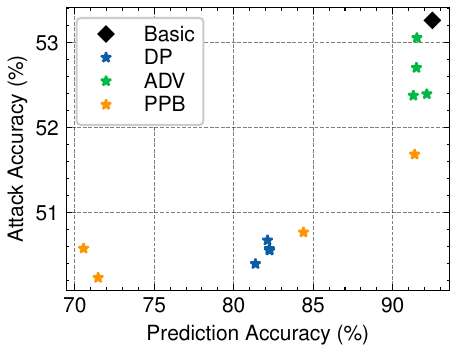}
\caption{L1 Structured}
\end{subfigure}
\begin{subfigure}[b]{0.24\textwidth}
\includegraphics[width=\linewidth]{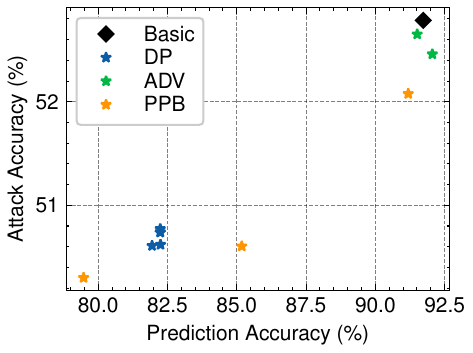}
\caption{L2 Structured}
\end{subfigure}
\begin{subfigure}[b]{0.24\textwidth}
\includegraphics[width=\linewidth]{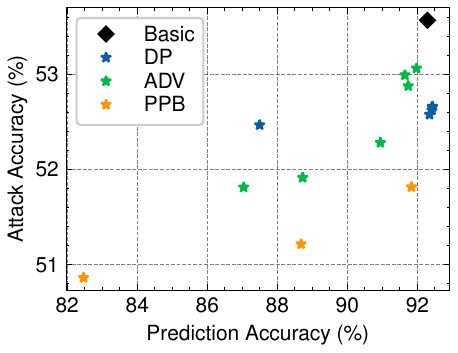}
\caption{Slimming}
\end{subfigure}
\caption{Performance of defenses against adaptive attacks (SVHN, DenseNet121, Sparsity 0.6).}
\end{figure*}

\begin{figure*}[!h]
\centering
\begin{subfigure}[b]{0.24\textwidth}
\includegraphics[width=\linewidth]{figs/exp/defend/defend_location_column_level_0.6_adp.pdf}
\caption{Sparsity 0.6}
\end{subfigure}
\begin{subfigure}[b]{0.24\textwidth}
\includegraphics[width=\linewidth]{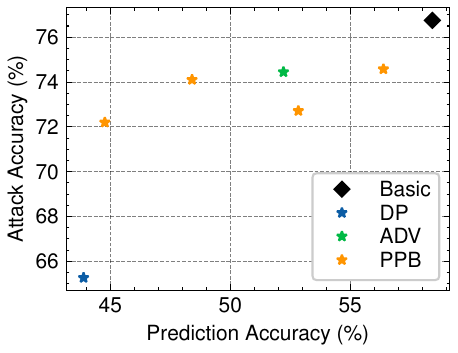}
\caption{Sparsity 0.7}
\end{subfigure}
\caption{Performance of defenses against adaptive attacks (Location, FC, L1 Unstructured).}
\end{figure*}

\begin{figure*}[!h]
\centering
\begin{subfigure}[b]{0.24\textwidth}
\includegraphics[width=\linewidth]{figs/exp/defend/defend_purchase_column_level_0.6_adp.pdf}
\caption{Sparsity 0.6}
\end{subfigure}
\begin{subfigure}[b]{0.24\textwidth}
\includegraphics[width=\linewidth]{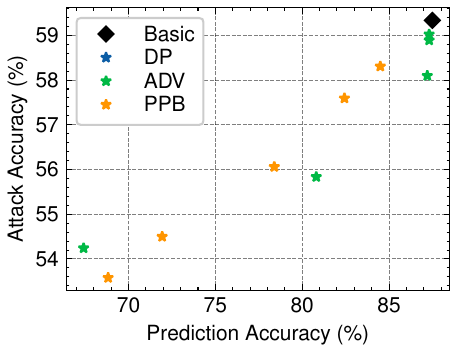}
\caption{Sparsity 0.7}
\end{subfigure}
\caption{Performance of defenses against adaptive attacks (Purchase, FC, L1 Unstructured).}
\end{figure*}

\begin{figure*}[!h]
\centering
\begin{subfigure}[b]{0.24\textwidth}
\includegraphics[width=\linewidth]{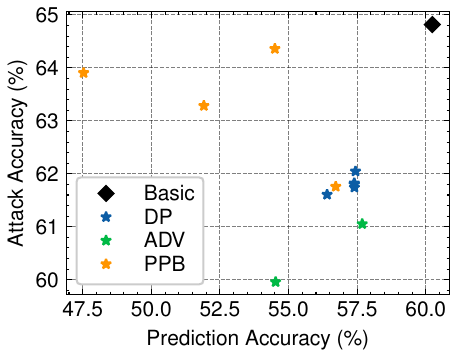}
\caption{Sparsity 0.6}
\end{subfigure}
\begin{subfigure}[b]{0.24\textwidth}
\includegraphics[width=\linewidth]{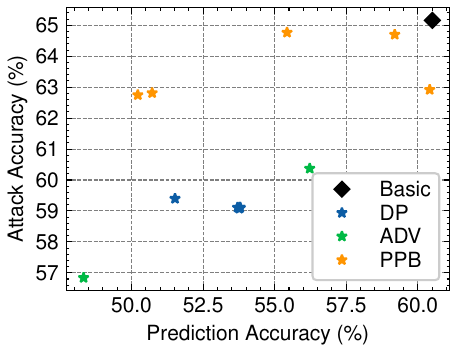}
\caption{Sparsity 0.7}
\end{subfigure}
\caption{Performance of defenses against adaptive attacks (Location, FC, L1 Unstructured).}
\end{figure*}

\end{document}